\documentclass{article}
\usepackage{arxiv}
\usepackage{amsthm,amsmath}
\usepackage[utf8]{inputenc}
\usepackage{hyperref}
\usepackage{mathpazo}
\usepackage{graphicx}
\usepackage{caption}
\usepackage{soul}

\newcommand{\R}{\mathbb{R}}

\newcommand{\N}{\mathcal{N}}

\renewcommand{\L}{\mathcal{L}}

\newcommand{\vv}{\mathbf{v}}
\newcommand{\y}{\mathbf{y}}
\newcommand{\x}{\mathbf{x}}

\newcommand{\X}{\mathbf{X}}

\DeclareMathOperator{\corr}{corr}

\newcommand{\rfurl}{\url{https://github.com/michoel-lab/Reverse-Pred-GWAS}}
\newcommand{\dreamurl}{\url{https://www.synapse.org/\#!Synapse:syn2820440/wiki/}}

\title{High-dimensional multi-trait GWAS by reverse prediction of genotypes using machine learning methods}

\author{Muhammad Ammar Malik\\
Department of Informatics\\
University of Bergen\\
PO Box 7803, 5020 Bergen, Norway\\
\texttt{muhammad.malik@uib.no}\\ 
\And Adriaan Ludl\\
Department of Informatics\\
University of Bergen\\
PO Box 7803, 5020 Bergen, Norway\\
\texttt{adriaan.ludl@uib.no} 
\And Tom Michoel\thanks{Corresponding author}\\ 
Department of Informatics\\
University of Bergen\\
PO Box 7803, 5020 Bergen, Norway\\
\texttt{tom.michoel@uib.no}\\}

\renewcommand{\shorttitle}{Reverse Prediction of Genotypes using ML}

\begin{document}

\maketitle


\begin{abstract}
\textbf{Motivation:} Multi-trait genome-wide association studies (GWAS) use multi-variate statistical methods to identify associations between genetic variants and multiple correlated traits simultaneously, and have higher statistical power than independent univariate analyses of traits. Reverse regression, where genotypes of genetic variants are regressed on multiple traits simultaneously, has emerged as a promising approach to perform multi-trait GWAS in high-dimensional settings where the number of traits exceeds the number of samples.

\textbf{Results:} We analyzed different machine learning methods (ridge regression, naive Bayes/independent univariate, random forests  and support vector machines) for  reverse regression in multi-trait GWAS, using genotypes, gene expression data and ground-truth transcriptional regulatory networks from the DREAM5 SysGen Challenge and from a cross between two yeast strains to evaluate methods. We found that genotype prediction performance, in terms of root mean squared error (RMSE), allowed to distinguish between genomic regions with high and low transcriptional activity. Moreover, model feature coefficients correlated with the strength of association between variants and individual traits, and were  predictive of true trans-eQTL target genes, with complementary findings across methods.

\textbf{Availability:} Code to reproduce the analysis is available at \rfurl.
\end{abstract}

\section{Background}

Genome-wide association studies (GWAS) aim to find statistical associations between genetic variants and traits of interest using data from a large number of individuals \cite{mccarthy2008genome,manolio2013bringing}. When multiple correlated traits are studied simultaneously, joint, multi-trait approaches can be more advantageous than studying the traits individually, due to increased power from taking into account cross-trait covariances and reduced multiple-testing burden by performing a single test for association to a set of traits \cite{allison1998multiple, ferreira2009multivariate, galesloot2014comparison, van2019genetic}.

The most commonly used multi-trait GWAS approaches are based on a multivariate analysis of variance (MANOVA) or canonical correlation analysis (CCA) \cite{ferreira2009multivariate}. However, these are applicable only to studies where the number of traits is relatively small, especially in comparison to the number of samples. When analyzing the effects of genetic variants on molecular traits (gene or protein expression levels, metabolite concentrations) or imaging features, we have to deal with a large number, often an order of magnitude or more greater than the sample size, of correlated traits simultaneously. For such studies, the standard procedure is still to conduct univariate linear regression or ANOVA tests for each genetic variant against each trait separately. While efficient algorithms exist to undertake this task \cite{shabalin2012matrix,qi2014,ongen2015fast}, the massive multiple-testing problem results in a significant loss of statistical power.

An alternative approach to multi-trait GWAS has been to reverse the functional relation between genotypes and traits, and fit a multivariate regression model that predicts genotypes from multiple traits simultaneously, instead of the usual approach to regress traits on genotypes. The first study to do this explicitly used logistic regression and showed a significant increase in power compared to univariate methods, without being dependent on assuming normally distributed genotypes like MANOVA or CCA \cite{o2012multiphen}. Although the method as presented in \cite{o2012multiphen} is still only valid when the number of traits is small, extending multivariate regression methods to high-dimensional settings is  straightforward. Thus a recent study used L2-regularized linear regression of single nucleotide polymorphisms (SNPs) on gene expression traits to identify trans-acting expression quantitative trait loci (trans-eQTLs), and showed that this approach aggregates evidence from many small trans-effects while being unaffected by strong expression correlations \cite{banerjee2020reverse}. In a very different application domain, regularized regression of SNP genotypes on longitudinal image phenotypes was used to identify time-dependent genetic associations with imaging phenotypes \cite{wang2012phenotype}.

Despite these advances, several limitations and open questions remain unanswered in high-dimensional GWAS. Firstly, linear models search for the linear combination of traits that is most strongly associated to the genetic variant, but there is no \textit{a priori} biological reason why only linear combinations should be considered. Secondly, while L2-regularization allows to deal with high-dimensional traits, it does not address the problem of variable selection. For instance, in the case of gene expression, we expect that trans-eQTLs are potentially associated with \emph{many}, but not \emph{all} genes. Indeed, in \cite{banerjee2020reverse} a secondary set of univariate tests is carried out to select genes associated to trans-eQTLs identified by the initial multi-variate regression. Thirdly, a systematic biological validation and comparison of the available methods is lacking.

Here we address these questions by considering a wider range of machine learning methods (in particular, random forests (RF) and support vector machines (SVM)) for reverse genotype prediction from gene expression traits. Hypothesizing that true trans-eQTL associations are mediated by transcription regulatory networks, we use simulated data from the DREAM5 Systems Genetics Challenge, and real data from 1,012 segregants of a cross between two budding yeast strains \cite{albert2018genetics} together with the YEASTRACT database of known transcriptional interactions \cite{monteiro2020yeastract}, to validate and compare these methods against univariate and L2-regularized linear regression.

\section{Approach}
As in other multi-trait GWAS methods, we consider one genetic variant at a time, and represent it by a random variable $Y$. We consider $p$ traits represented by random variables $X_1, X_2,\dots, X_p$ taking real values. We define  a ``forward" multi-trait association model probabilistically through a conditional distribution $p(X_1,\dots,X_p \mid Y)$, which corresponds to the natural direction where variation in $Y$ causes variation in the $X_i$. Using Bayes' formula, we can write the same model in the reverse causal direction using $Y$ as the dependent variable:
\begin{equation}\label{eq:1}
  P(Y\mid X_1,\dots,X_p) = P(X_1,\dots,X_p\mid Y) \frac{P(Y)}{P(X_1,\dots,X_p)}
\end{equation}
where $P(Y)$ and $P(X_1,\dots,X_p)$ are prior distributions. Conversely, a forward model $P(X_1,\dots,X_p\mid Y)$ can be obtained from a reverse model $P(Y\mid X_1,\dots,X_p)$ using the same formula.

We have data in the form of independent random samples from the joint distribution $P(Y,X_1,\dots,X_p)$ in $n$ individuals, represented by a genotype  vector $\y\in\R^n$ and trait vectors $\x_1,\x_2,\dots,\x_p \in\R^n$, which we gather in a matrix $\X=(\x_1,\x_2,\dots,\x_p)\in\R^{n\times p}$. The log-likelihood of observing the data is the log-probability density
\begin{align*}
  \L = \log p(\y,\X) &=  \log \prod_{j=1}^n p(y_j,x_{j1},\dots,x_{jp}) \\
  &= \sum_{j=1}^n \log p(y_j,x_{j1},\dots,x_{jp}),
\end{align*}
which can be expressed in terms of the forward or reverse conditional probabilities depending on the type of model being fit. We now review how existing as well as newly proposed, and low-dimensional as well as high-dimensional multi-trait GWAS methods fit within this framework.

\subsection{\textbf{Univariate tests}}
\label{sec:univariate-tests}

The simplest method for multi-trait GWAS in the high-dimensional setting consists of testing each trait for association with the genetic variant  independently. In this case we fit, by maximum-likelihood, a model $p(x_i\mid y)$ for each trait $X_i$ independently using a linear model
\begin{align*}
  p(x_i\mid y) = \N(\mu_y,\sigma^2)
\end{align*}
a normal distribution with mean $\mu_y$ dependent on the genotype value $y$. This corresponds to the multi-trait model
\begin{align*}
  p(x_1,\dots,x_p\mid y) = \prod_{i=1}^p p(x_i\mid y)
\end{align*}
Using Bayes' rule eq.~\eqref{eq:1}, we obtain
\begin{align*}
  P(y\mid x_1,\dots,x_p) &= p(x_1,\dots,x_p\mid y) \frac{P(y)}{p(x_1,\dots,x_p)}\\
  &\propto P(y) \prod_{i=1}^p p(x_i\mid y)
\end{align*}
where $P(y)$ is the prior probability (background frequency) of observing genotype class $y$. This is the formula for a \emph{naive Bayes classifier} of the genotype $y$ given features $x_i$. In the univariate approach, statistical tests are carried out to determine whether a genotype-dependent model $p(x_i\mid y)$ is more likely or not than a model where the trait is independent of the genotype. This is equivalent to doing a feature selection to determine which traits to include in the naive Bayes classifier.

\subsection{\textbf{Canonical correlation analysis}}
\label{sec:mv-plink-method}

MV-PLINK \cite{ferreira2009multivariate} is a multivariate method based on Canonical Correlation Analysis (CCA). Given two sets of random variables $(X_1,X_2,\dots, X_p)$ and \newline $(Y_1,Y_2,\dots,Y_q)$, CCA finds linear coefficients $\mathbf{a}\in\R^p$ and $\mathbf{b}\in\R^q$ that maximize the correlation
\begin{align*}
  \rho(\mathbf{a},\mathbf{b}) = \corr\left(\sum_{i=1}^p a_iX_i,\sum_{j=1}^qb_jY_j\right)
\end{align*}
It can be shown (see SI Section~ S1) that if $q=1$, then the maximizing coefficients $\mathbf{\hat a}$ are given by $\mathbf{\hat a} = (\X^T\X)^{-1} \X^T\y$, 
where $\X$ and $\y$ are the data sampled from the joint distribution $P(Y,X_1,X_2,\dots, X_p)$. These are the same coefficients that would be obtained from a \emph{linear regression} model where $Y$ is modelled as a linear function of the predictors $(X_1,X_2,\dots, X_p)$, or from the maximum-likelihood solution of a reverse probabilistic model
\begin{equation}\label{eq:2}
  p(y\mid x_1,\dots,x_p) = \N\left(\sum_{i=1}^pa_ix_i,\sigma^2\right).
\end{equation}

\subsection{\textbf{Reverse logistic regression}}
\label{sec:multiphen-method}

MultiPhen \cite{o2012multiphen} is a method that is described directly in terms of a model to predict genotypes from multiple traits, using proportional odds \emph{logistic regression}, that is, instead of fitting the genotype class probabilities $P(y=m\mid x_1,\dots, x_p)$, for $m=0,1,2$ (for biallelic data), the method fits
\begin{align*}
  P(y\leq m\mid x_1,\dots, x_p) = \frac1{1+e^{-\alpha_m-\sum_{i=1}^p \beta_i x_i}}
\end{align*}
 Then a likelihood ratio test is used to determine if this model fits the data better than a model where $\beta_1=\dots=\beta_p=0$, thus carrying out a single test for each genetic variant, testing whether the variant is associated with \emph{any} of the traits using the logistic regression model.

\subsection{\textbf{L2-Regularized reverse regression}}
\label{sec:reverse-regression}

Expressing CCA for multi-trait GWAS as a linear regression of the variant genotype on the trait values [eq.~(\ref{eq:2})] immediately leads to a generalization to the high-dimensional setting in the form of regularizing the regression coefficients, that is, augmenting eq.~(\ref{eq:2}) with a prior distribution $p(a_i) = \N(0,\alpha)$, $i=1,\dots,p$.
Finding the maximum-likelihood values of the regression coefficients is equivalent to $L_2$-regularized or ridge regression. This is the approach followed by \cite{banerjee2020reverse}, who combined it with a likelihood ratio test to determine whether the fitted model is more likely than a model where the genotype is independent of the traits ($a_i=0$ for all $i$) and obtain a single association $p$-value for each variant.

\subsection{\textbf{Reverse genotype prediction using machine learning methods}}
\label{sec:reverse-genotype-pred}

From the above, we conclude that existing multi-trait GWAS methods can be described as reverse genotype prediction methods. From this perspective, $L_2$-regularized linear regression is but one of many established machine learning methods that could be used to predict an outcome variable $Y$ from a high number of predictors or features $X_i$, $i=1,\dots,p$. Hence we propose to consider a wider range of machine learning methods such as random forest regression (RFR) and support vector regression (SVR) \cite{friedman2001elements}. Our overall hypothesis is that genetic variants whose genotypes can be predicted with higher accuracy are more likely to affect some or all of the traits under consideration than variants whose genotypes cannot be predicted well, and that feature weights in the fitted models measure the strength of biological association between a variant and a trait.

\section{Methods}

\subsection{\textbf{Reverse genotype prediction}}
For genotype prediction using machine learning models, the expression values were treated as explanatory variables whereas the genotype value of a variant was treated as a response variable. The prediction performance was measured by computing the root mean squared error (RMSE) between the predicted and the actual genotype value of variants.

\subsection{\textbf{Trans-eQTL target prediction}}
Trans-eQTL target prediction was done using weights assigned to the features by the machine learning methods: feature importance in case of random forest regression (RFR), and coefficients for support vector regression (SVR) and ridge regression (RR). We computed the area under the receiver operating characteristic (AUROC) curve to measure prediction performance by comparing the weights against the true targets in the ground truth for each variant.

\subsection{\textbf{Datasets}}
\subsubsection{Simulated data}
The simulated data for our experiments was obtained from 
DREAM5 Systems Genetic Challenge A (\dreamurl), generated by the Sys\-Gen\-SIM software \cite{pinna2011simulating}. The DREAM data consists of simulated genotype and transcriptome data of synthetic gene regulatory networks. The dataset consists of 15 sub-datasets, where 5 different networks are provided and for each network 100, 300 and 999 samples are simulated. Every sub-dataset contains 1000 genes. We used the networks with 999 samples only.

In the DREAM data, each genetic variant is associated to a unique causal gene that mediates its effect. We therefore defined ground-truth trans-eQTL targets for each variant as the causal gene's direct targets in the ground-truth network.

In the DREAM data 25\% of the variants acted in \textit{cis}, meaning they affected expression of their causal gene directly. The remaining 75\% of the variants acted in \textit{trans}. Since the identities of the \textit{cis} and \textit{trans} eQTLs are unknown, we computed the P-values of genotype-gene expression associations between matching variant-gene pairs using Pearson correlation and selected all genes with P-values less than 1/750 to identify cis-acting eQTLs.

\subsubsection{Yeast data}

The yeast data used in this paper was obtained from \cite{albert2018genetics}. The expression data contains expression values for 5,720 genes in 1,012 segregants.  The genotype data consists of binary genotype values for 42,052 genetic markers in the same 1,012 segregants.

Batch and optical density (OD) effects, as given by the covariates provided in \cite{albert2018genetics}, were removed from the expression data using categorical regression, as implemented in the \textit{statsmodels} python package. The expression data was then normalized to have zero mean and unit standard deviation.

To match variants to genes, we considered the list of genome-wide significant eQTLs provided by  \cite{albert2018genetics} whose confidence interval (of variable size) overlapped with an interval covering a gene plus 1,000 bp upstream and 200 bp downstream of the gene position. This resulted in a list of 2,884 genes and for each of these genes we defined its matching variant as the most strongly associated variant from the list.

Networks of known transciptional regulatory interactions in yeast (S. cerevisiae) were obtained from the YEASTRACT (Yeast Search for Transcriptional Regulators And Consensus Tracking) \cite{monteiro2020yeastract}. Regulation matrices were obtained from \url{http://www.yeastract.com/formregmatrix.php}.
We re\-trieved the ground-truth matrix containing all reported interactions of the type \emph{DNA binding and expression evidence}.  Self regulation was removed from the ground-truths. The Ensembl database (release 83, December 2015) \cite{yates2020ensembl} was used to map gene names to their identifiers. After overlaying the ground-truth with the set of genes with matching cis-eQTL,  a ground-truth network of 80 transcription factors (TFs) with matching cis-eQTL and 3,394 target genes was obtained.

The expression dataset was then filtered to contain only the genes present in the ground truth network, and ground-truth trans-eQTL sets for the 80 TF-associated cis-eQTL genetic variants were defined as direct targets of the corresponding TFs in the ground-truth network.

\subsection{\textbf{Experimental settings}}
In all sets of experiments  we used 5-fold cross-validation, except for the feature selection experiment in yeast data where we used 80-20 train-test split due to time constraints.

Ridge Regression (RR), Random Forest Regression (RFR), Support Vector Regression (SVR), and Naive Bayes (NB) were implemented using the Python library \textit{scikit-learn}. For RR, the regularization strength ($\alpha$) was set to 100 and other parameters were set to their defaults. For RR and SVR, the default parameters were used. For NB, we used the Gaussian Naive Bayes from \textit{scikit-learn} library. For  trans-eQTL predictions, univariate linear correlation was also used to compare with the regression methods mentioned above.

\subsubsection{Feature selection}
For each method we took the absolute values of the feature importances/coefficients,  scaled so that their sum equals to 1, and sorted these in descending order. These scaled values represent the relative contribution of each feature to the prediction of each variant. We selected the top-scoring features which together contributed 50\% of the feature weight sum.

\subsection{\textbf{Code}}
The scripts to reproduce the analysis are available at \newline \rfurl.

\begin{figure}[t]

\begin{minipage}[t]{0.50\linewidth}
  \textbf{A}\\
      \includegraphics[width=\linewidth]{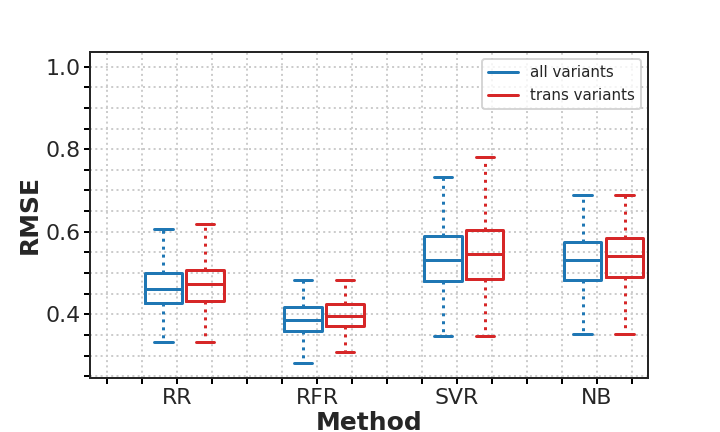}
      
  \end{minipage}
  \hfill
  \begin{minipage}[t]{0.50\linewidth}
  \textbf{B}\\
      \includegraphics[width=1.1\linewidth, height = 0.6\textwidth]{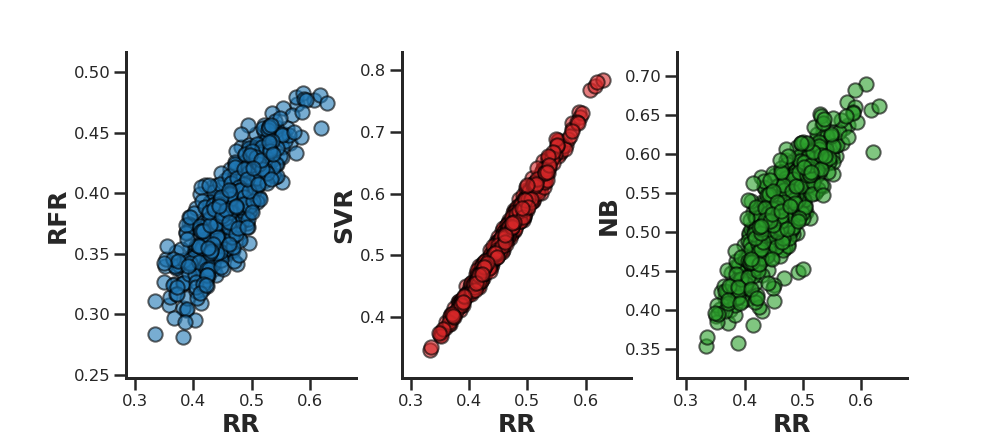}
\end{minipage}

  \caption{RMSE values for genotype prediction on DREAM5 simulated data. (\textbf{A}) Boxplots show the distribution of the RMSE values for all variants (blue) and for trans-acting-only variants (red) for random forest regression (RFR), support vector regression (SVR), ridge regression (RR), and naive Bayes (NB). (\textbf{B}) Scatter plots show RMSE values of RFR, SVR, and NB vs RR for all variants. The data shown are for DREAM Network 1. The results for Network 2-5 are shown in Supp. Figs. S1-S4.}
  \label{fig:genotype-prediction-dream}
\end{figure}

\begin{figure}[h!]
  \begin{minipage}[t]{0.50\linewidth}
  \textbf{A}\\
      \includegraphics[width=\linewidth]{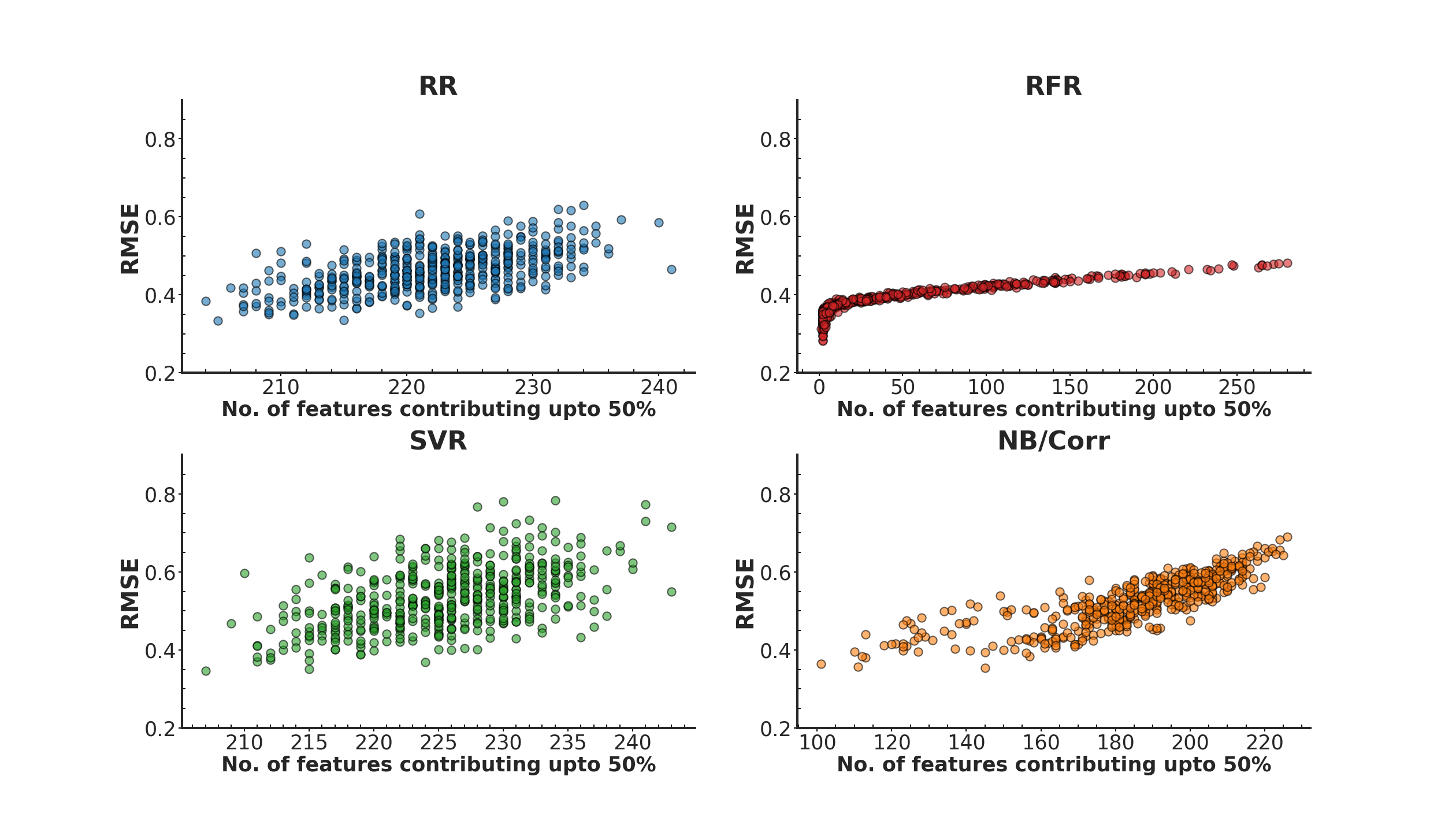}
  \end{minipage}
  \hfill
  \begin{minipage}[t]{0.50\linewidth}
  \textbf{B}\\
      \includegraphics[width=\linewidth]{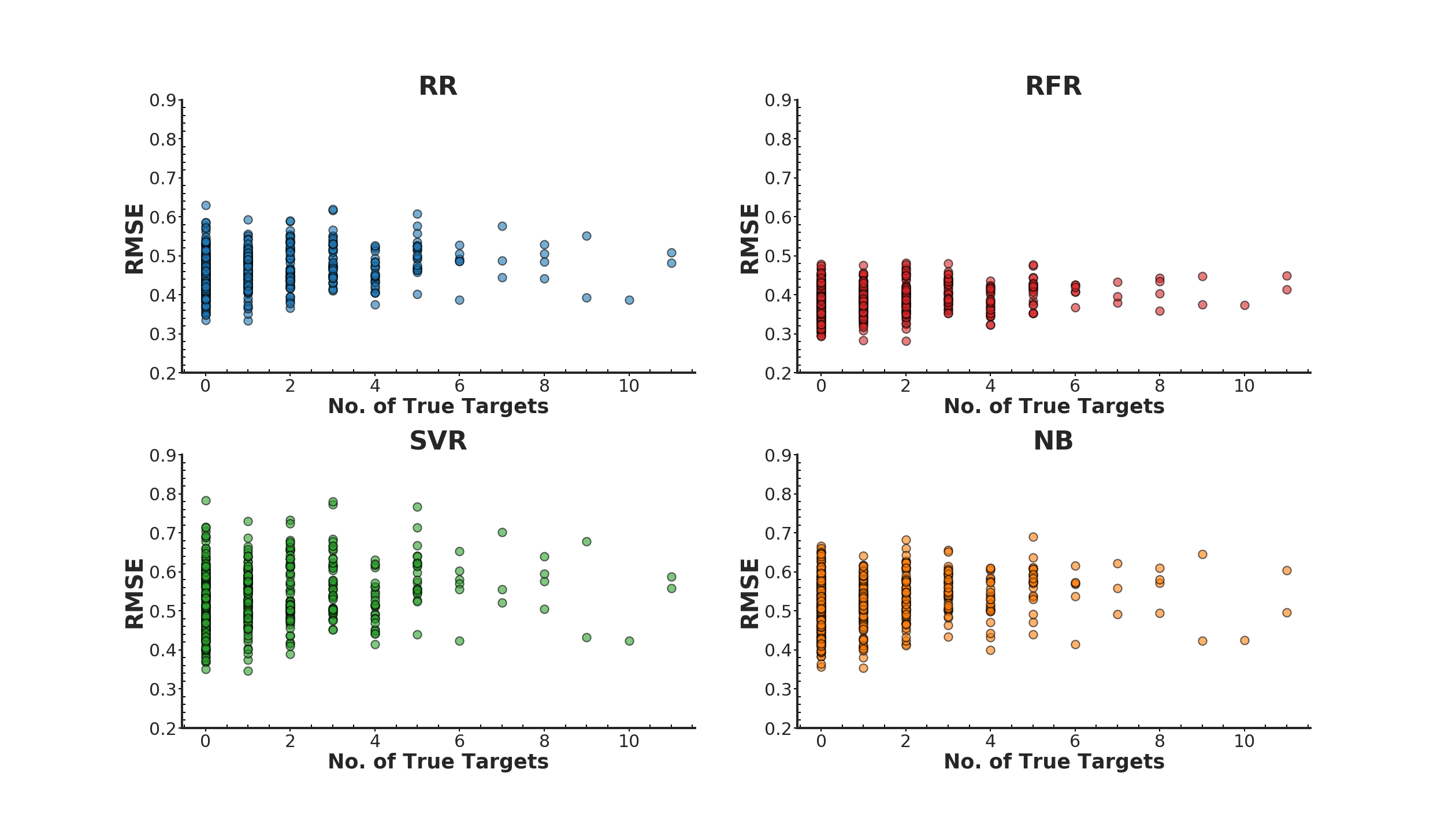}
  \end{minipage}
  \hfill
  \begin{center}
        \begin{minipage}[t]{0.50\linewidth}
        \textbf{C}\\
      \includegraphics[width=\linewidth]{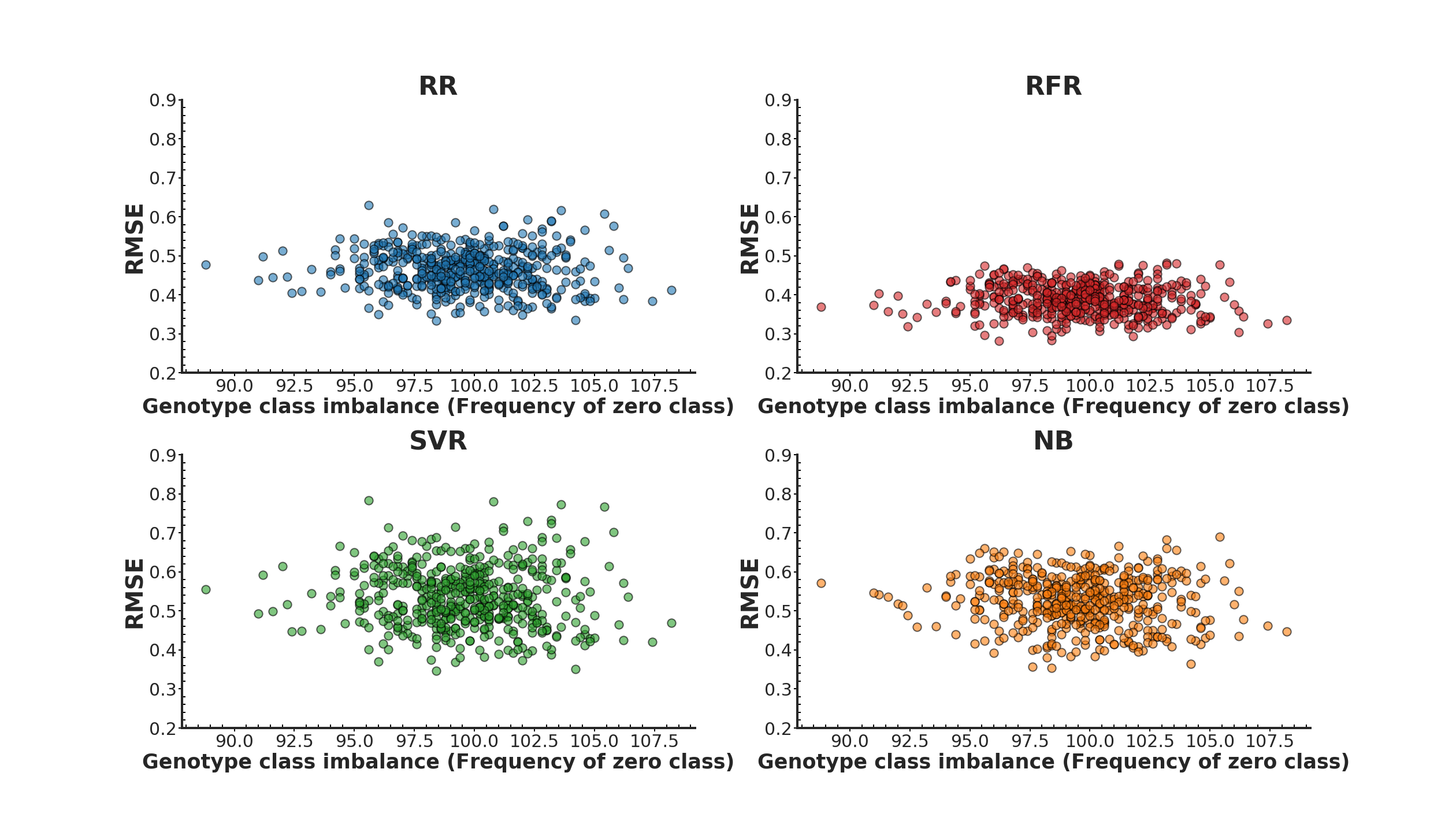}
  \end{minipage}
  \end{center}
  \caption{Scatter plots of genotype RMSE values on DREAM5 simulated data against the number of selected model features (\textbf{A}), the number of true trans-eQTL targets in the ground-truth network (\textbf{B}), and the genotype class balance (frequency of the zero class) (\textbf{C}), for random forest regression (RFR), support vector regression (SVR), ridge regression (RR), and naive Bayes (NB). The data shown are for DREAM Network 1. The results for Network 2-5 are shown in Supp. Figs. S5-S8.}
  \label{fig:genotype-prediction-dream-control}
\end{figure}

\section{Results}

\subsection{\textbf{Reverse genotype prediction and trans-eQTL analysis in simulated data}}
\label{sec:reverse-genotype-pre}

In the DREAM5 Systems Genetics Challenge, binary genotypes and steady-state gene expression data for 1,000 genes were simulated for a population of 999 individuals, based on a gene network topology and the individuals' genotypes at a set of genome-wide DNA variants, using non-linear ordinary differential equations (ODEs) \cite{pinna2011simulating}. In the simulations, there was a one-to-one mapping between genetic variants and genes, such that the effects of each variant are mediated by exactly one causal gene. 25\% of the variants acted in \emph{cis}, meaning they affected expression of their causal gene, but not the value of any of the parameters in the ODE model. The remaining 75\% of the variants acted in \emph{trans}, meaning they did not affect expression of their causal gene, but did affect the transcription rate of the causal gene's targets in the network. Simulated data for five networks are available.

\subsubsection{Genotype prediction accuracy varies across genetic variants}
\label{sec:genotype-pred-accur}
We trained models to predict the genotypes for variants whose causal gene had at least one target in the ground-truth network (covering between 491-644 genes depending on the network/dataset) using the expression data from all 1,000 genes as predictors, using  Random Forest Regression (RFR), Support Vector Regression (SVR), Ridge Regression (RR) and Naive Bayes classification (NB). RMSE was then measured for each predicted variant in the test data. Mean performance across the five train-test folds is reported here. 

RFR achieved the best prediction performance (lowest RMSE) overall (RMSE $\sim 0.3-0.5$). RR achieved RMSEs in the range of $\sim 0.6-0.8$. In contrast to RFR and RR, the RMSE varied widely for SVR and NB ($\sim 0.3-0.9$) (Fig.~\ref{fig:genotype-prediction-dream}A). We did not observe a significant change in the distribution of RMSE values for all the variants versus keeping only \textit{trans}-acting variants (Fig.~\ref{fig:genotype-prediction-dream}A), i.e. \textit{cis}-acting variants are not significantly easier to predict (by virtue of having a highly correlated \textit{cis}-gene) than variants that only have \textit{trans}-associated genes. While RMSE values are correlated between the methods (Fig.~\ref{fig:genotype-prediction-dream}B), the correlation is imperfect (with the exception of SVR-RR), such that there is considerable variation in the RMSE-based ranking of variants between the methods.

Taken together these result show that prediction performance varies across genetic variants within each method (i.e. variants can be ranked according to their RMSE) and that RFR can be preferred over the others in terms of average prediction performance, but with considerable variation in relative performance across methods for individual variants.

Next, we compared the genotype prediction performance for the different methods with the number of features contributing upto $50\%$ of the total sum of feature weights (cf.\ Methods). In general, variants that were more predictable had models with fewer features, and vice versa, irrespective of the prediction method used (Fig.~\ref{fig:genotype-prediction-dream-control}A). On the other hand, we did not observe any significant relation between the prediction performance and the number of true targets in the ground truth network (Fig.~\ref{fig:genotype-prediction-dream-control}B). We also tested whether RMSE was influenced by the genotype class imbalance. This was not the case for the regression-based methods used here (Fig.~\ref{fig:genotype-prediction-dream-control}C). 

\begin{figure}[h!]
    \centering
    \includegraphics[width=0.7\linewidth]{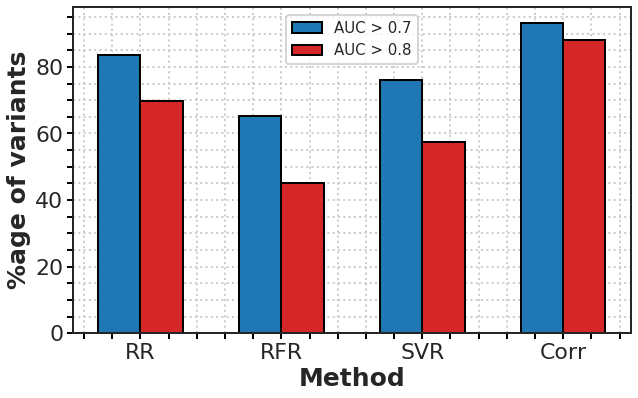}
    \caption{Bar plots show the proportion of variants with trans-eQTL target prediction AUROC $> 0.7$ (blue) and $> 0.8$ (red) for random forest regression (RFR), support vector regression (SVR), ridge regression (RR), and univariate correlation (Corr). The data shown are for DREAM Network 1. The results for Network 2-5 are shown in Supp. Fig. S9.}
    \label{fig:eqtl-prediction-dream-a}
\end{figure}

\begin{figure}[h!]
  \begin{minipage}[t]{.50\linewidth}
  \textbf{(A)}\\
      \includegraphics[width=\linewidth]{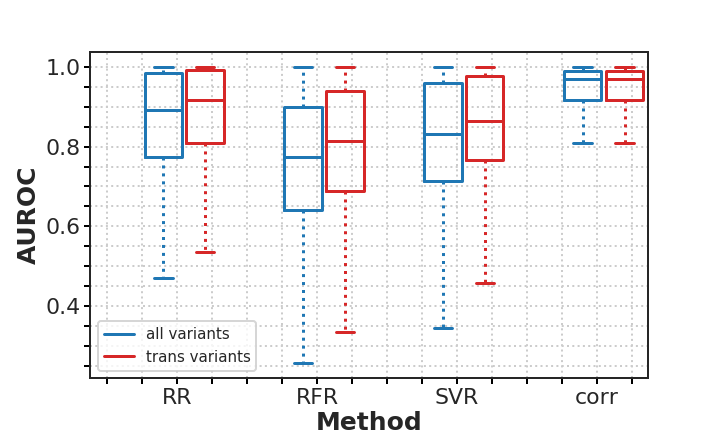}
  \end{minipage}
  \hfill
  \begin{minipage}[t]{.50\linewidth}
  \textbf{(B)}\\
      \includegraphics[width=1.1\linewidth, height=0.6\textwidth]{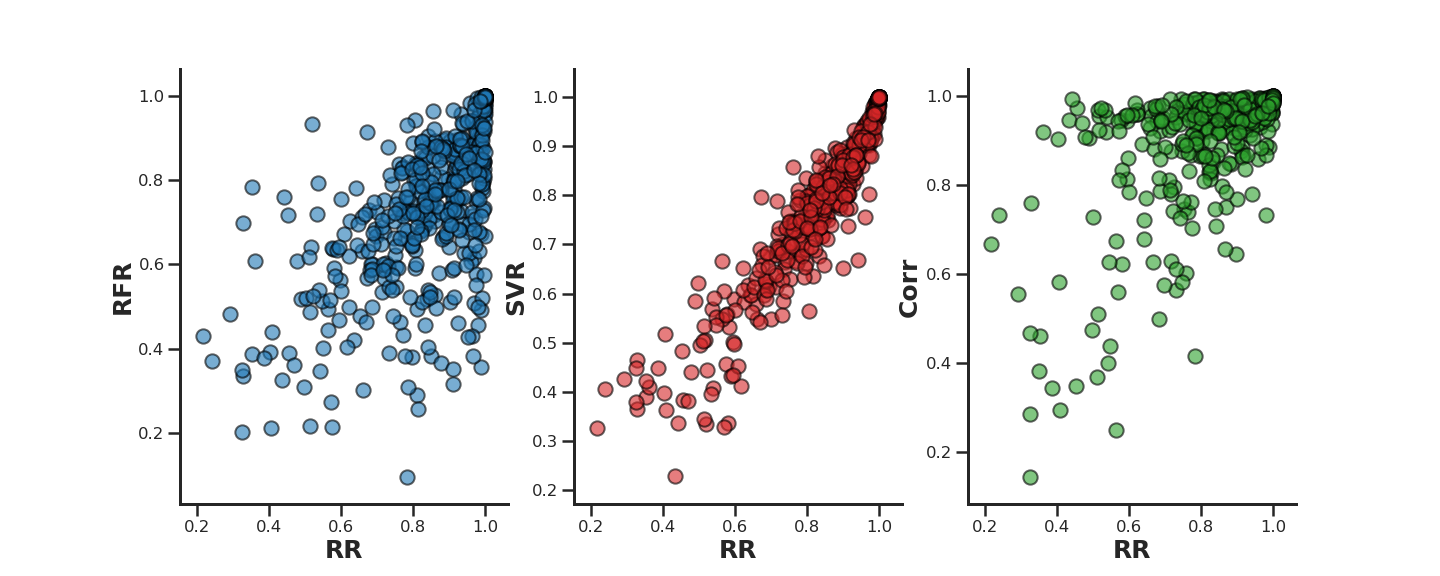}
  \end{minipage}
  \caption{Trans-eQTL target prediction performance on DREAM5 simulated data. \textbf{(A)} Boxplots show the distribution of AUROC values for all variants (blue) and for  trans-acting-only variants (red) for random forest regression (RFR), support vector regression (SVR), ridge regression (RR), and univariate correlation (Corr). \textbf{(B)} Scatter plots show AUROC values of classification methods RFR, SVR, and Corr vs RR for all variants. The data shown are for DREAM Network 1. The results for Network 2-5 are shown in Supp. Figs. S10-S13.}
  \label{fig:eqtl-prediction-dream}
\end{figure}

\begin{figure}[t!]
  \begin{minipage}[t]{.50\linewidth}
  \textbf{A}\\
      \includegraphics[width=\linewidth]{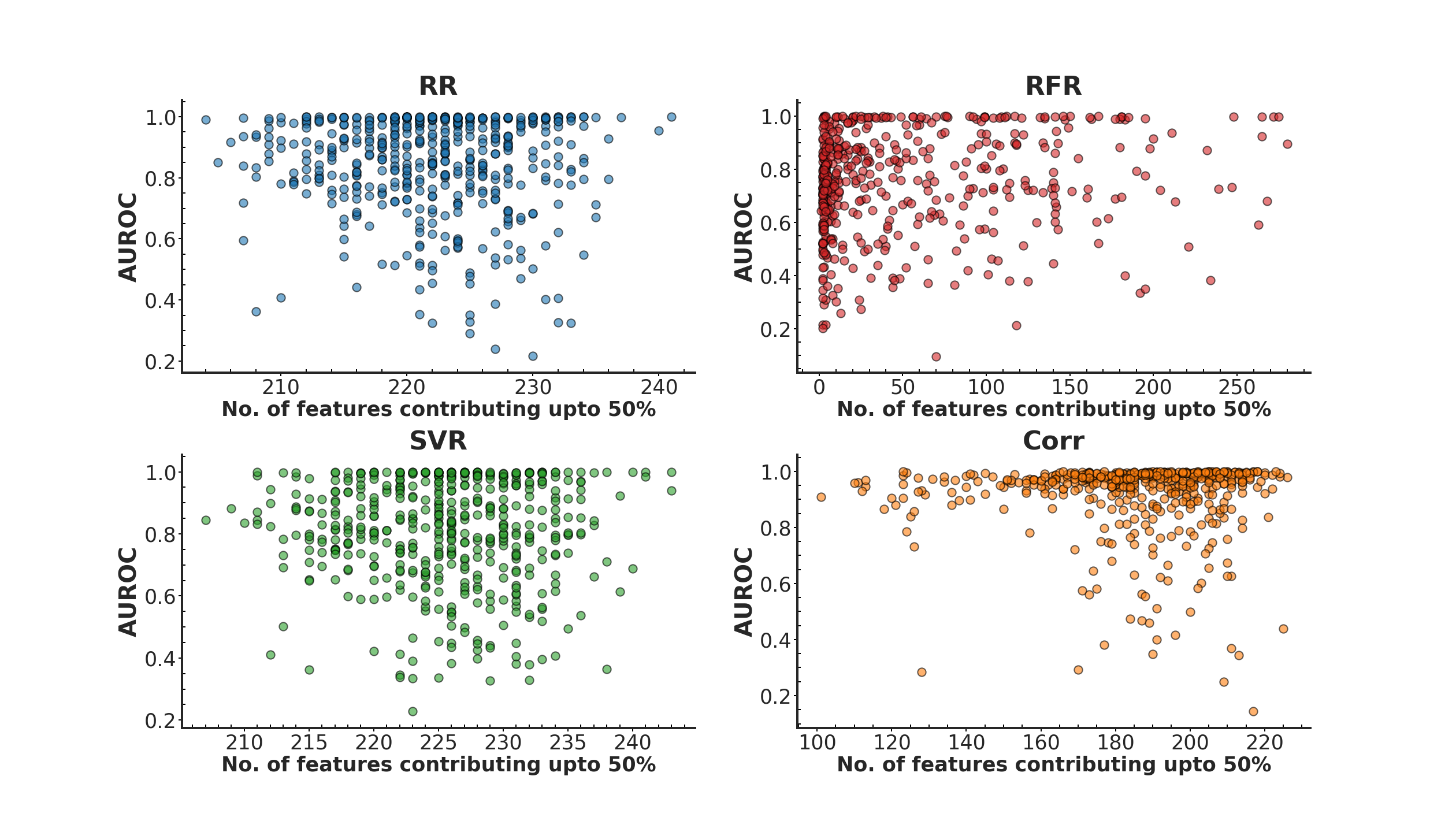}
  \end{minipage}
  \hfill
  \begin{minipage}[t]{.50\linewidth}
  \textbf{B}\\
      \includegraphics[width=\linewidth]{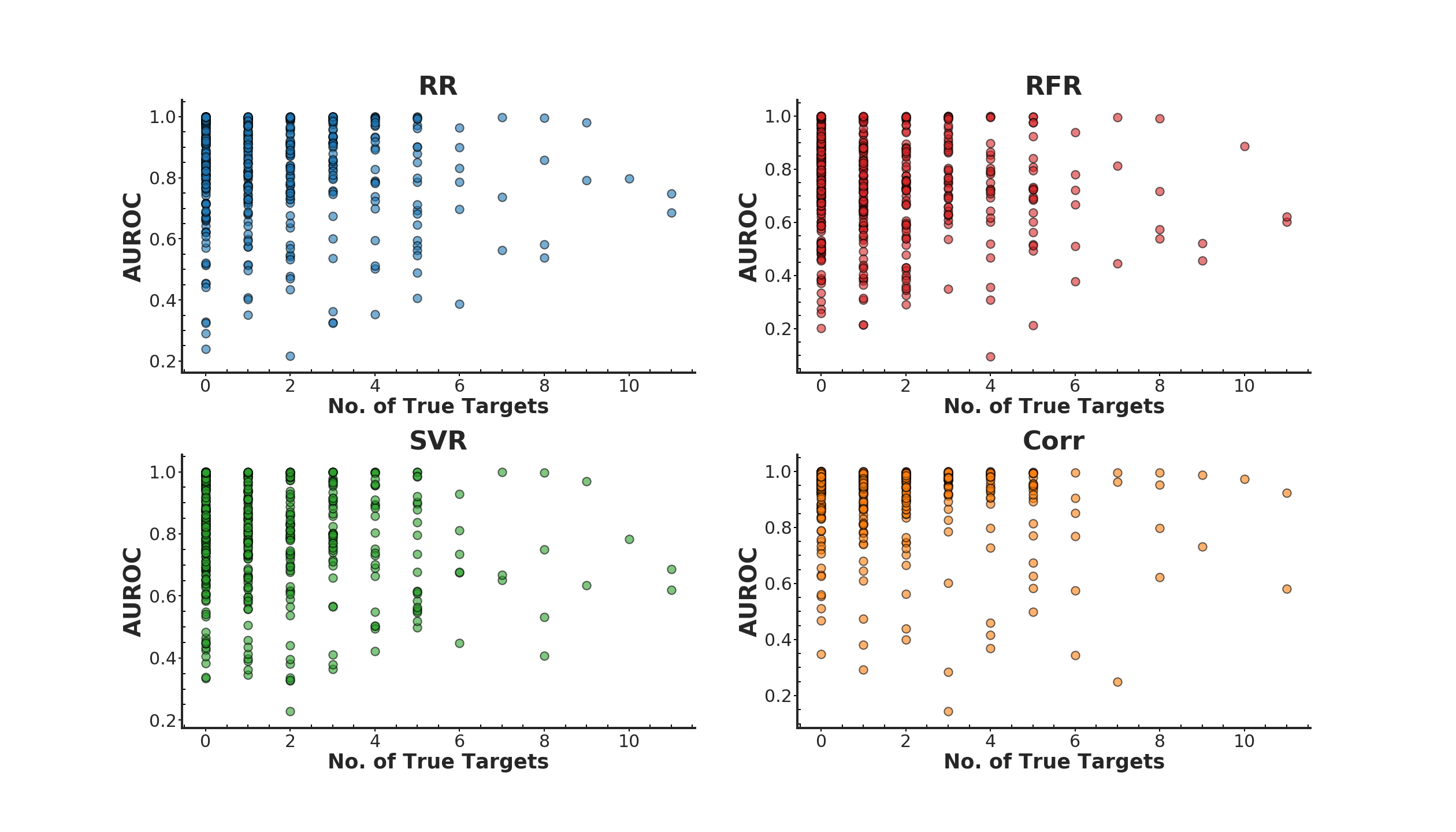}
  \end{minipage}
  \hfill
  \begin{center}
        \begin{minipage}[t]{.50\linewidth}
        \textbf{C}\\
      \includegraphics[width=\linewidth]{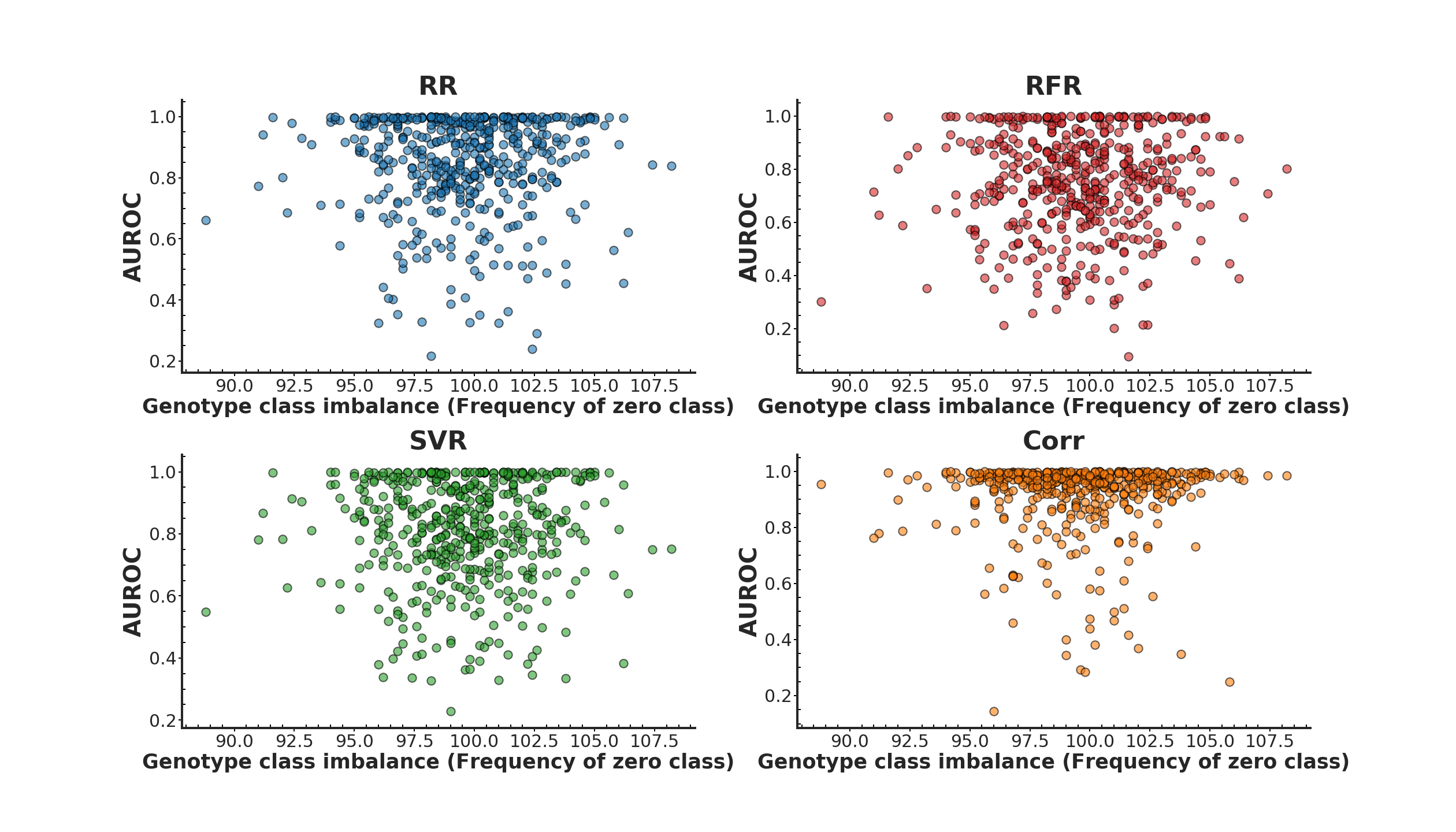}
  \end{minipage}
  \end{center}
  \caption{Scatter plots of trans-eQTL target prediction performance (AUROC) on DREAM5 simulated data against the number of selected model features (\textbf{A}), the number of true trans-eQTL targets in the ground-truth network (\textbf{B}), and the genotype class balance (frequency of the zero class) (\textbf{C}), for random forest regression (RFR), support vector regression (SVR), ridge regression (RR), and univariate correlation/naive Bayes (NB). The data shown are for DREAM Network 1. The results for Network 2-5 are shown in Supp. Figs. S14-S17.}
  \label{fig:eqtl-prediction-dream-control}
\end{figure}

\begin{figure}[h!]
    \centering
    \includegraphics[width=1.00\linewidth]{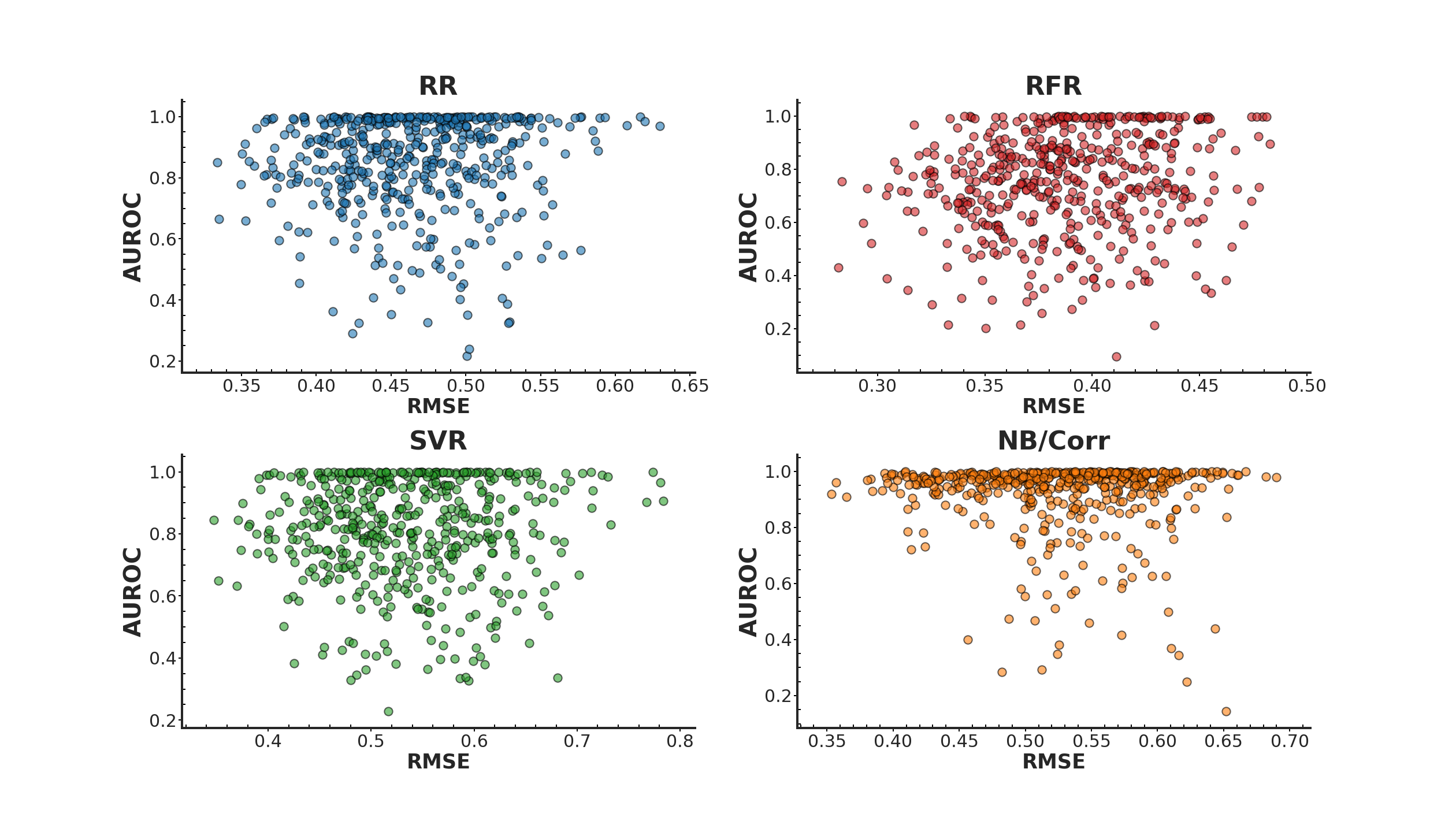}
    \caption{Scatter plots show trans-eQTL target prediction performance (AUROC) vs genotype prediction performance (RMSE) on DREAM5 simulated data for all genetic variants for random forest regression (RFR), support vector regression (SVR), ridge regression (RR), and univariate correlation/naive Bayes (NB/Corr). The data shown are for DREAM Network 1. The results for Network 2-5 are shown in Supp. Figs. S18-S21.}
  \label{fig:genotype-target-comparison-dream}
\end{figure}

\subsubsection{Feature importances are predictive of true trans-eQTL associations}
\label{sec:feat-import}
To evaluate the ability of reverse genotype prediction methods to identify true trans-eQTL targets of a given variant, we defined true trans associations as direct target genes of a variant's causal gene in the ground-truth network and used feature importances/coefficients in the genotype prediction model to predict how likely a gene is to be a trans-eQTL of a given variant (see Methods). Performance was measured using the area under the receiver operating curve (AUROC). 

For all methods, more than $\sim 55\%$, resp. $ \sim 65\%$ of variants with at least one trans-eQTL target in the ground-truth network had AUROC$>0.8$, resp. $0.7$, with univariate linear correlation and ridge regression performing somewhat better than random forest and SVR (Fig.~\ref{fig:eqtl-prediction-dream-a}). Ridge regression and univariate correlation methods also had less variation in terms of AUROCs when compared with RFR and SVR, and no significant difference in terms of AUROC was observed when using all the variants versus using only \textit{trans}-acting variants  (Fig.~\ref{fig:eqtl-prediction-dream}A). Interestingly, the variants for which high AUROCs were obtained differed between RFR, RR and univariate correlation methods, whereas RR and SVR obtained nearly identical performance on all variants. (Fig.~\ref{fig:eqtl-prediction-dream}B).

When compared to potential explanatory factors, no significant relation was observed between AUROC values and number of selected model features  (Fig.~\ref{fig:eqtl-prediction-dream-control}A), number of ground-truth targets  (Fig.~\ref{fig:eqtl-prediction-dream-control}B), or the genotype class balance (Fig. \ref{fig:eqtl-prediction-dream-control}C).

\subsubsection{Genotype and trans-eQTL prediction performance do not correlate}

Finally we tested whether genotype prediction accuracy can be used as a proxy for trans-eQTL prediction accuracy, that is, in the absence of ground-truth networks, can we use genotype prediction accuracy to filter variants whose model feature weights are indicative of true trans-eQTL targets? However, we did not observe any correlation between the genotype prediction performance and trans-eQTL target prediction performance for any of the methods (Fig. \ref{fig:genotype-target-comparison-dream}).

\subsection{\textbf{Reverse genotype prediction and trans-eQTL analysis in yeast}}
\label{sec:reverse-genotype-pre-yeast}

\begin{figure}[h!]
  \begin{minipage}[t]{0.5\linewidth}
  \textbf{A}\\
      \includegraphics[width=\linewidth]{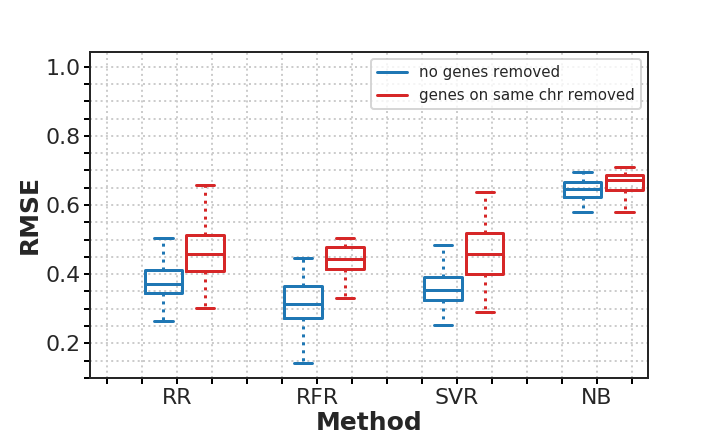}
      
  \end{minipage}
  \hfill
  \begin{minipage}[t]{0.5\linewidth}
  \textbf{B}\\
      \includegraphics[width=1.0\linewidth, height = 0.6\textwidth]{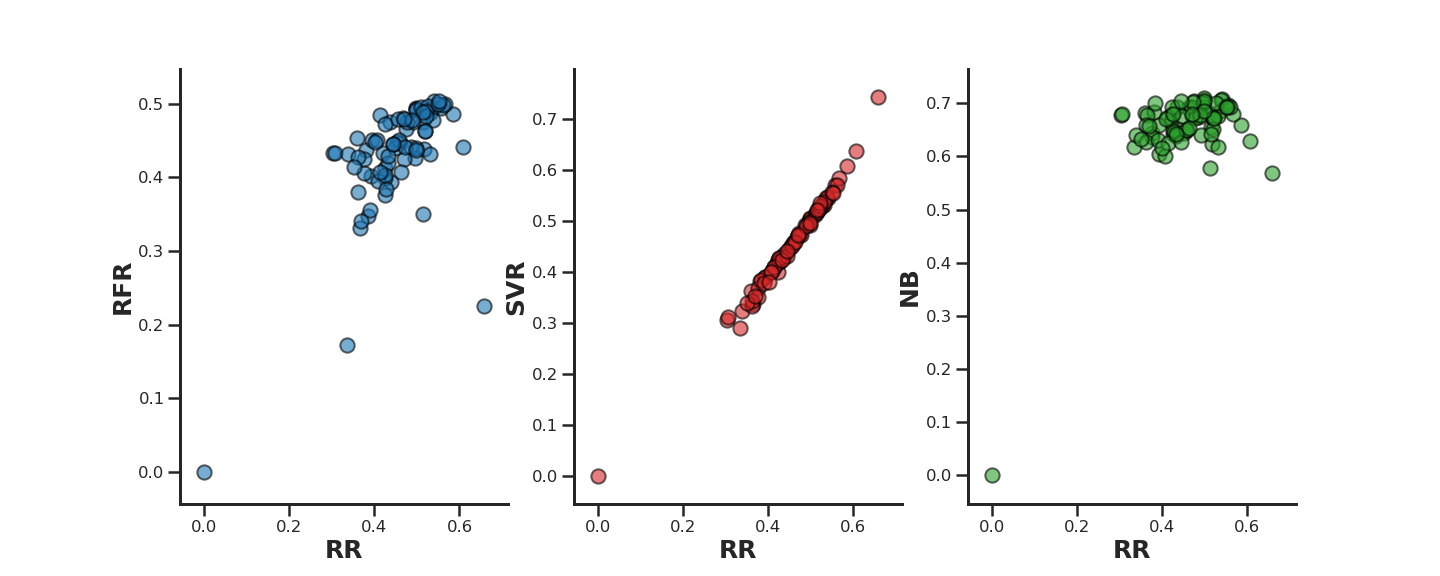}
  \end{minipage}
  \caption{Genotype prediction performance on yeast data. Genotype prediction performance on yeast data. (\textbf{A}) Boxplots show the distribution of the performance for all variants using all genes (blue) and excluding genes on the same chromosome as the variant (red) as predictors, for random forest regression (RFR), support vector regression (SVR), ridge regression (RR), and naive Bayes (NB). (\textbf{B}) Scatter plots show RMSE values of classification methods RFR, SVR, and NB vs RR for all variants. Genes on the same chromosome were excluded as predictors for each SNP.}
  \label{fig:genotype-prediction-yeast}
\end{figure}

\begin{figure}[h!]
  \begin{minipage}[t]{0.5\linewidth}
  \textbf{A}\\
      \includegraphics[width=\linewidth]{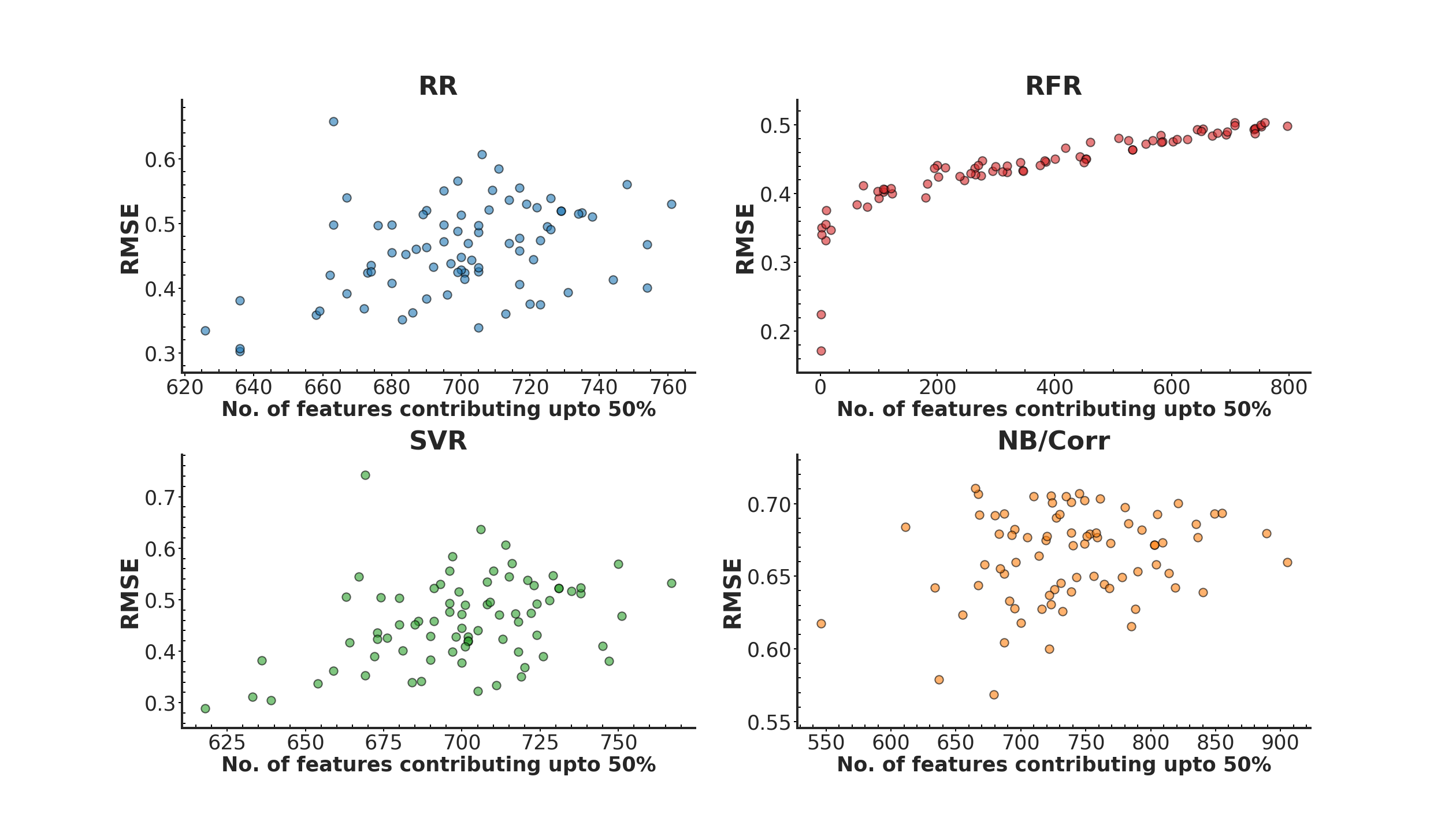}
  \end{minipage}
  \hfill
  \begin{minipage}[t]{0.5\linewidth}
  \textbf{B}\\
      \includegraphics[width=\linewidth]{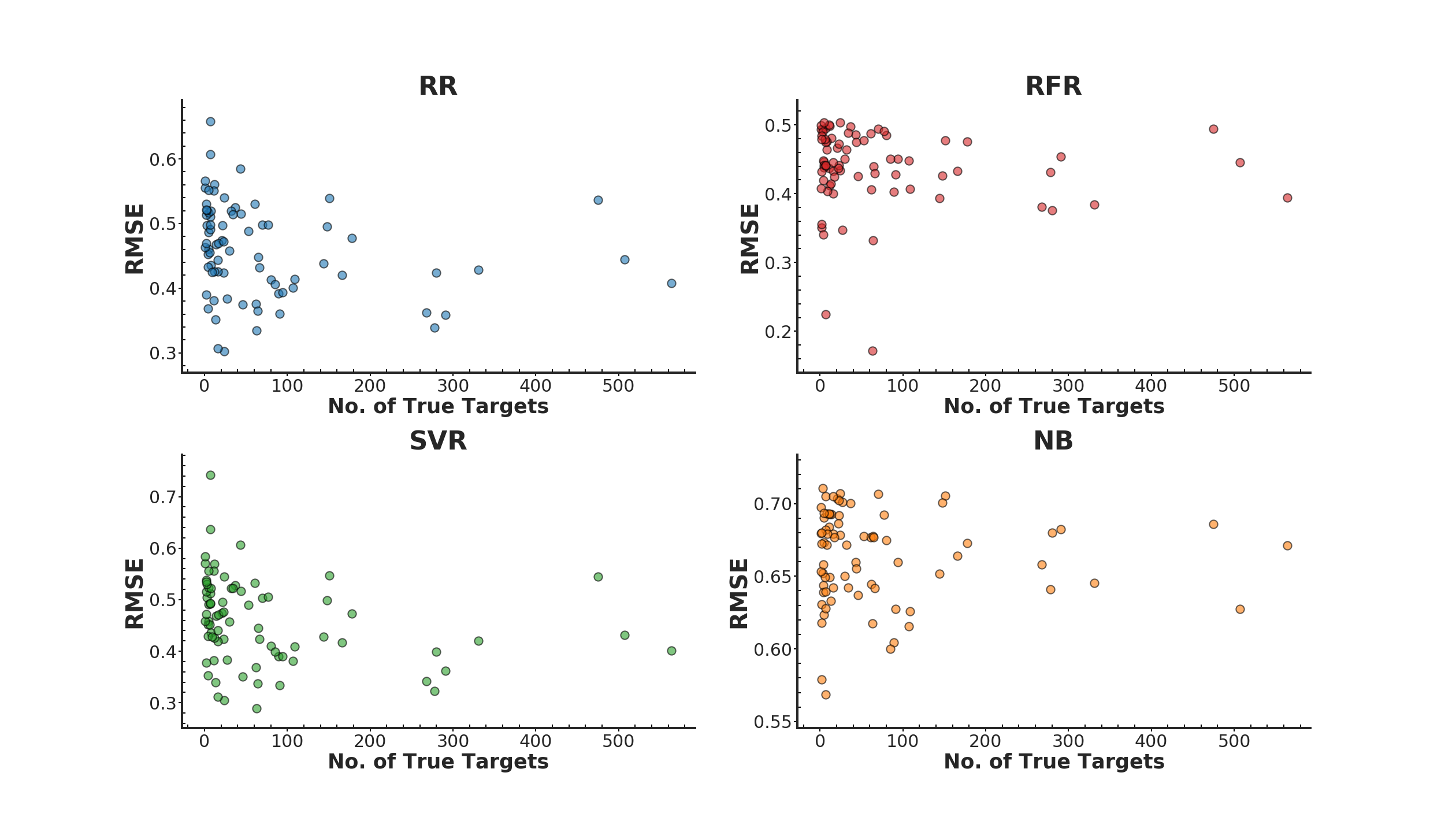}
  \end{minipage}
  \hfill
  \begin{center}
        \begin{minipage}[t]{0.5\linewidth}
        \textbf{C}\\
      \includegraphics[width=\linewidth]{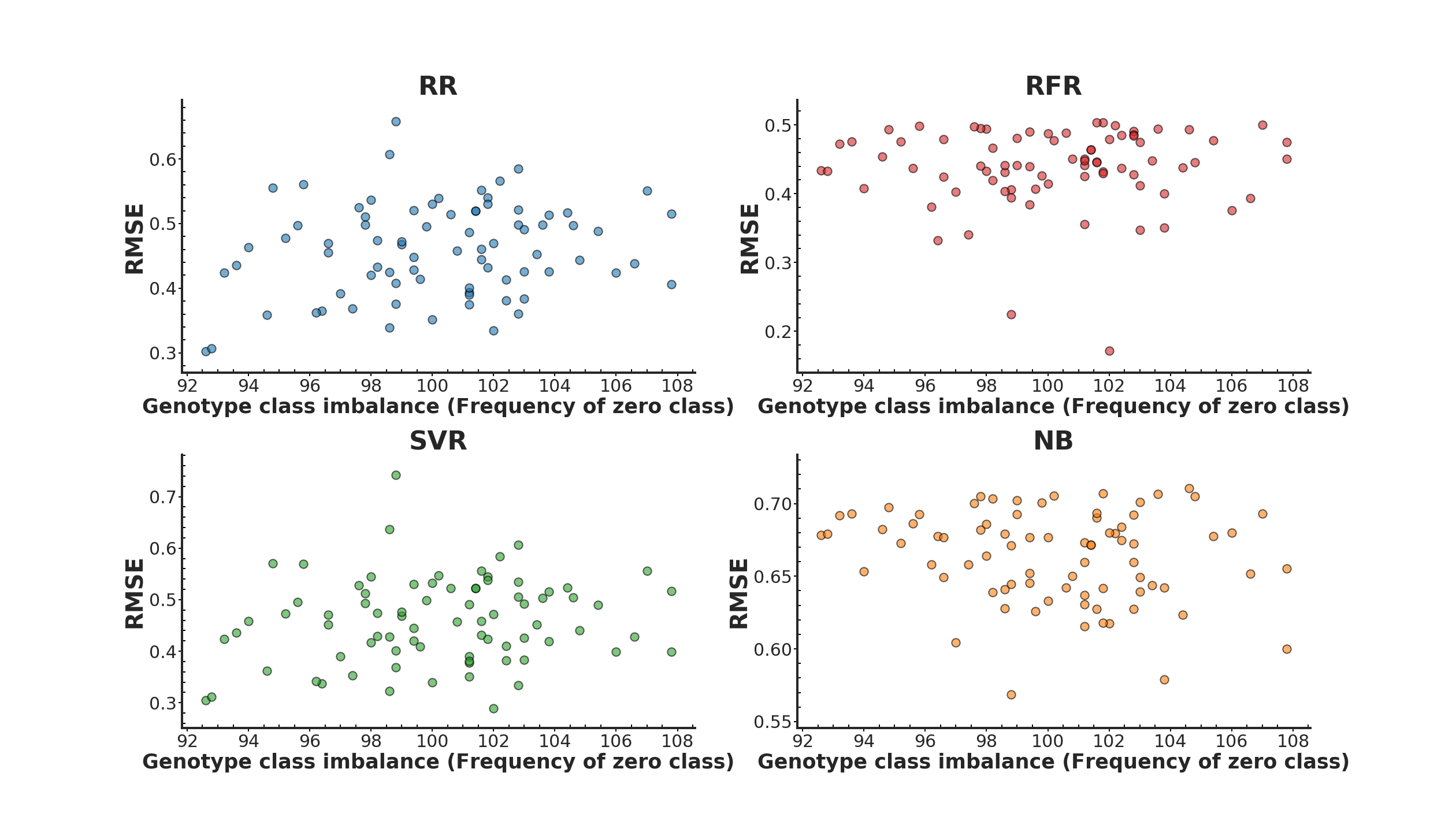}
  \end{minipage}
  \end{center}
 \caption{Scatter plots of genotype prediction performance on yeast data against the number of selected model features (\textbf{A}), the number of true trans-eQTL targets in the ground-truth network  (\textbf{B}) and the genotype class balance (frequency of the zero class) (\textbf{C}), for random forest regression (RFR), support vector regression (SVR), ridge regression (RR), and naive Bayes (NB). Genes on the same chromosome were excluded as predictors for each SNP.}
  \label{fig:genotype-prediction-yeast-control}
\end{figure}

In the next set of experiments we repeated the same analysis on yeast dataset. Compared to the simulated data, the yeast data differs in two important aspects. First, ground-truth target information is available for a small set of transcription factors (TFs) only. Secondly, we have no knowledge of the causal gene(s) corresponding to each variant, and need to rely on a local \emph{cis}-association between a variant and a TF to define a ground-truth set of trans-eQTL targets to a variant (cf.\ Methods).

\subsubsection{Genotype prediction accuracy varies across genetic variants}
\label{sec:genotype-pred-accur-yeast}

Genotype prediction performance for the yeast data also varied across genetic variants. Similar to DREAM data RFR achieved lowest RMSE values in the yeast data as well (Fig.~\ref{fig:genotype-prediction-yeast}A). We tested whether prediction performance may be explained by local \emph{cis}-associations by removing genes on the same chromosome as the test variant from the list of predictors. In this case we did observe that RMSE increased significantly (i.e. prediction performance decreased) when removing local genes, except for NB, and that after removing local genes, RR, RFR, and SVR have similar average prediction performance (Fig.~\ref{fig:genotype-prediction-yeast}A).

Correlations of RMSEs between methods showed a similar pattern as in the simulated data, with RR and SVR RMSEs being particularly strongly correlated (Fig.~\ref{fig:genotype-prediction-yeast}B).

As in the simulated data, genotype prediction performance decreased (i.e. RMSE increased) with increasing number of model features (Fig.~\ref{fig:genotype-prediction-yeast-control}A), but did not depend significantly on the number of true targets (Fig.~\ref{fig:genotype-prediction-yeast-control}B) or genotype class balance (Fig.~\ref{fig:genotype-prediction-yeast-control}C).

\begin{figure}
    \centering
    \includegraphics[width=0.7\linewidth]{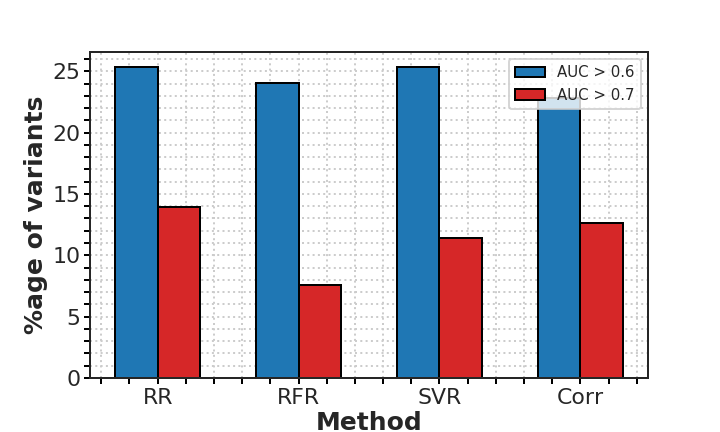}
    \caption{Bar plots show the number of variants with trans-eQTL target prediction AUROC $\geq 0.6$ (blue) and $\geq 0.7$ (red) for random forest regression (RFR), support vector regression (SVR), ridge regression (RR), and univariate correlation (Corr). Genes on the same chromosome were excluded as predictors for each SNP. }
    \label{fig:eqtl-prediction-yeast-a}
\end{figure}

Next we tested whether feature importance weights were predictive of true trans-eQTL associations, defined as genes that were bound by and differentially expressed upon perturbation of a TF for which a given variant is a cis-eQTL (cf.\ Methods). In this case, feature importances were only modestly predictive, with 20-30\%, resp. 10-15\%, of TF cis-eQTLs obtaining AUROCs $>0.6$, resp. $>0.7$, and, as in the simulated data, there were fewer variants with high AUROC for RFR, compared to the other methods (Fig. \ref{fig:eqtl-prediction-yeast-a}).  

\begin{figure}[t!]
  \begin{minipage}[t]{0.50\linewidth}
  \textbf{(A)}\\
      \includegraphics[width=\linewidth]{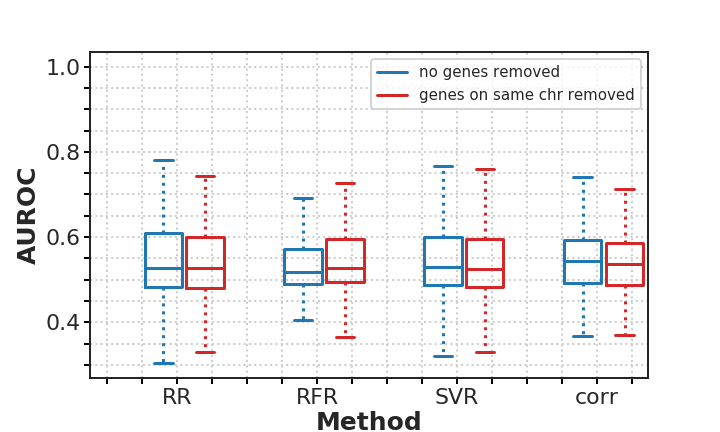}
  \end{minipage}
  \hfill
  \begin{minipage}[t]{0.5\linewidth}
  \textbf{(B)}\\
      \includegraphics[width=1.1\linewidth, height=0.6\textwidth]{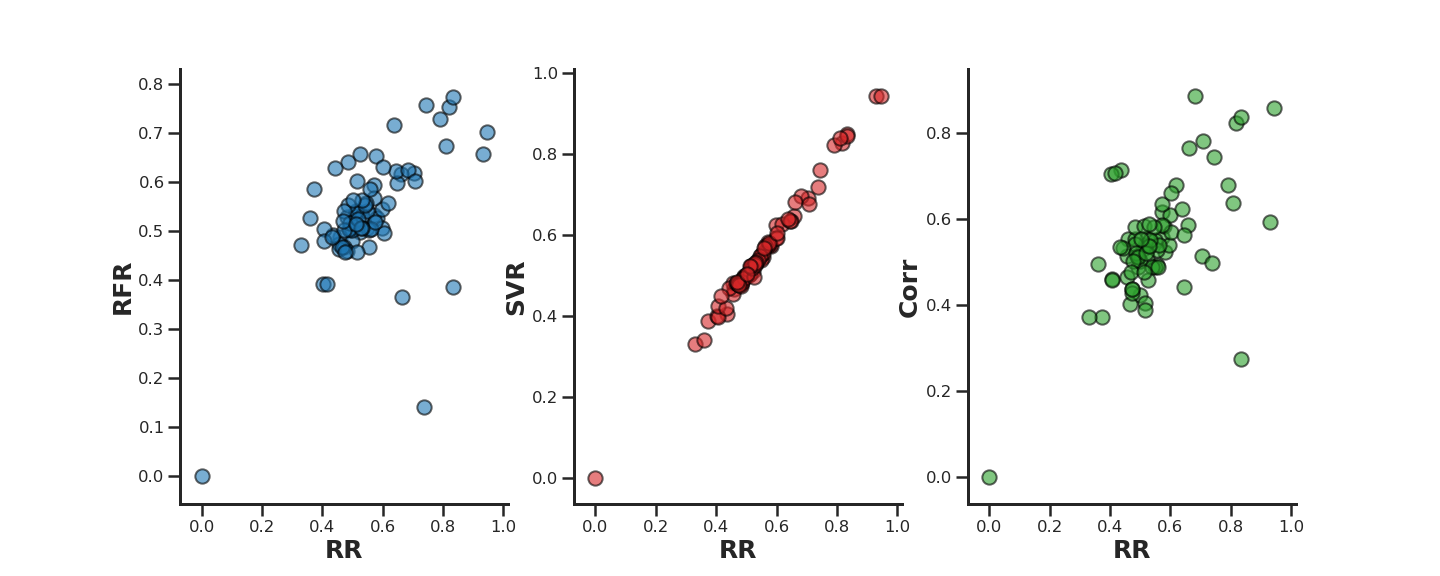}
  \end{minipage}
  \caption{Trans-eQTL target prediction performance on yeast data. \textbf{(A)} Boxplots show the distribution of AUROC values using all genes (blue) and excluding genes on the same chromosome (red) as predictors, for random forest regression (RFR), support vector regression (SVM), ridge regression (RR), and  univariate correlation (Corr). \textbf{(B)} Scatter plots show AUROC values of classification methods RFR, SVR, and univariate correlation (Corr) vs RR for all variants. Genes on the same chromosome were excluded as predictors for each SNP.}
  \label{fig:eqtl-prediction-yeast}
\end{figure}

\begin{figure}[h!]
  \begin{minipage}[t]{.50\linewidth}
  \textbf{A}\\
      \includegraphics[width=\linewidth]{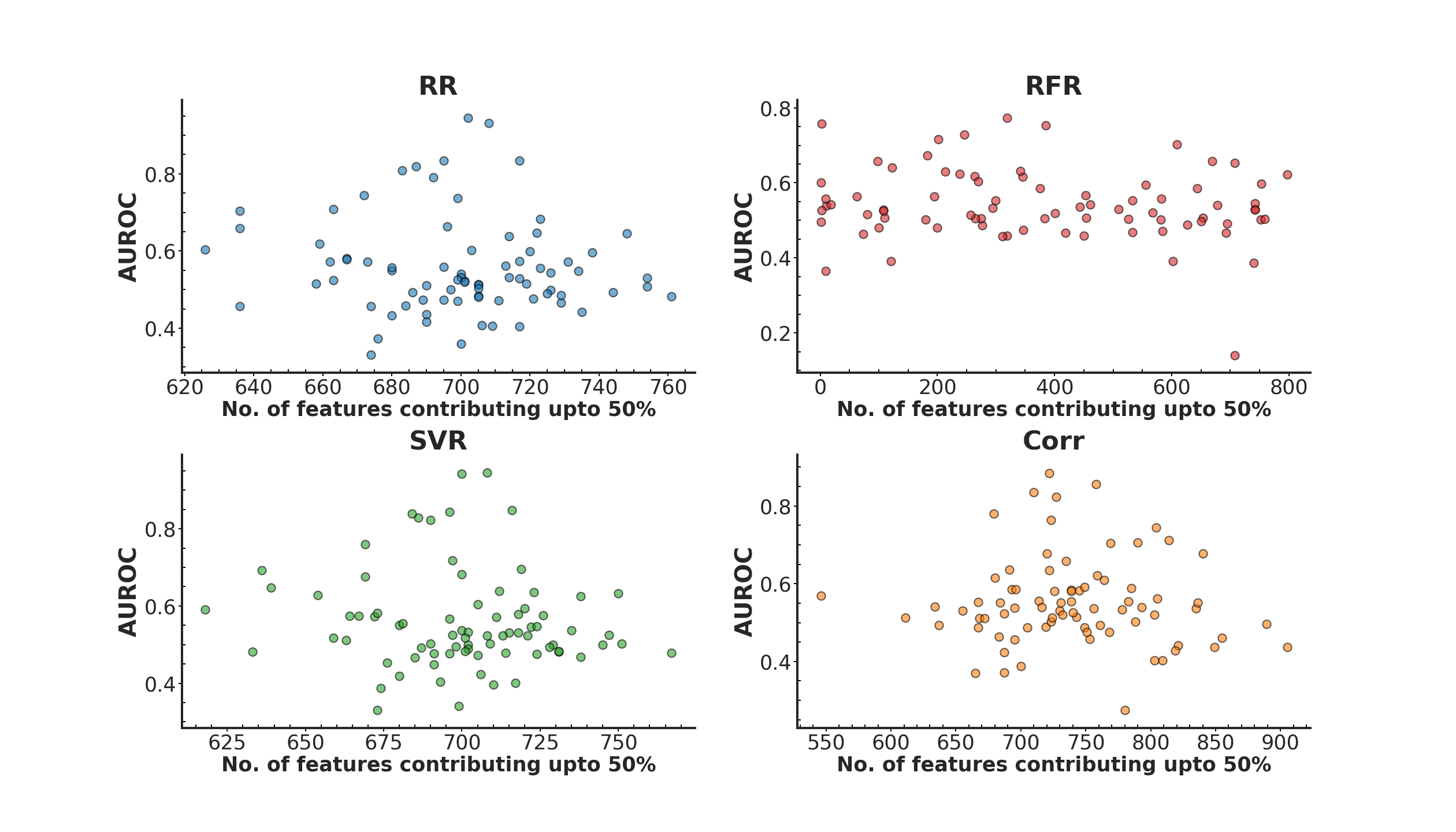}
  \end{minipage}
  \hfill
  \begin{minipage}[t]{.50\linewidth}
  \textbf{B}\\
      \includegraphics[width=\linewidth]{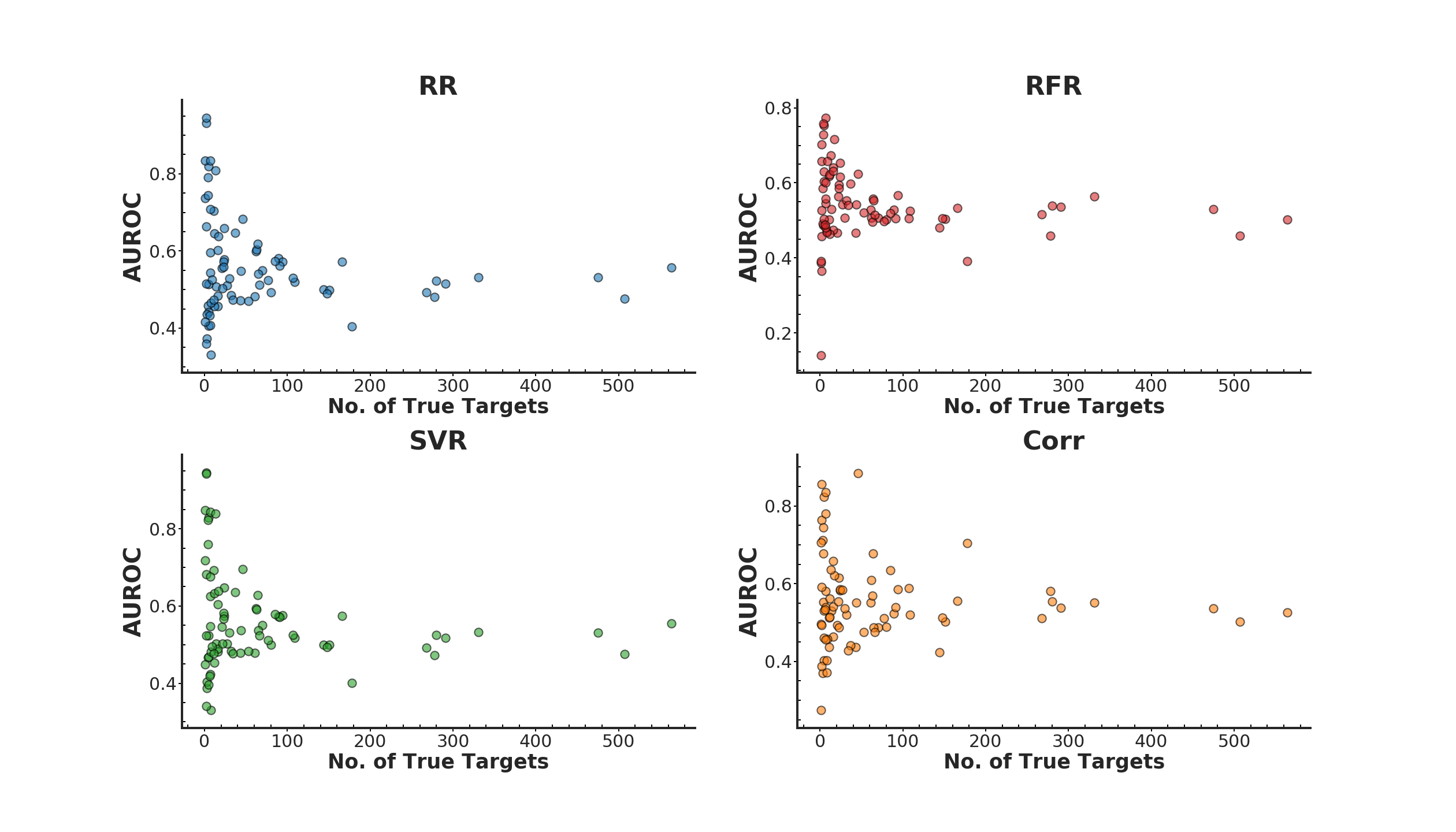}
  \end{minipage}
  \hfill
  \begin{center}
        \begin{minipage}[t]{.50\linewidth}
        \textbf{C}\\
      \includegraphics[width=\linewidth]{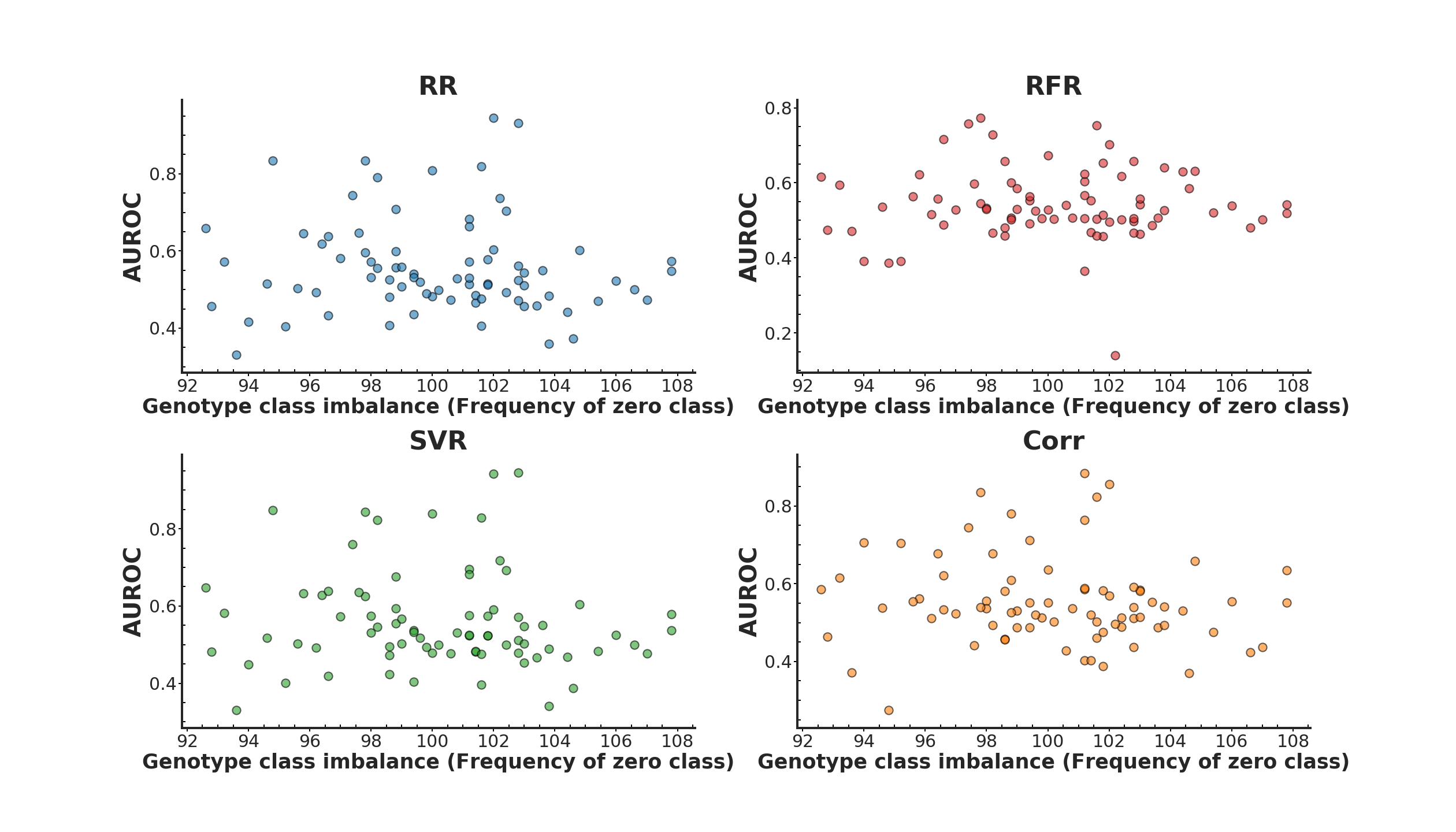}
  \end{minipage}
  \end{center}
\caption{Scatter plots of trans-eQTL target prediction performance (AUROC) on yeast data against the number of selected model features (\textbf{A}), the number of true trans-eQTL targets in the ground-truth network  (\textbf{B}) and the genotype class balance (frequency of the zero class) (\textbf{C}), for random forest regression (RFR), support vector regression (SVR), ridge regression (RR), and univariate correlation (Corr). Genes on the same chromosome were excluded as predictors for each SNP.}
  \label{fig:eqtl-prediction-yeast-control}
\end{figure}

We confirmed that the distribution of AUROC values was not affected by removing genes on the same chromosome as a variant of interest from the list of predictors (Fig. \ref{fig:eqtl-prediction-yeast}A). Furthermore, the AUROC values showed no relation with the number of selected features, the number of true targets, or the genotype class balance (Fig. \ref{fig:eqtl-prediction-yeast-control}).

\begin{figure}[b!]
    \centering
     \includegraphics[width=1.10\linewidth, height=0.30\textheight]{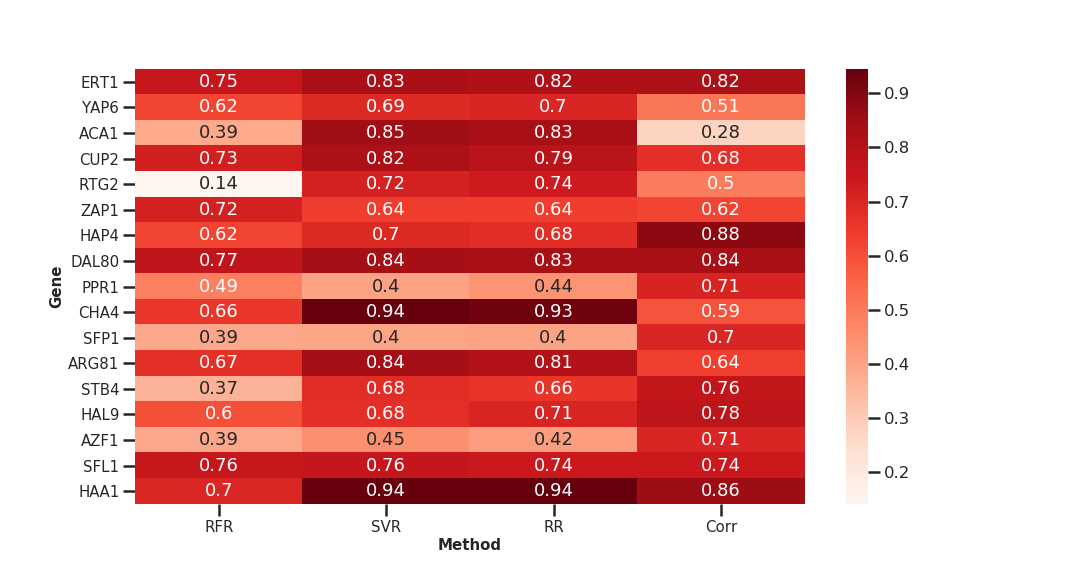}
    \caption{AUROC values for genes where at least one of the four methods (RFR, SVR, RR, Corr) gives AUROC above 0.7.}
    \label{tab:AUROC-above-07-yeast}
\end{figure}

Although AUROCs generally correlated between methods (Fig. \ref{fig:eqtl-prediction-yeast}B), in line with the correlation of RMSE values, AUROC values tended to be systematically higher for SVR and RR compared to RFR and univariate correlations. Interestingly, univariate correlation and SVR share the same number of TF eQTLs with AUROC $>0.70$ (10), only 5 were common and each method had five TFs not found by the other method. (Fig.\ref{tab:AUROC-above-07-yeast}).

\begin{figure}[h!]
    \centering
    \includegraphics[width=1.00\linewidth]{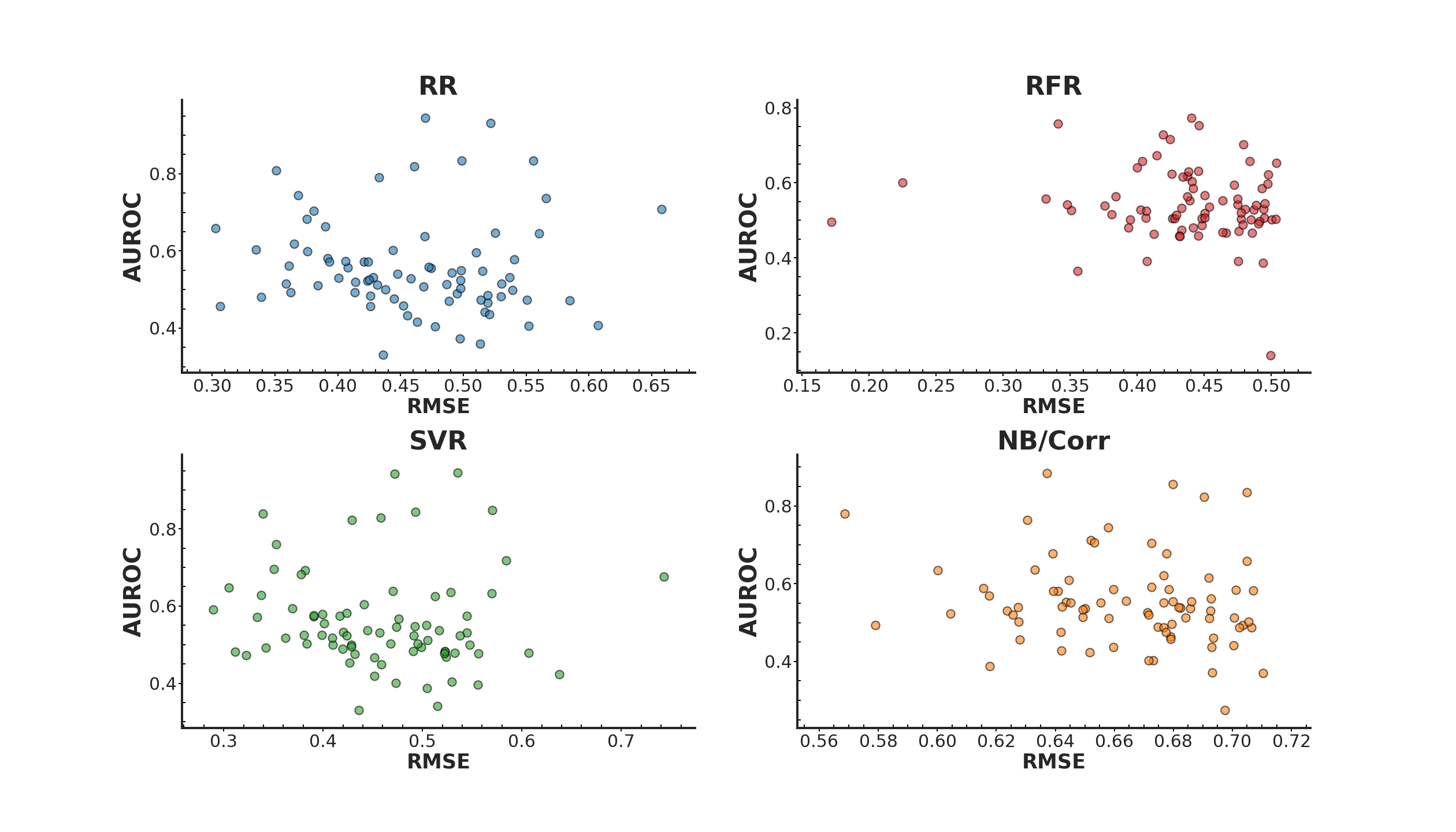}
    \caption{Scatter plots show trans-eQTL target prediction accuracy (AUROC) vs genotype prediction accuracy for random forest regression (RFR), support vector regression (SVR), ridge regression (RR), and naive Bayes/univariate correlation (NB/Corr) on yeast data. Genes on the same chromosome were excluded as predictors for each SNP.}
    \label{fig:genotype-target-comparison-yeast}
\end{figure}

\subsubsection{Genotype and trans-eQTL prediction performance do not correlate}

Similar to the DREAM data we again observed poor correlation between genotype and trans-eQTL prediction performance  (Fig. \ref{fig:genotype-target-comparison-yeast}).

\subsubsection{Feature selection in random forest produces a map of transcriptional hotspots}
\label{sec:feature-selection-ml}

\begin{figure}[h!]
  \centering
  \includegraphics[width=1.00\linewidth]{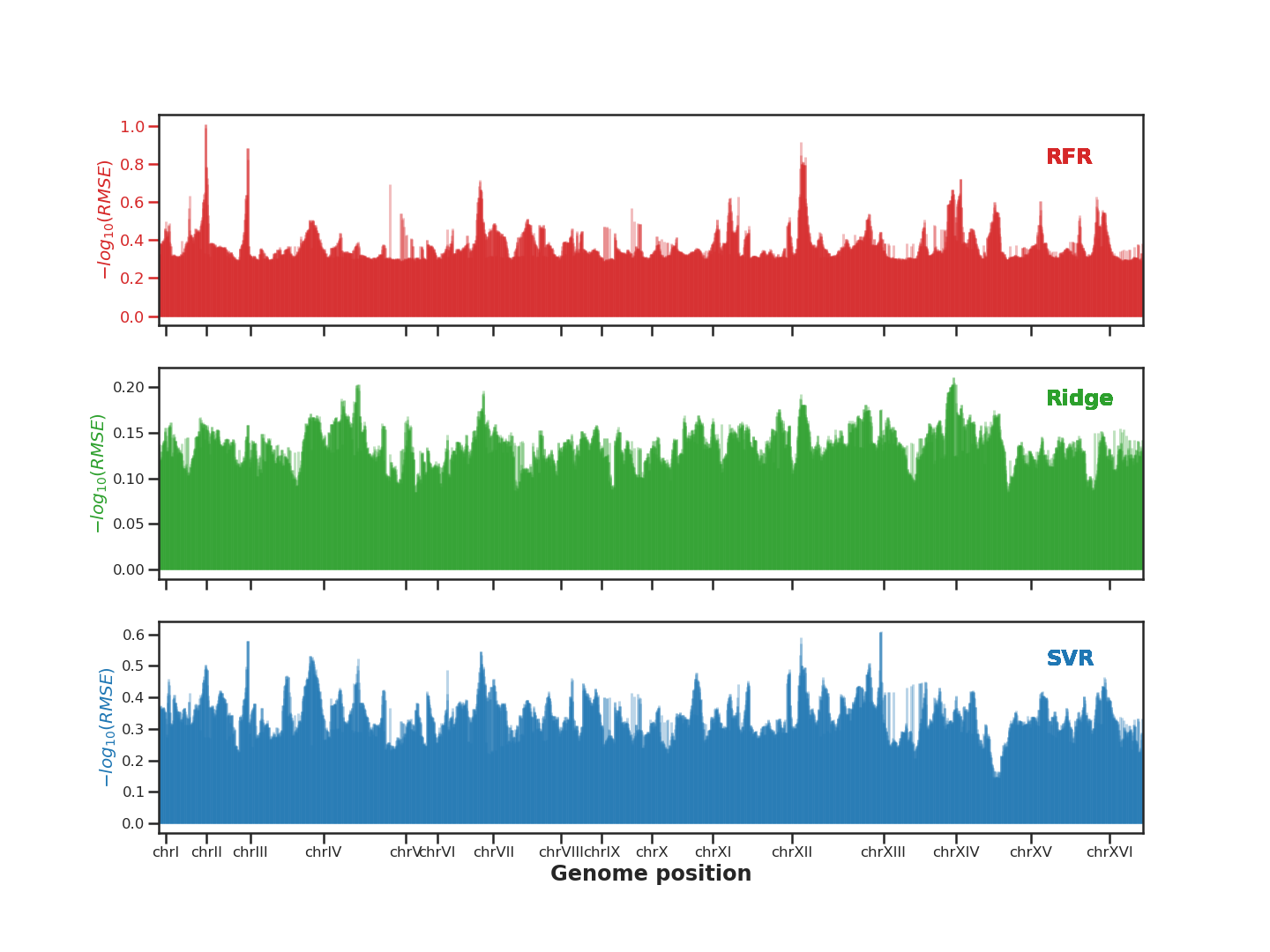}
  \caption{Expression hotspot maps showing the negative log transformed RMSE values vs genome position for 2884 SNPs in the yeast genome, for random forest (RF, top), ridge regression (Ridge, middle), and support vector regression (SVR, bottom). Genes on the same chromosome were excluded as predictors for each SNP.}
  \label{fig:hotspots-rmse-yeast}
\end{figure}

Transcriptional hotspots are regions of the genome associated with widespread changes in gene expression \cite{albert2018genetics}. We learned prediction models for all 2,884 SNPs in the yeast genome that were associated with local changes in gene expression and plotted the RMSE for each predicted SNP against its genome position.
RFR showed a wide variation in RMSE values for SNPs, across the whole genome, allowing to delineate genomic ranges with high and low regulatory activity. Whereas RR and SVR showed much less variation, and did not allow to separate high and low activity regions on most chromosomes (Fig. \ref{fig:hotspots-rmse-yeast}). Interestingly, the regions detected by RFR overlapped only partially with traditional hotspot maps based on univariate correlations (Supp. Fig. S22), again suggesting that non-linear methods like random forest may detect biological signals missed by traditional methods.

\section{Discussion}
\label{sec:discussion}

In this study we analyzed the use of machine learning methods for genotype prediction in high-dimensional multi-trait GWAS. The basic hypotheses of reverse genotype prediction from multiple trait combinations are that variants whose genotypes can be predicted with higher accuracy are more likely to have a true effect on a large number of the measured traits, and that feature importances or coefficients in the trained models indicate the strength of association between variants and individual traits. However, existing studies have not presented conclusive evidence for these hypotheses, because they only performed downstream analysis for the highest scoring variants, and only considered linear models. Here we performed an in-depth validation of various machine learning methods for reverse genotype prediction in the context of trans-eQTL analysis, including univariate, ridge regression, random forest, and support vector regression,  using both simulated and real transcriptional regulatory networks to define ground-truth sets of trans-eQTL target sets.

Our results support the basic hypotheses only partially. In particular, although genotype prediction performance indeed varied across genetic variants, there was no relation between genotype prediction performance and the number of gene expression traits affected by a variant, nor with the accuracy of predicting individual trans-eQTL target genes from model feature importances or coefficients. This is important, because it shows that in the absence of ground-truth information, we cannot use RMSE to select variants for which model features will overlap best with true trans-associated genes. This was further illustrated by the fact that random forest regression performed best at the genotype prediction task, but performed worst on the trans-eQTL prediction task. 


The only systematic relation we observed, both in the simulated and the yeast data, was a negative correlation between genotype prediction performance and number of model features, suggesting that if a variant can be predicted well, it can be done with a relatively small number of traits. 

While RMSE cannot be used to select variants with good trans-eQTL prediction performance, we did observe that model feature importances or coefficients were generally predictive of how likely a given gene is a true trans-eQTL target of a given variant. Predictive performance was very strong in simulated data, with more than 75\% of variants obtaining an AUROC  greater than 80\%, but also in yeast, 15-20\% of variants obtained an AUROC greater than 70\%.

An important goal of multi-trait GWAS is to distinguish between variants that are associated with high vs low number of traits. Interestingly, we found that only random forest, but not SVR or ridge regression, resulted in models with a wide variation in the number of selected features across variants. However this involved use of a simple, heuristic strategy for feature selection, and further research to finetune this result will be required.

One aspect of multi-trait GWAS not considered in this study is statistical inference. For linear methods, the null distribution of the model fit score under the assumption of no association can be approximated analytically to obtain a p-value for the significance of any observed score. Non-linear methods such as random forest or SVM require a large number of permutations for each variant separately to obtain a p-value, which becomes computationally infeasible for a large number of variants. However approximate methods may yet overcome this hurdle \cite{knijnenburg2009fewer}. More importantly though, since our results indicate that model fit is not related to either the strength or extent of true biological relations, the relevance of performing statistical inference on this test statistic is in doubt.

Another area of future research concerns the generalization to other organisms, in particular human. We focused on realistic simulated data from the DREAM project and data from the eukaryotic model organism yeast, due to the availability of data from a study with extraordinarily large sample size and extensive, high-quality ground-truth transcriptional interaction data. The availability of ground-truth associations also motivated our choice of studying gene expression traits. It will be of interest to expand this work to other types of traits, including protein and metabolite levels, as well as high-dimensional phenotypic traits such as images. 

In summary, feature importance weights in machine learning models that predict genotypes from high-dimensional sets of traits identify biologically relevant variant-trait associations, but comparing the relative importance of variants through these models in a GWAS-like manner using a single test statistic remains an open challenge.

\bibliographystyle{ieeetr}
\bibliography{refs}

\newpage

\renewcommand\thesection{S\arabic{section}}
\renewcommand\thefigure{S\arabic{figure}}
\renewcommand\thetable{S\arabic{table}}
\renewcommand\theequation{S\arabic{equation}}
\setcounter{figure}{0}
 \setcounter{table}{0}
 \setcounter{section}{0}
\setcounter{equation}{0}

\begin{center}
  {\LARGE \textbf{Supplementary Information}}
\end{center}

\section{Canonical Correlation Analysis}
\label{sec:canon-corr-analys}
Given two sets of random variables $(X_1,X_2,\dots, X_p)$ and $(Y_1,Y_2,\dots,Y_q)$, CCA finds linear coefficients $a\in\R^p$ and $b\in\R^q$ that maximize the correlation
\begin{align*}
  \rho(a,b) = \corr\left(\sum_{i=1}^p a_iX_i,\sum_{j=1}^qb_jY_j\right)
\end{align*}

It can be shown \footnotemark[1] that the optimal vector $\mathbf{a}$ is an eigenvector of the matrix $\Sigma_{XX}^{-1}\Sigma_{XY}\Sigma_{YY}^{-1}\Sigma_{YX}$, where $\Sigma_{XX}$, $\Sigma_{XY}$ and $\Sigma_{YY}$ are the covariance matrices among the $X$ and $Y$ variables.  In the special case where $q=1$ (one SNP), $\Sigma_{YY}$ is a number and $\Sigma_{XY}$ a column vector, and this matrix takes the form $\Sigma_{XX}^{-1}\vv\vv^T$, where $\vv=\Sigma_{YY}^{-1}\Sigma_{XY}$. The (only) eigenvector of such a matrix is $\mathbf{a}=\Sigma_{XX}^{-1}\vv$.

To estimate the coefficients $\mathbf{a}$  from data,  assume that we have standardized data $\X\in\R^{n\times p}$ and $\y\in\R^n$, such that
\begin{align*}
  \sum_{k=1}^n x_{ik} = \sum_{k=1}^n y_k = 0
  && \frac1{n-1}\sum_{k=1}^n x_{ik}^2 = \frac1{n-1}\sum_{k=1}^n y_k^2 = 1.
\end{align*}
Then the estimates for the covariances are
\begin{align*}
  \hat\Sigma_{XX} = \frac{\X^T\X}{n-1} && \hat\Sigma_{XY}=\frac{\X^T\y}{n-1} && \hat\Sigma_{YY}=1,
\end{align*}
and hence
\begin{align*}
  \mathbf{\hat a} = (\X^T\X)^{-1} \X^T\y.
\end{align*}

\section{Supplementary Figures}
\label{sec:supp_figs}

\begin{figure}[h!]
\begin{minipage}[t]{0.50\linewidth}
  \textbf{A}\\
      \includegraphics[width=\linewidth]{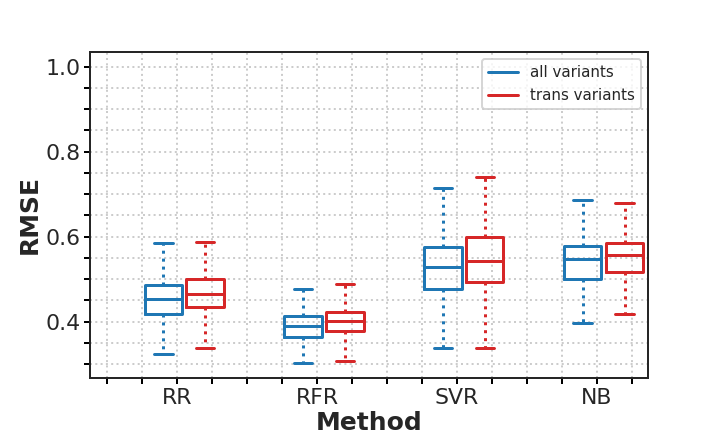}
      
  \end{minipage}
  \hfill
  \begin{minipage}[t]{0.50\linewidth}
  \textbf{B}\\
      \includegraphics[width=1.1\linewidth, height = 0.6\textwidth]{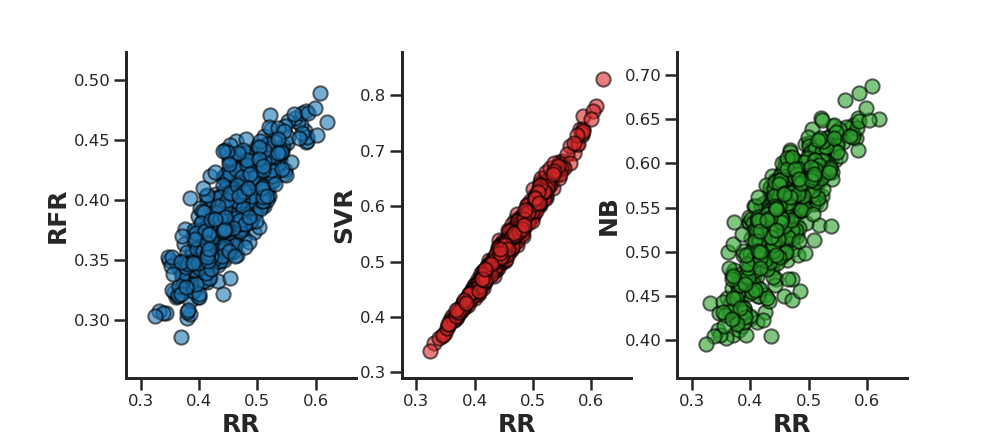}
\end{minipage}
  \caption{RMSE values for genotype prediction on DREAM5 simulated data. \textbf{A.} Boxplots show the distribution of the RMSE values for all variants (blue) and for trans-acting-only variants (red) for random forest regression (RFR), support vector regression (SVR), ridge regression (RR), and naive Bayes (NB). \textbf{B.} Scatter plots show RMSE values of RFR, SVR, and NB vs RR for all variants. The data shown are for \textbf{DREAM Network 2}.}
  \label{fig:genotype-prediction-dream2}
\end{figure}

\begin{figure}[t]
\begin{minipage}[t]{0.50\linewidth}
  \textbf{A}\\
      \includegraphics[width=\linewidth]{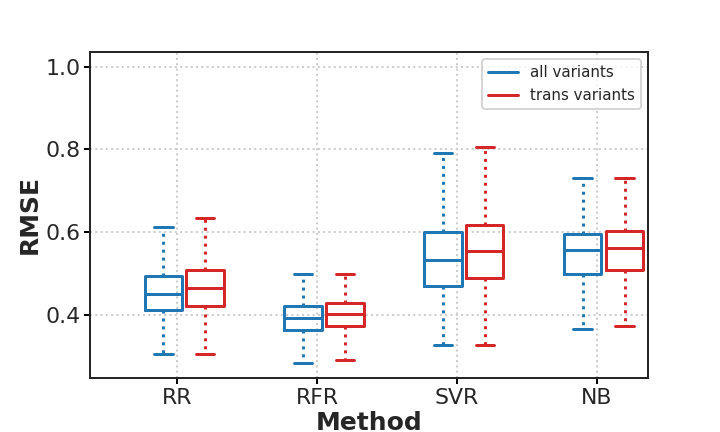}
      
  \end{minipage}
  \hfill
  \begin{minipage}[t]{0.50\linewidth}
  \textbf{B}\\
      \includegraphics[width=1.1\linewidth, height = 0.6\textwidth]{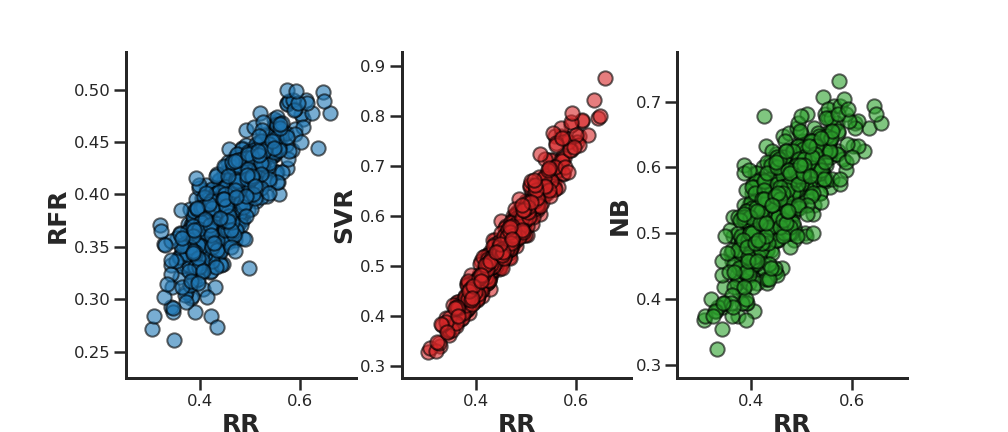}
\end{minipage}
  \caption{RMSE values for genotype prediction on DREAM5 simulated data. \textbf{A.} Boxplots show the distribution of the RMSE values for all variants (blue) and for trans-acting-only variants (red) for random forest regression (RFR), support vector regression (SVR), ridge regression (RR), and naive Bayes (NB). \textbf{B.} Scatter plots show RMSE values of RFR, SVR, and NB vs RR for all variants. The data shown are for \textbf{DREAM Network 3}.}
  \label{fig:genotype-prediction-dream3}
\end{figure}

\begin{figure}[t]
\begin{minipage}[t]{0.50\linewidth}
  \textbf{A}\\
      \includegraphics[width=\linewidth]{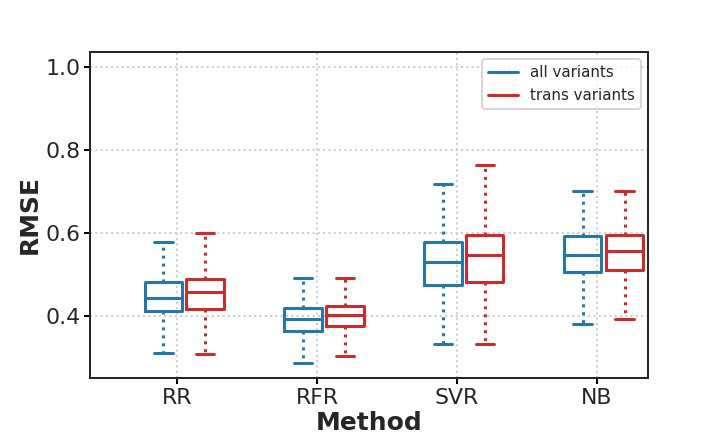}
      
  \end{minipage}
  \hfill
  \begin{minipage}[t]{0.50\linewidth}
  \textbf{B}\\
      \includegraphics[width=1.1\linewidth, height = 0.6\textwidth]{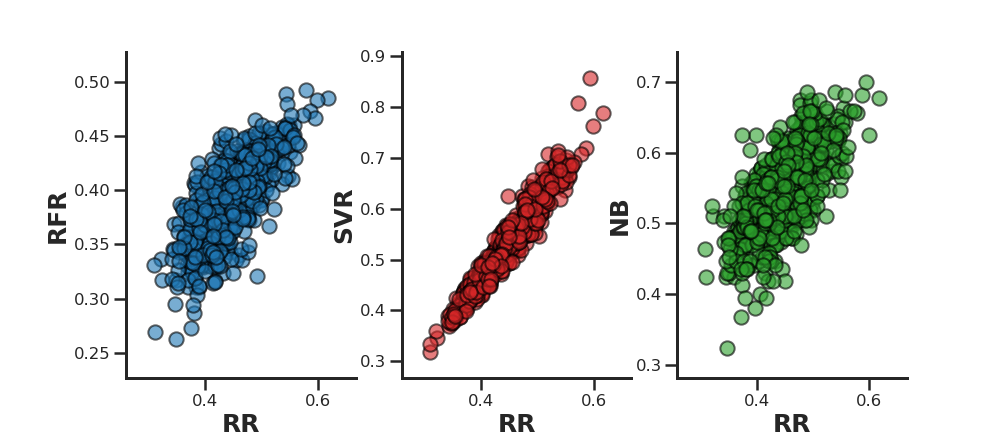}
\end{minipage}
  \caption{RMSE values for genotype prediction on DREAM5 simulated data. \textbf{A.} Boxplots show the distribution of the RMSE values for all variants (blue) and for trans-acting-only variants (red) for random forest regression (RFR), support vector regression (SVR), ridge regression (RR), and naive Bayes (NB). \textbf{B.} Scatter plots show RMSE values of RFR, SVR, and NB vs RR for all variants. The data shown are for \textbf{DREAM Network 4}.}
  \label{fig:genotype-prediction-dream4}
\end{figure}

\begin{figure}[t]
\begin{minipage}[t]{0.50\linewidth}
  \textbf{A}\\
      \includegraphics[width=\linewidth]{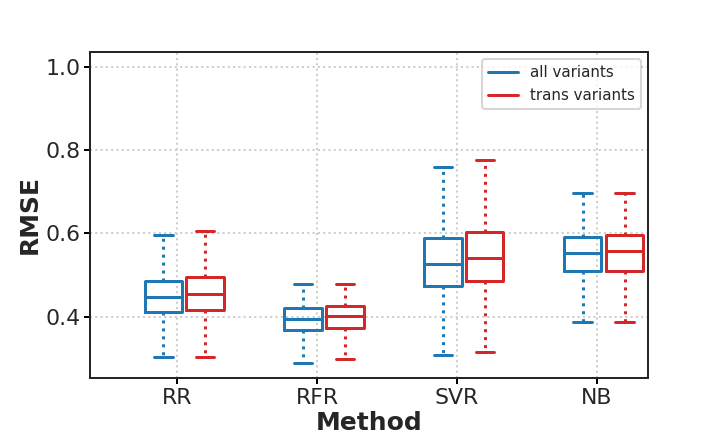}
      
  \end{minipage}
  \hfill
  \begin{minipage}[t]{0.50\linewidth}
  \textbf{B}\\
      \includegraphics[width=1.1\linewidth, height = 0.6\textwidth]{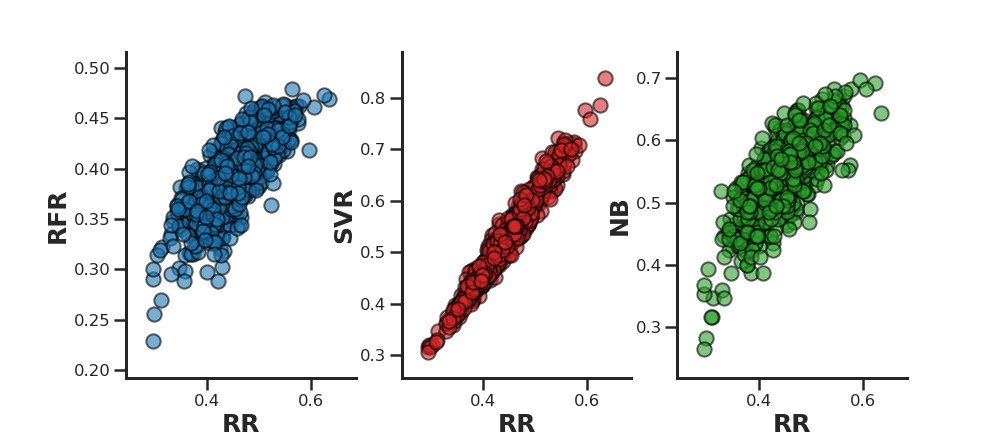}
\end{minipage}
  \caption{RMSE values for genotype prediction on DREAM5 simulated data. \textbf{A.} Boxplots show the distribution of the RMSE values for all variants (blue) and for trans-acting-only variants (red) for random forest regression (RFR), support vector regression (SVR), ridge regression (RR), and naive Bayes (NB). \textbf{B.} Scatter plots show RMSE values of RFR, SVR, and NB vs RR for all variants. The data shown are for \textbf{DREAM Network 5}.}
  \label{fig:genotype-prediction-dream5}
\end{figure}

\begin{figure}[t!]
  \begin{minipage}[t]{0.50\linewidth}
  \textbf{A}\\
      \includegraphics[width=\linewidth]{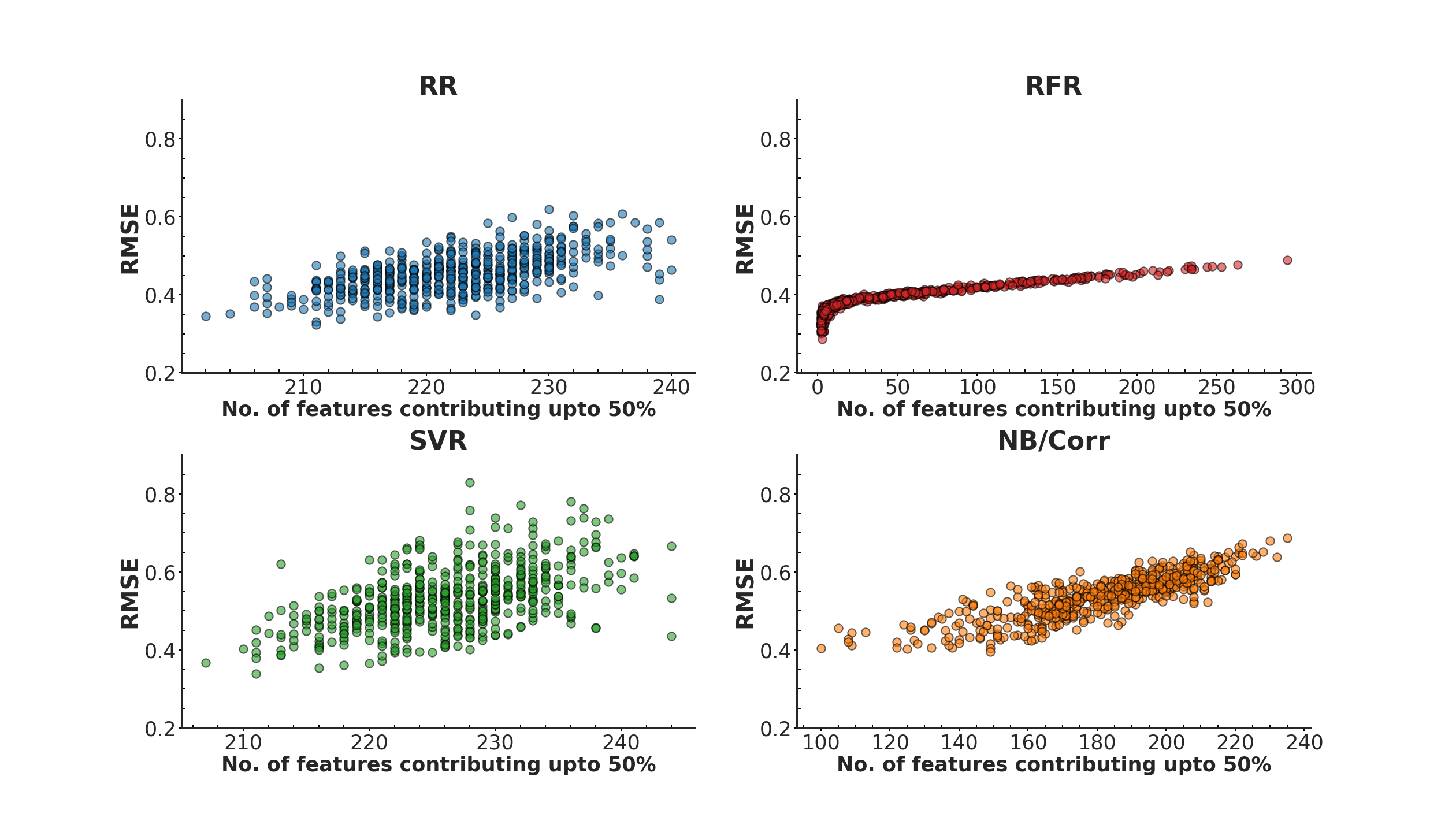}
  \end{minipage}
  \hfill
  \begin{minipage}[t]{0.50\linewidth}
  \textbf{B}\\
      \includegraphics[width=\linewidth]{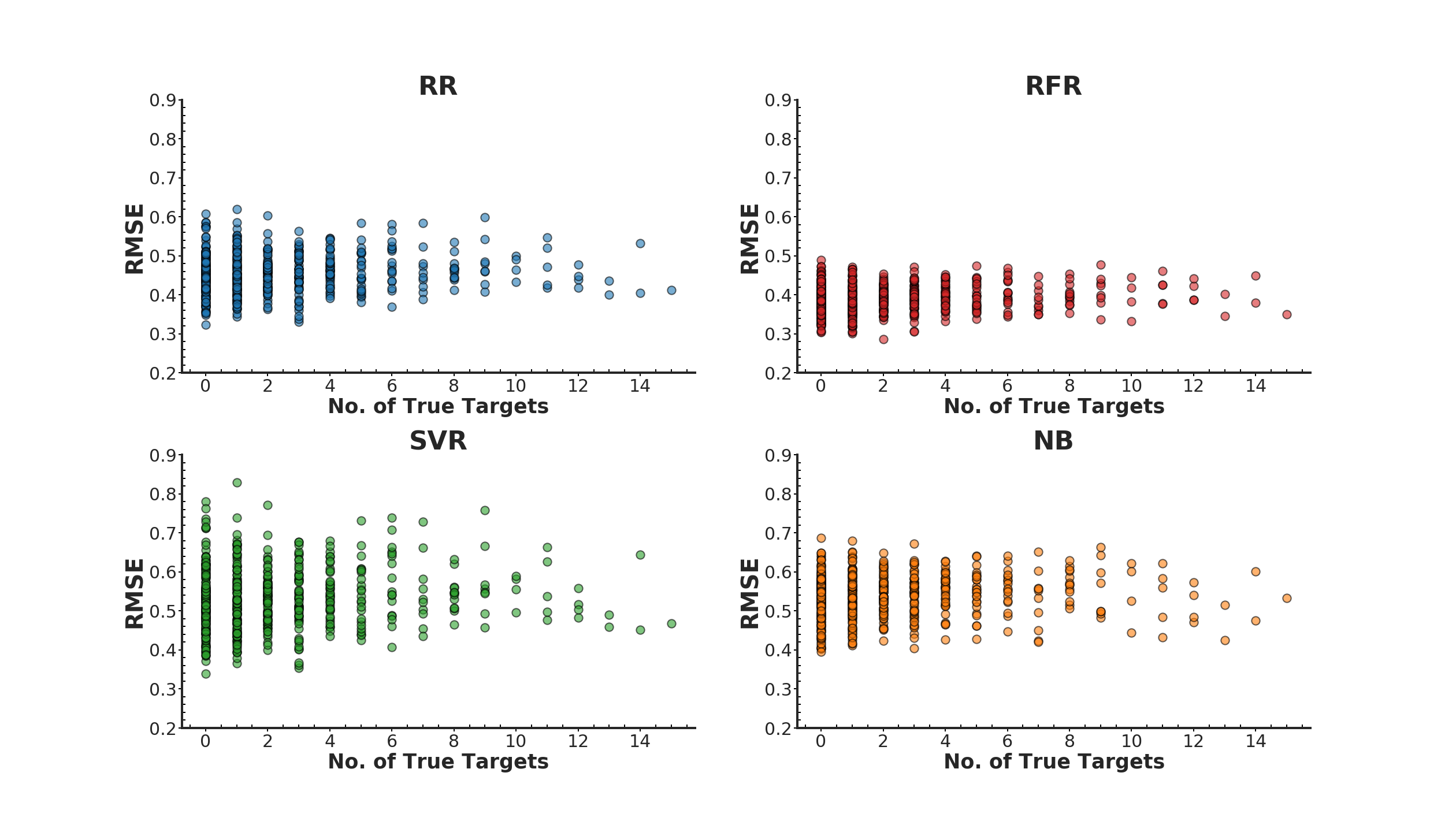}
  \end{minipage}
  \hfill
  \begin{center}
        \begin{minipage}[t]{0.50\linewidth}
        \textbf{C}\\
      \includegraphics[width=\linewidth]{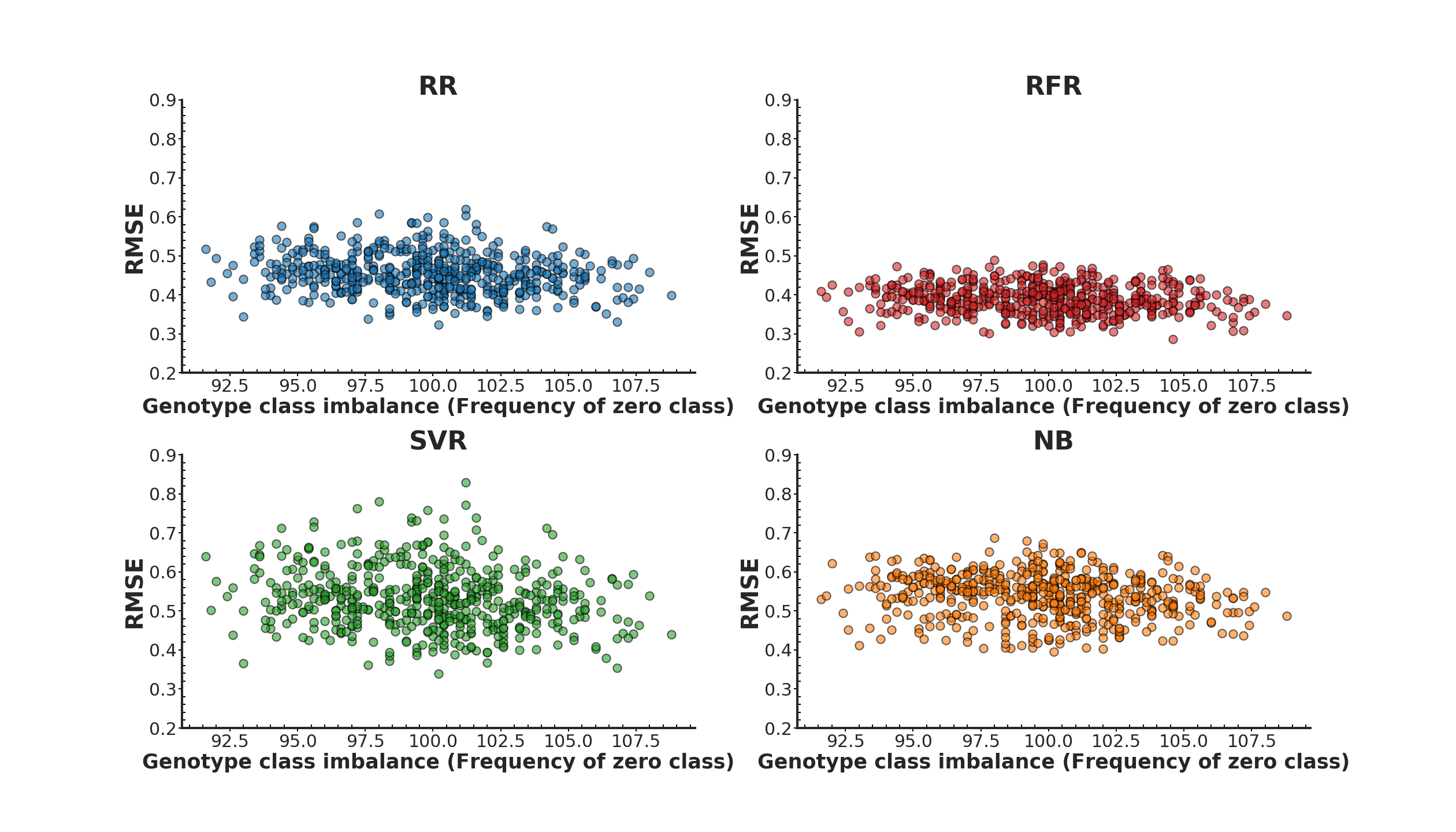}
  \end{minipage}
  \end{center}
  \caption{Scatter plots of genotype RMSE values on DREAM5 simulated data against the number of selected model features (\textbf{A}), the number of true trans-eQTL targets in the ground-truth network (\textbf{B}), and the genotype class balance (frequency of the zero class) (\textbf{C}), for random forest regression (RFR), support vector regression (SVR), ridge regression (RR), and naive Bayes (NB). The data shown are for \textbf{DREAM Network 2}.}
  \label{fig:genotype-prediction-dream-control2}
\end{figure}

\begin{figure}[t!]
  \begin{minipage}[t]{0.50\linewidth}
  \textbf{A}\\
      \includegraphics[width=\linewidth]{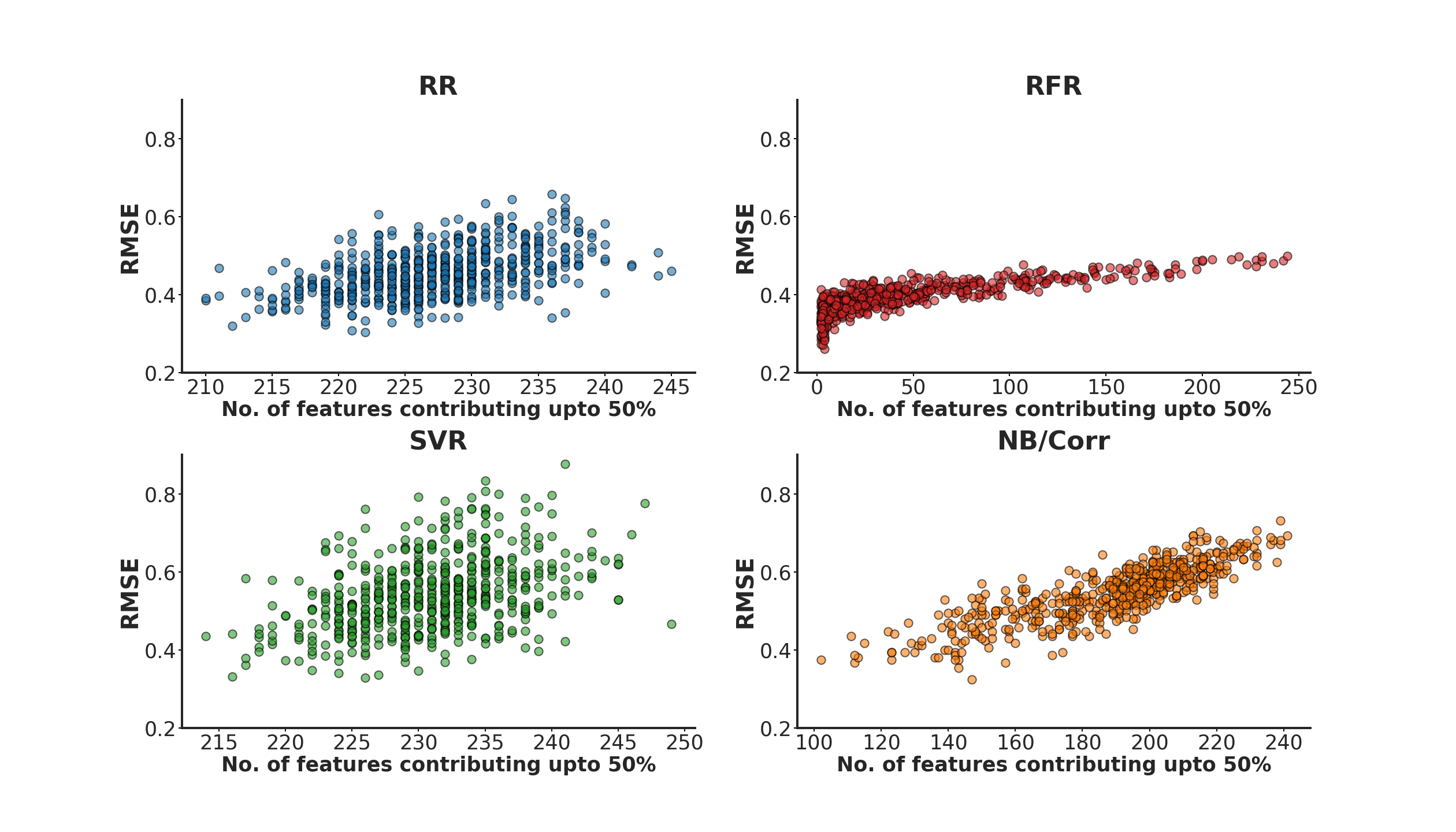}
  \end{minipage}
  \hfill
  \begin{minipage}[t]{0.50\linewidth}
  \textbf{B}\\
      \includegraphics[width=\linewidth]{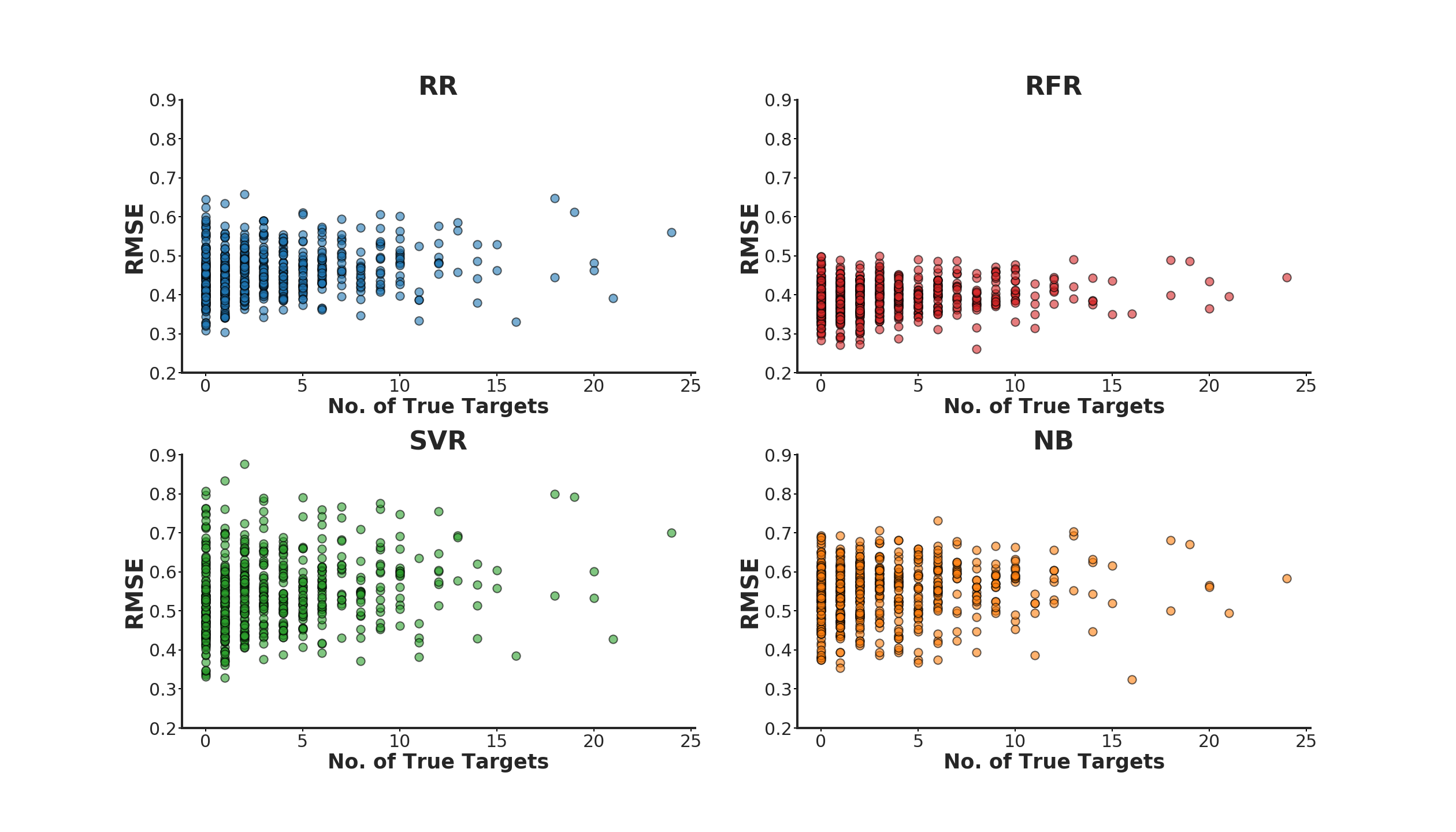}
  \end{minipage}
  \hfill
  \begin{center}
        \begin{minipage}[t]{0.50\linewidth}
        \textbf{C}\\
      \includegraphics[width=\linewidth]{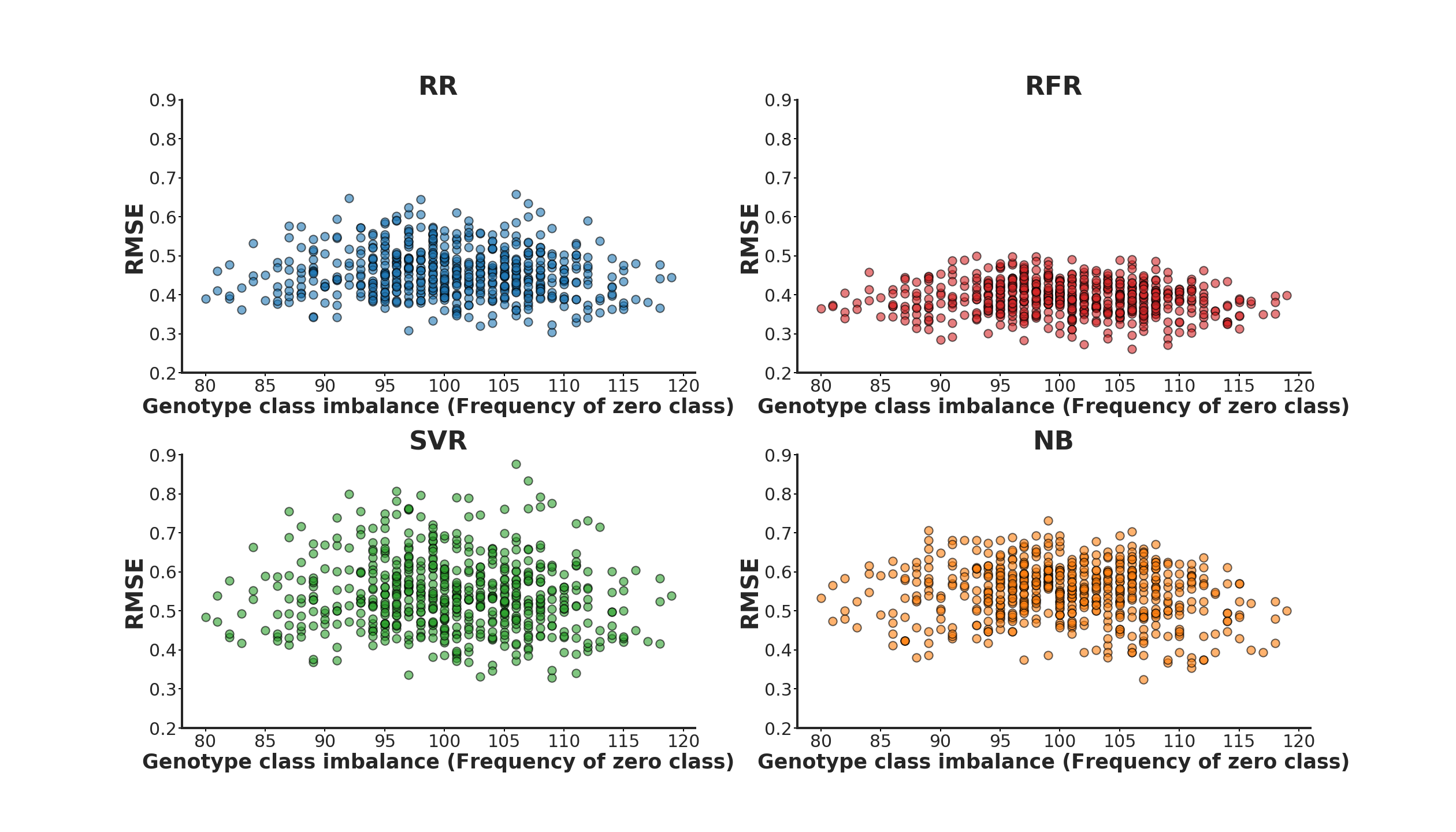}
  \end{minipage}
  \end{center}
  \caption{Scatter plots of genotype RMSE values on DREAM5 simulated data against the number of selected model features (\textbf{A}), the number of true trans-eQTL targets in the ground-truth network (\textbf{B}), and the genotype class balance (frequency of the zero class) (\textbf{C}), for random forest regression (RFR), support vector regression (SVR), ridge regression (RR), and naive Bayes (NB). The data shown are for \textbf{DREAM Network 3}.}
  \label{fig:genotype-prediction-dream-control3}
\end{figure}

\begin{figure}[t!]
  \begin{minipage}[t]{0.50\linewidth}
  \textbf{A}\\
      \includegraphics[width=\linewidth]{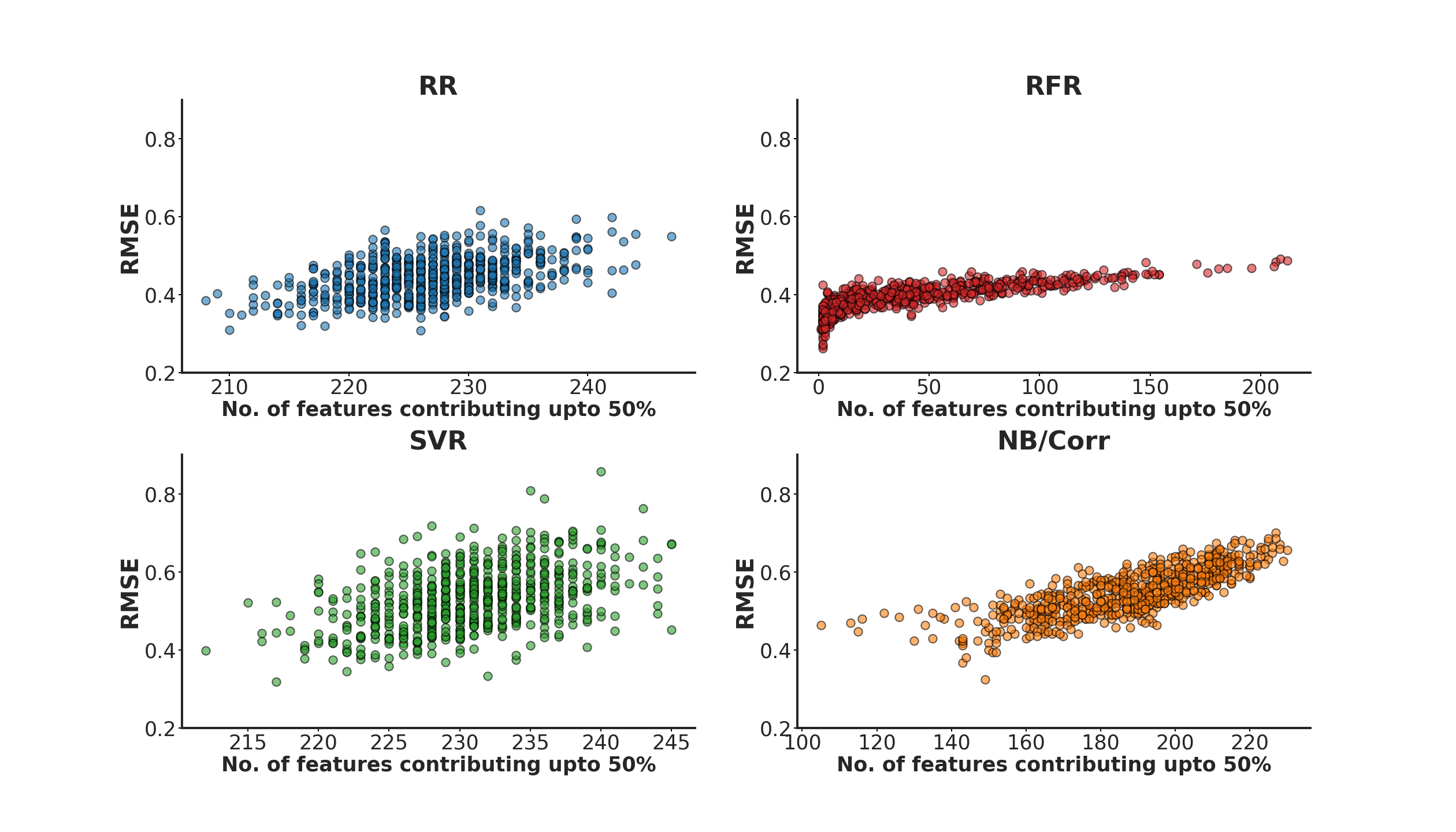}
  \end{minipage}
  \hfill
  \begin{minipage}[t]{0.50\linewidth}
  \textbf{B}\\
      \includegraphics[width=\linewidth]{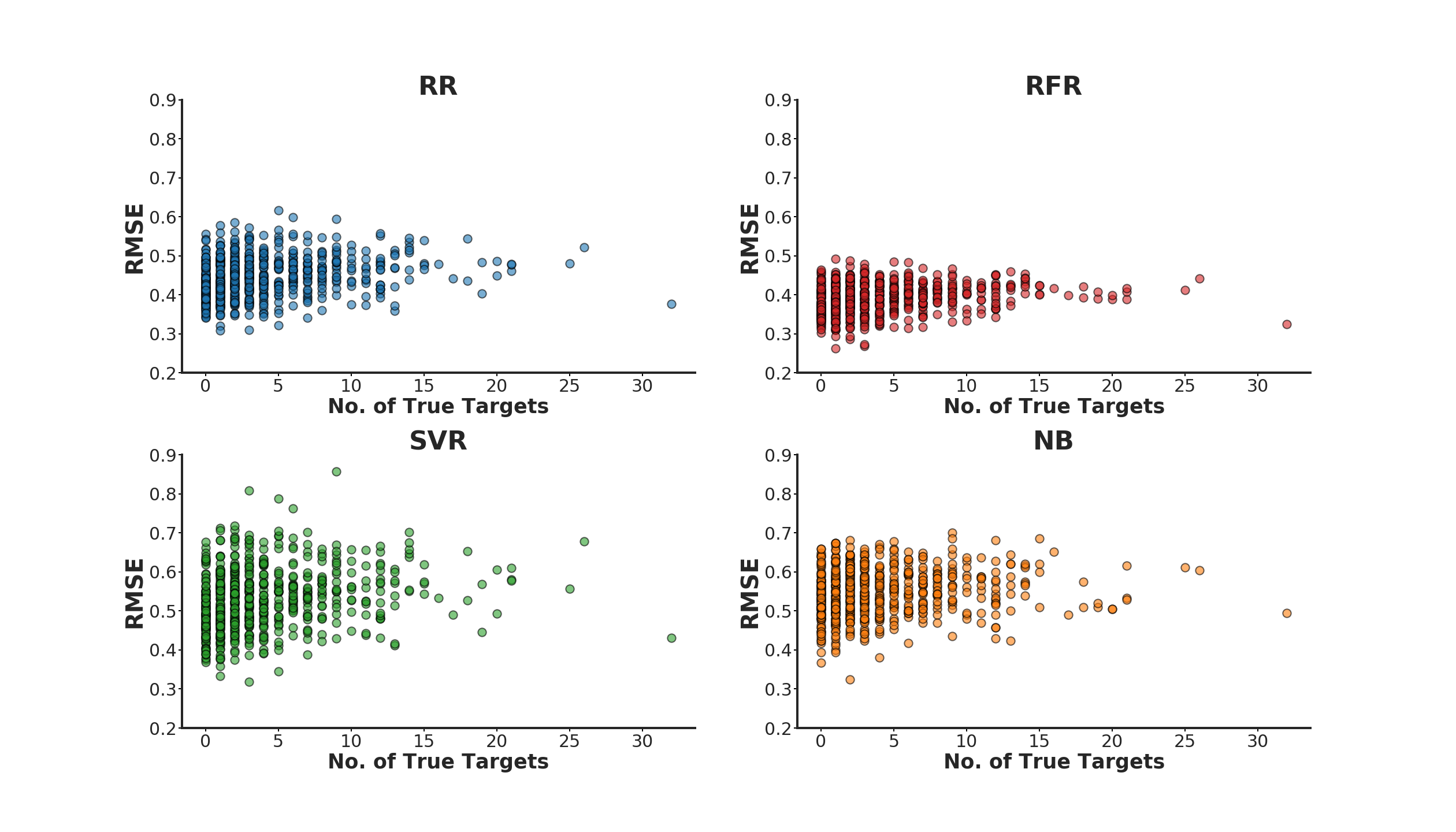}
  \end{minipage}
  \hfill
  \begin{center}
        \begin{minipage}[t]{0.50\linewidth}
        \textbf{C}\\
      \includegraphics[width=\linewidth]{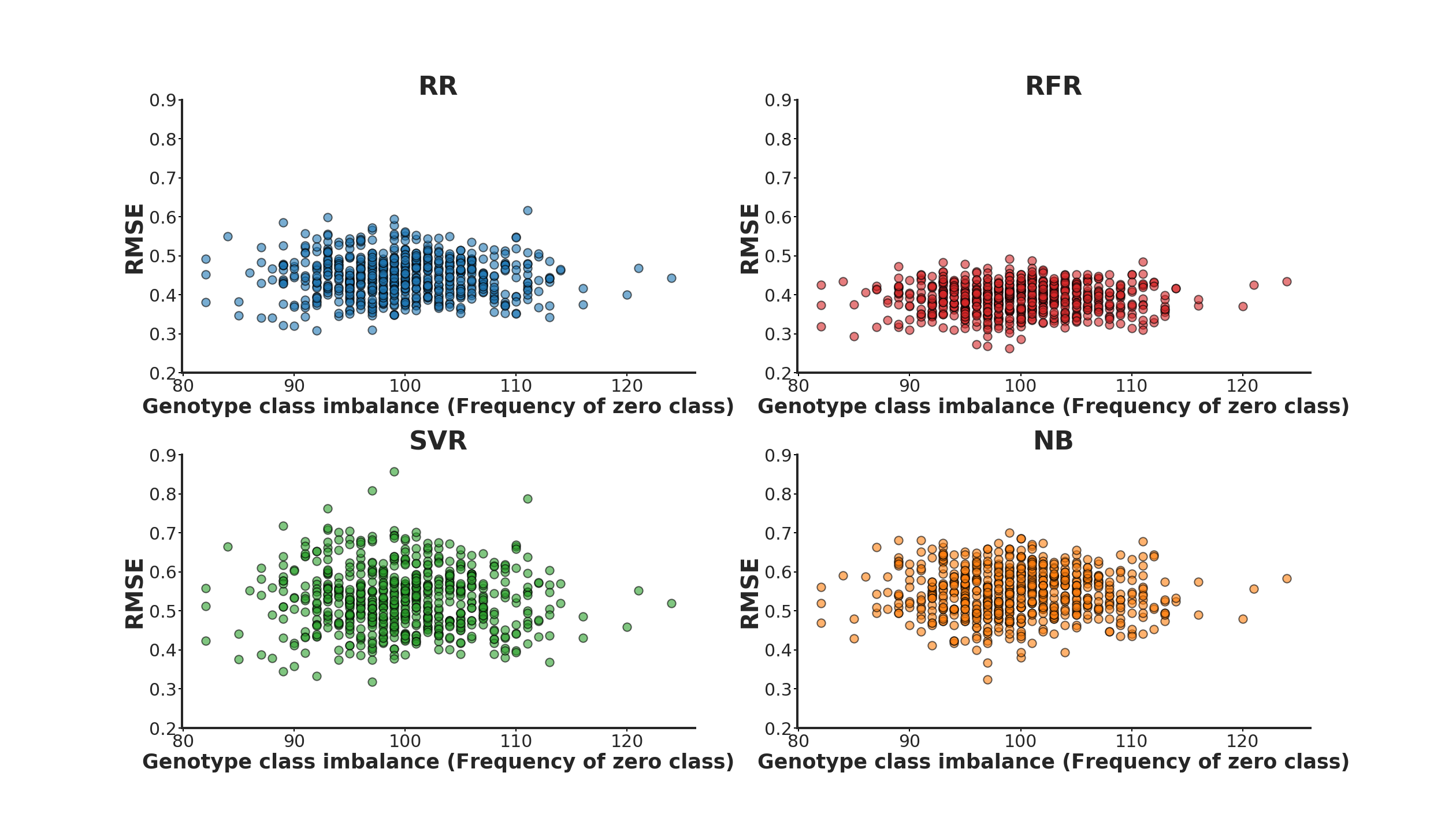}
  \end{minipage}
  \end{center}
  \caption{Scatter plots of genotype RMSE values on DREAM5 simulated data against the number of selected model features (\textbf{A}), the number of true trans-eQTL targets in the ground-truth network (\textbf{B}), and the genotype class balance (frequency of the zero class) (\textbf{C}), for random forest regression (RFR), support vector regression (SVR), ridge regression (RR), and naive Bayes (NB). The data shown are for \textbf{DREAM Network 4}.}
  \label{fig:genotype-prediction-dream-control4}
\end{figure}

\begin{figure}[t!]
  \begin{minipage}[t]{0.50\linewidth}
  \textbf{A}\\
      \includegraphics[width=\linewidth]{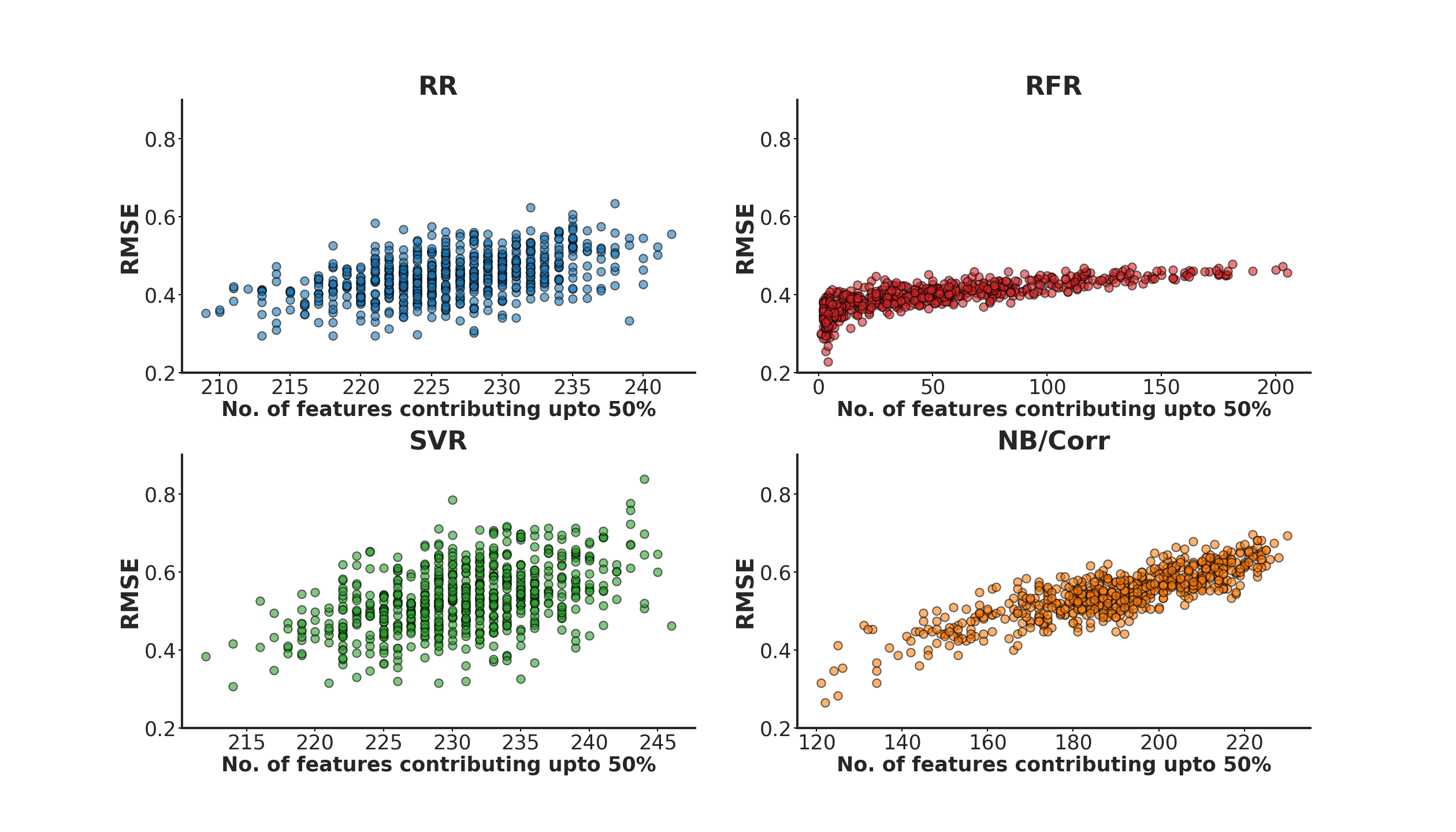}
  \end{minipage}
  \hfill
  \begin{minipage}[t]{0.50\linewidth}
  \textbf{B}\\
      \includegraphics[width=\linewidth]{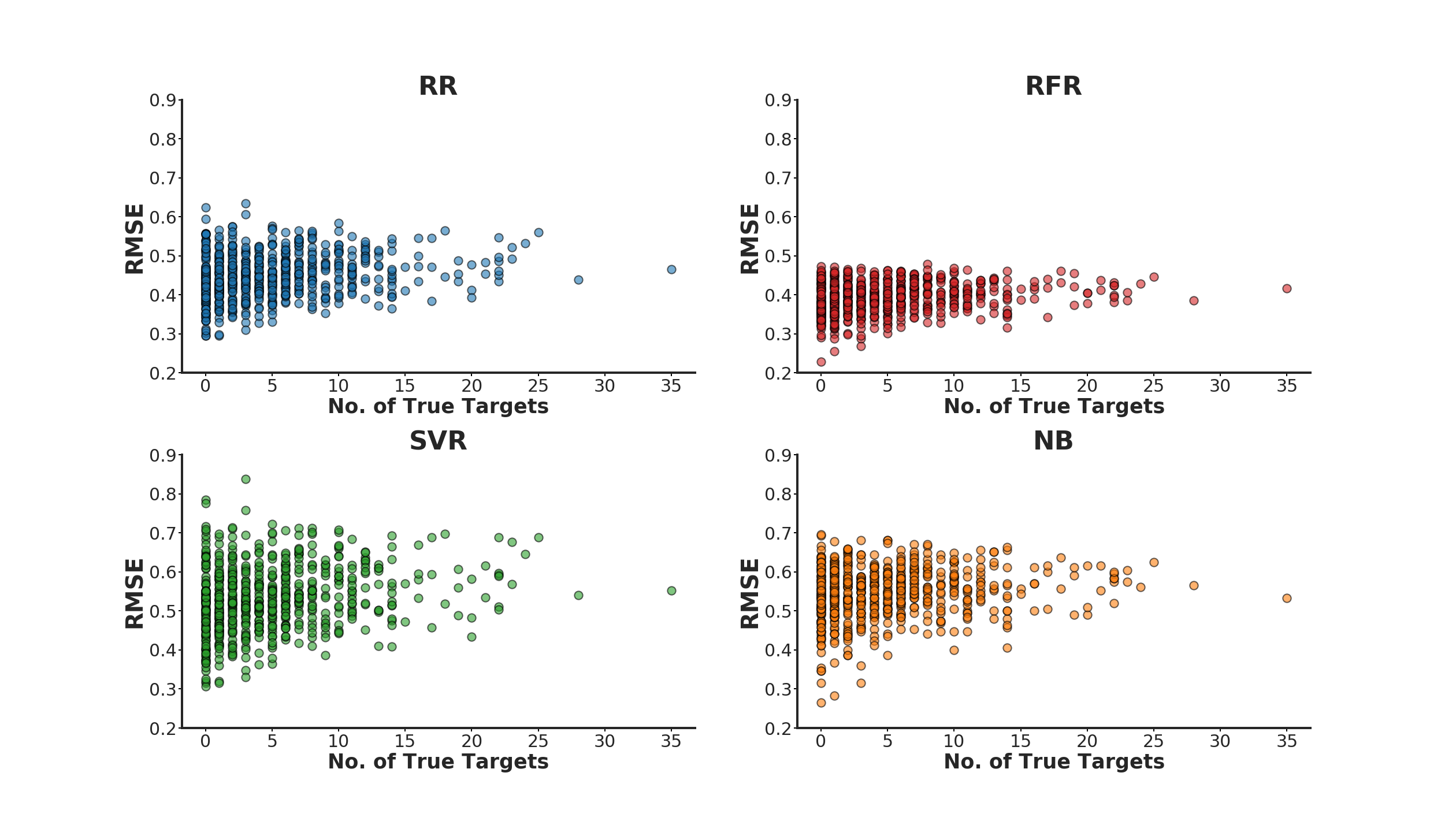}
  \end{minipage}
  \hfill
  \begin{center}
        \begin{minipage}[t]{0.50\linewidth}
        \textbf{C}\\
      \includegraphics[width=\linewidth]{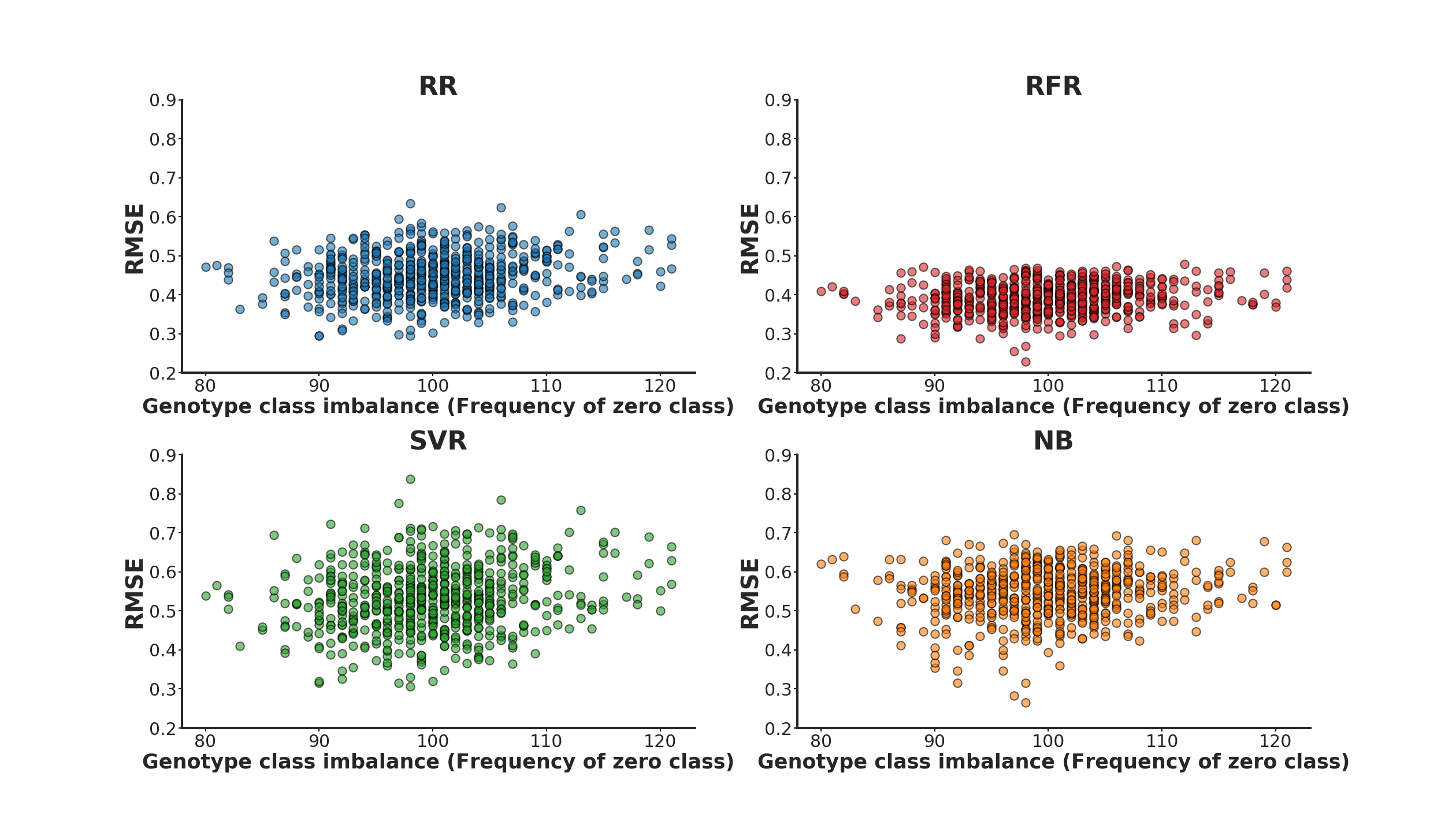}
  \end{minipage}
  \end{center}
  \caption{Scatter plots of genotype RMSE values on DREAM5 simulated data against the number of selected model features (\textbf{A}), the number of true trans-eQTL targets in the ground-truth network (\textbf{B}), and the genotype class balance (frequency of the zero class) (\textbf{C}), for random forest regression (RFR), support vector regression (SVR), ridge regression (RR), and naive Bayes (NB). The data shown are for \textbf{DREAM Network 5}.}
  \label{fig:genotype-prediction-dream-control5}
\end{figure}

\begin{figure}[h!]
\begin{minipage}[t]{0.50\linewidth}
\textbf{A}\\
\includegraphics[width=0.9\linewidth]{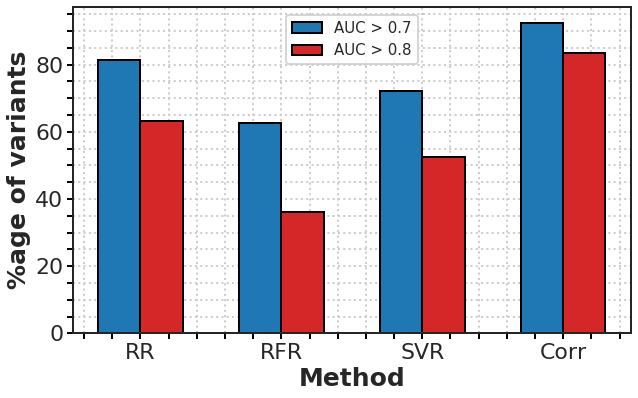}
\end{minipage}
\begin{minipage}[t]{0.50\linewidth}
\textbf{B}\\
\includegraphics[width=0.9\linewidth]{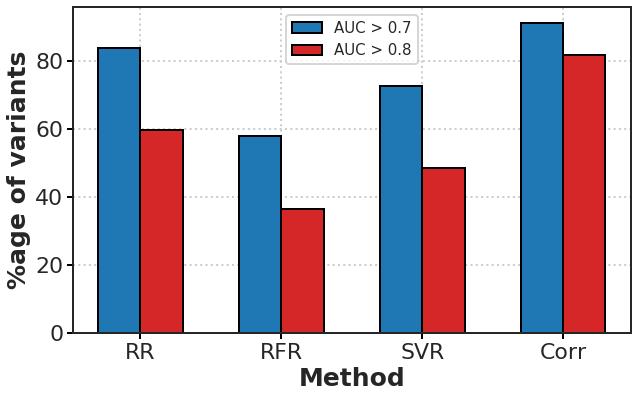}
\end{minipage}
\begin{minipage}[t]{0.50\linewidth}
\textbf{C}\\
\includegraphics[width=0.9\linewidth]{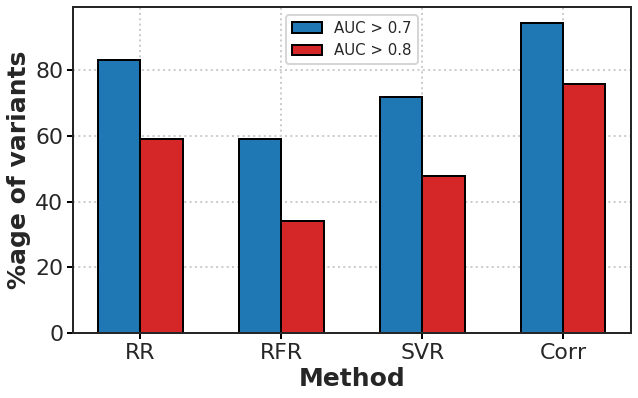}
\end{minipage}
\begin{minipage}[t]{0.50\linewidth}
\textbf{D}\\
\includegraphics[width=0.9\linewidth]{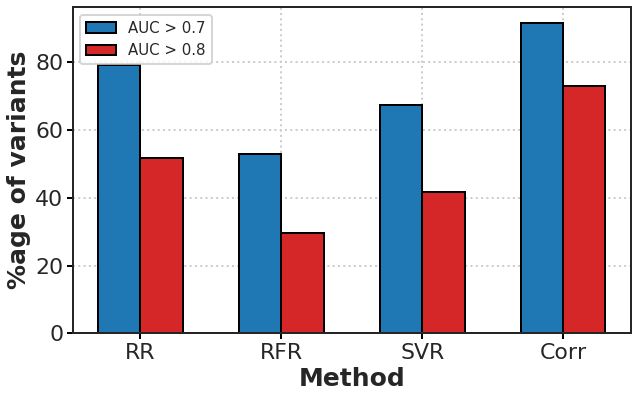}
\end{minipage}

\caption{Bar plots show the proportion of variants with trans-eQTL target prediction AUROC $> 0.7$ (blue) and $> 0.8$ (red) for random forest regression (RFR), support vector regression (SVR), ridge regression (RR), and univariate correlation (Corr). (\textbf{A}) DREAM Network 2, (\textbf{B}) DREAM Network 3, (\textbf{C}) DREAM Network 4, (\textbf{D}) DREAM Network 5.}
\end{figure}

\begin{figure}[h!]
  \begin{minipage}[t]{.50\linewidth}
  \textbf{(A)}\\
      \includegraphics[width=\linewidth]{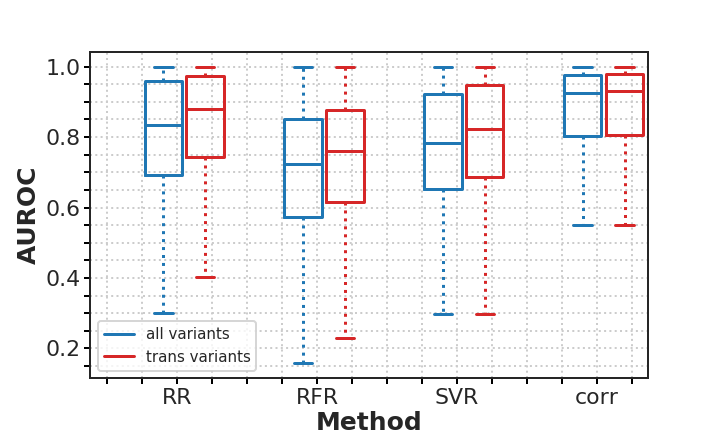}
  \end{minipage}
  \hfill
  \begin{minipage}[t]{.50\linewidth}
  \textbf{(B)}\\
      \includegraphics[width=1.1\linewidth, height=0.6\textwidth]{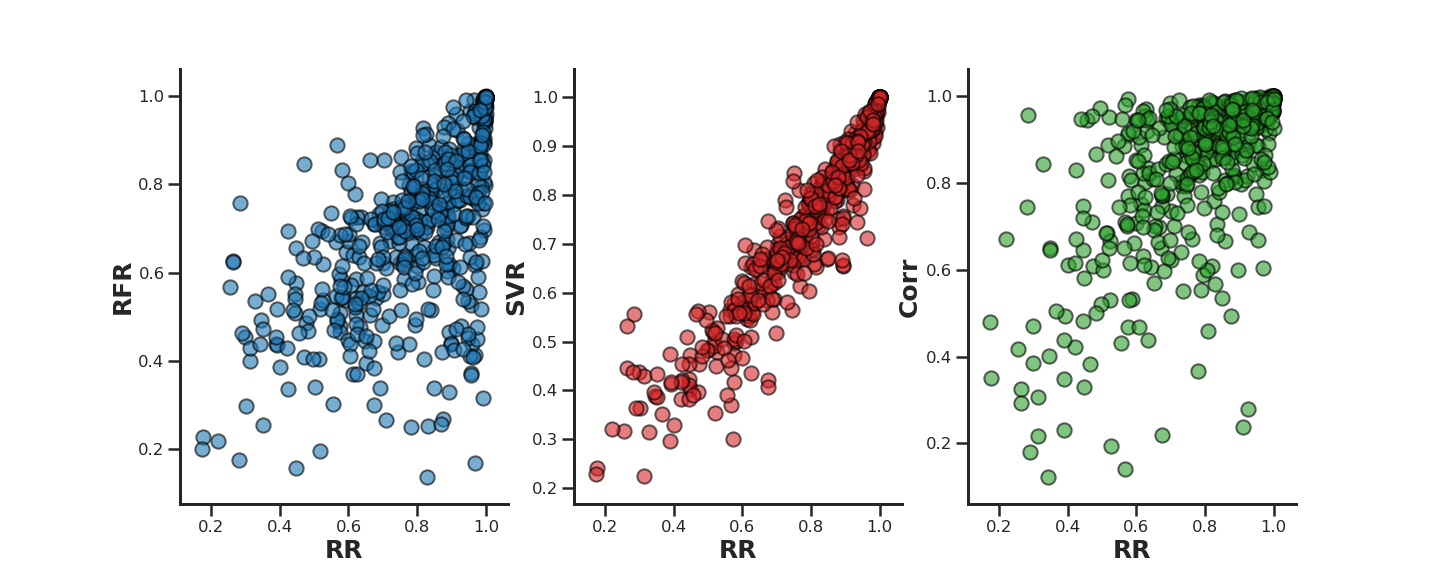}
  \end{minipage}
  \caption{Trans-eQTL target prediction performance on DREAM5 simulated data. \textbf{(A)} Boxplots show the distribution of AUROC values for all variants (blue) and for  trans-acting-only variants (red) for random forest regression (RFR), support vector regression (SVR), ridge regression (RR), and univariate correlation (Corr). \textbf{(B)} Scatter plots show AUROC values of classification methods RFR, SVR, and Corr vs RR for all variants. The data shown are for \textbf{DREAM Network 2}.}
  \label{fig:eqtl-prediction-dream2}
\end{figure}

\begin{figure}[h!]
  \begin{minipage}[t]{.50\linewidth}
  \textbf{(A)}\\
      \includegraphics[width=\linewidth]{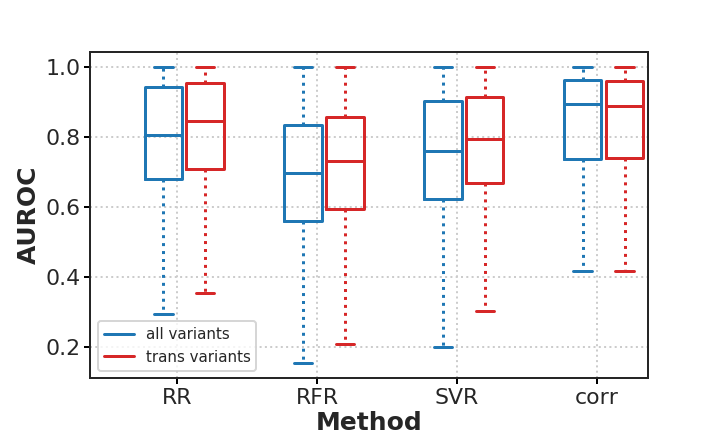}
  \end{minipage}
  \hfill
  \begin{minipage}[t]{.50\linewidth}
  \textbf{(B)}\\
      \includegraphics[width=1.1\linewidth, height=0.6\textwidth]{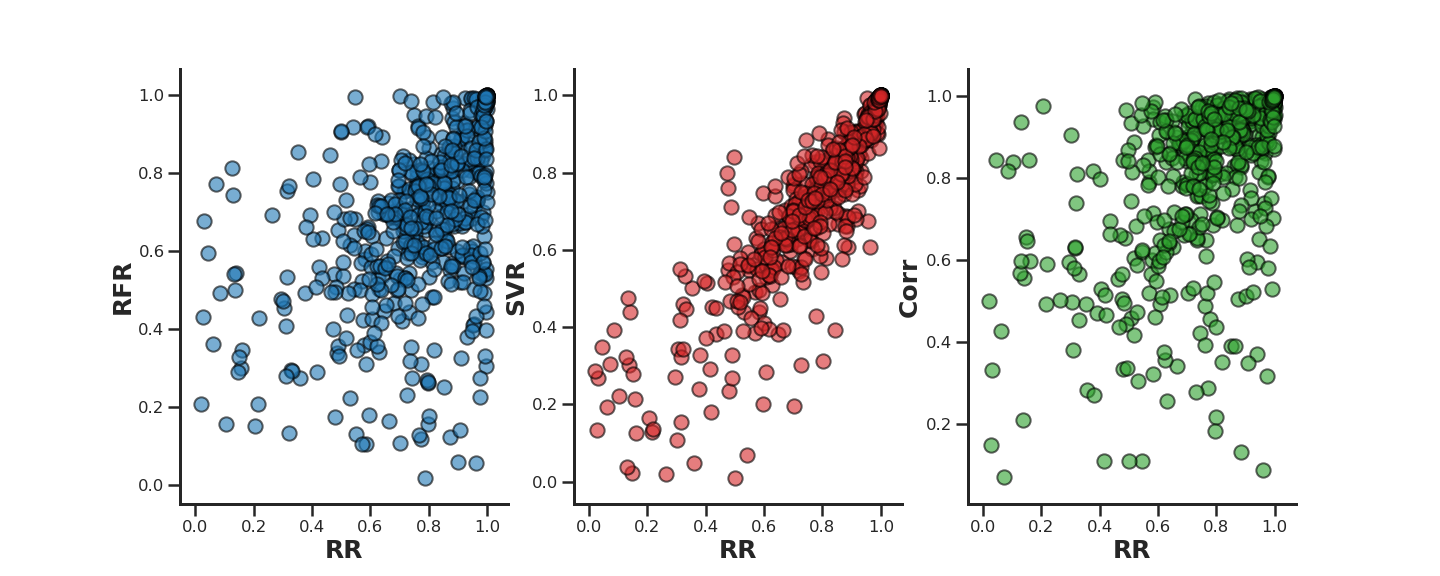}
  \end{minipage}
  \caption{Trans-eQTL target prediction performance on DREAM5 simulated data. \textbf{(A)} Boxplots show the distribution of AUROC values for all variants (blue) and for  trans-acting-only variants (red) for random forest regression (RFR), support vector regression (SVR), ridge regression (RR), and univariate correlation (Corr). \textbf{(B)} Scatter plots show AUROC values of classification methods RFR, SVR, and Corr vs RR for all variants. The data shown are for \textbf{DREAM Network 3}.}
  \label{fig:eqtl-prediction-dream3}
\end{figure}

\begin{figure}[h!]
  \begin{minipage}[t]{.50\linewidth}
  \textbf{(A)}\\
      \includegraphics[width=\linewidth]{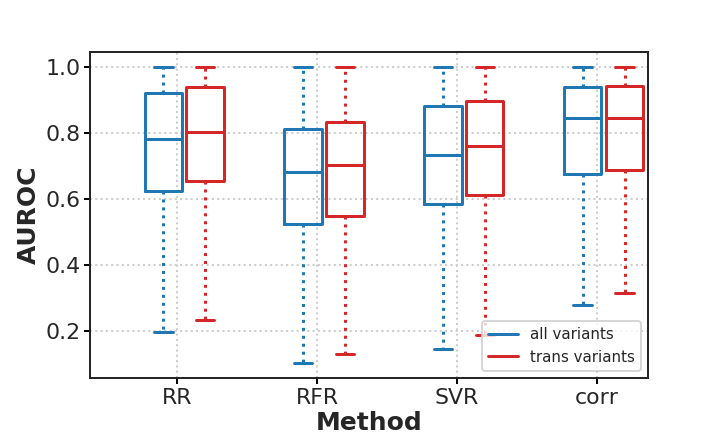}
  \end{minipage}
  \hfill
  \begin{minipage}[t]{.50\linewidth}
  \textbf{(B)}\\
      \includegraphics[width=1.1\linewidth, height=0.6\textwidth]{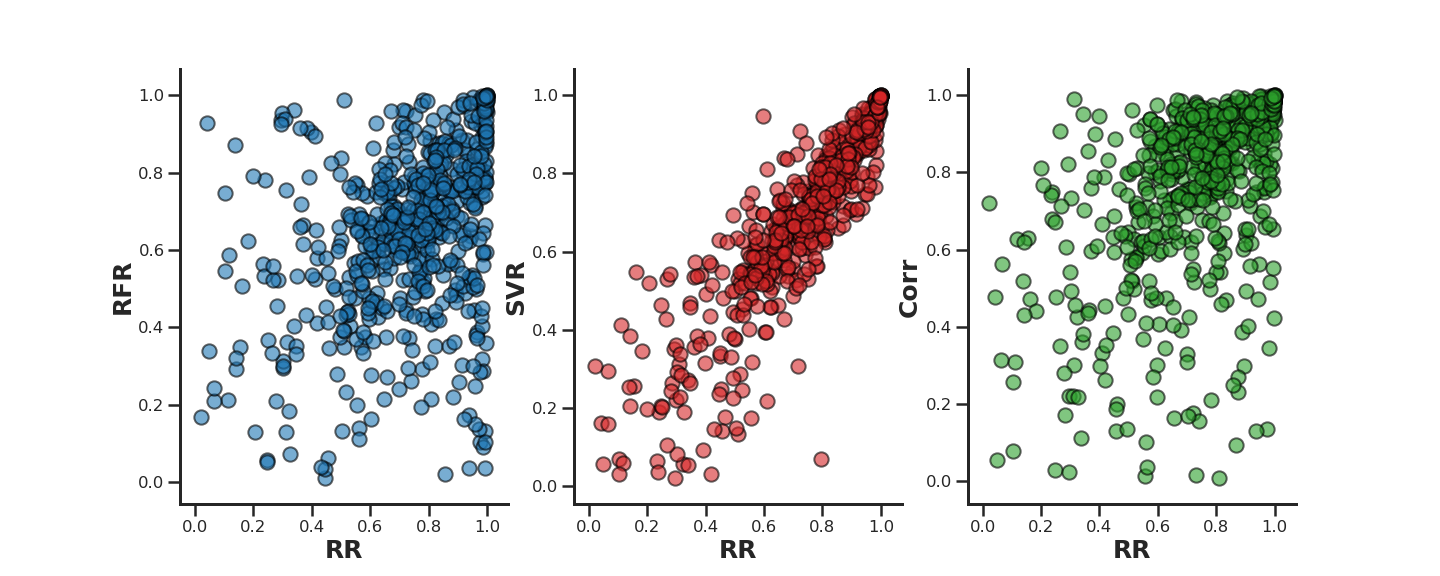}
  \end{minipage}
  \caption{Trans-eQTL target prediction performance on DREAM5 simulated data. \textbf{(A)} Boxplots show the distribution of AUROC values for all variants (blue) and for  trans-acting-only variants (red) for random forest regression (RFR), support vector regression (SVR), ridge regression (RR), and univariate correlation (Corr). \textbf{(B)} Scatter plots show AUROC values of classification methods RFR, SVR, and Corr vs RR for all variants. The data shown are for \textbf{DREAM Network 4}.}
  \label{fig:eqtl-prediction-dream4}
\end{figure}

\begin{figure}[h!]
  \begin{minipage}[t]{.50\linewidth}
  \textbf{(A)}\\
      \includegraphics[width=\linewidth]{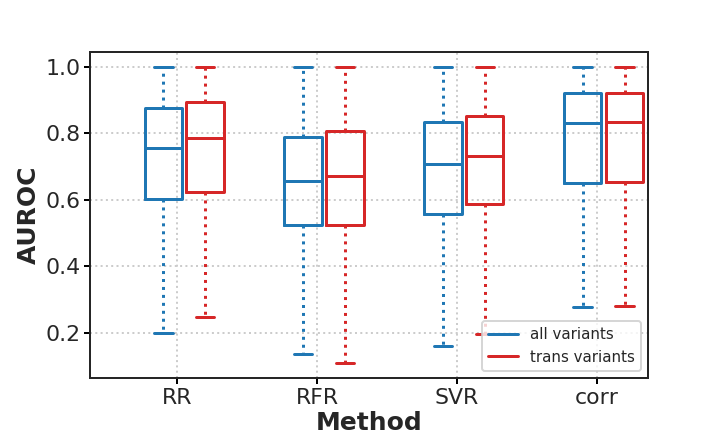}
  \end{minipage}
  \hfill
  \begin{minipage}[t]{.50\linewidth}
  \textbf{(B)}\\
      \includegraphics[width=1.1\linewidth, height=0.6\textwidth]{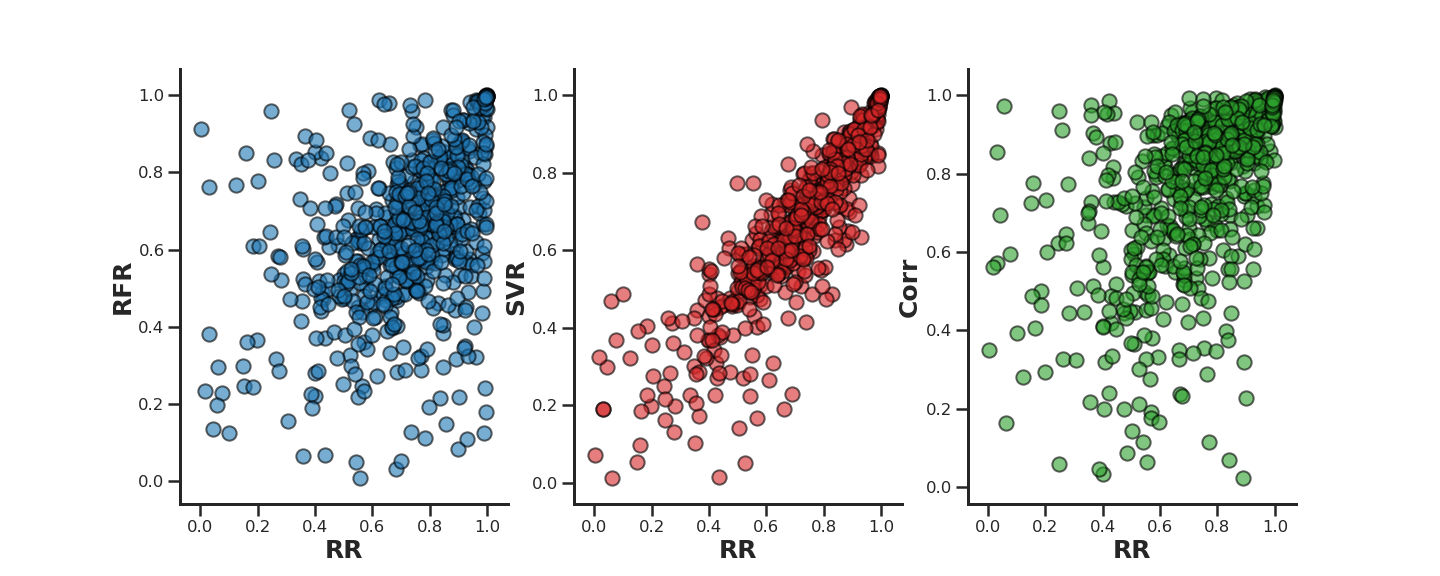}
  \end{minipage}
  \caption{Trans-eQTL target prediction performance on DREAM5 simulated data. \textbf{(A)} Boxplots show the distribution of AUROC values for all variants (blue) and for  trans-acting-only variants (red) for random forest regression (RFR), support vector regression (SVR), ridge regression (RR), and univariate correlation (Corr). \textbf{(B)} Scatter plots show AUROC values of classification methods RFR, SVR, and Corr vs RR for all variants. The data shown are for \textbf{DREAM Network 5}.}
  \label{fig:eqtl-prediction-dream5}
\end{figure}

\begin{figure}[t!]
  \begin{minipage}[t]{.50\linewidth}
  \textbf{A}\\
      \includegraphics[width=\linewidth]{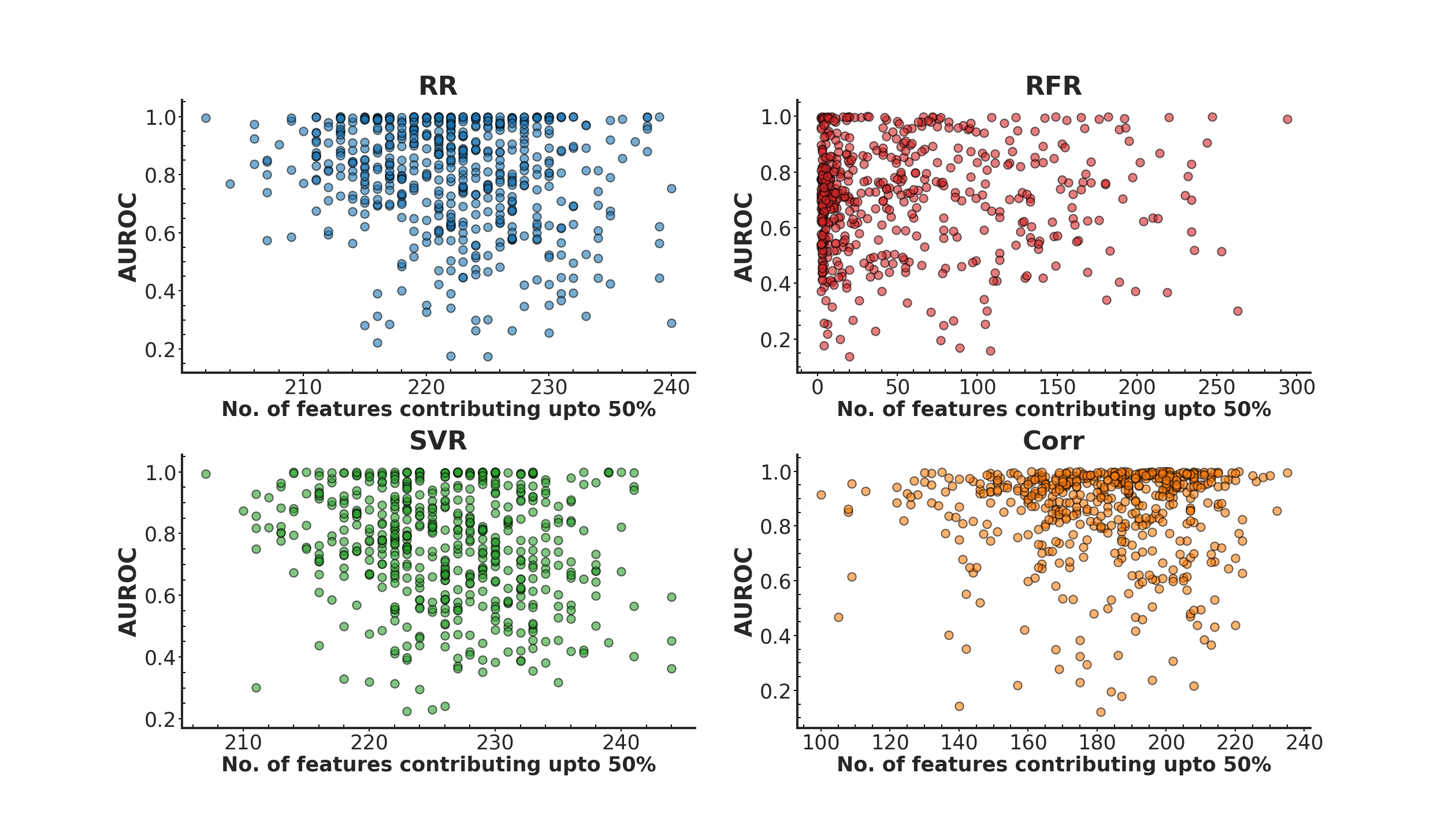}
  \end{minipage}
  \hfill
  \begin{minipage}[t]{.50\linewidth}
  \textbf{B}\\
      \includegraphics[width=\linewidth]{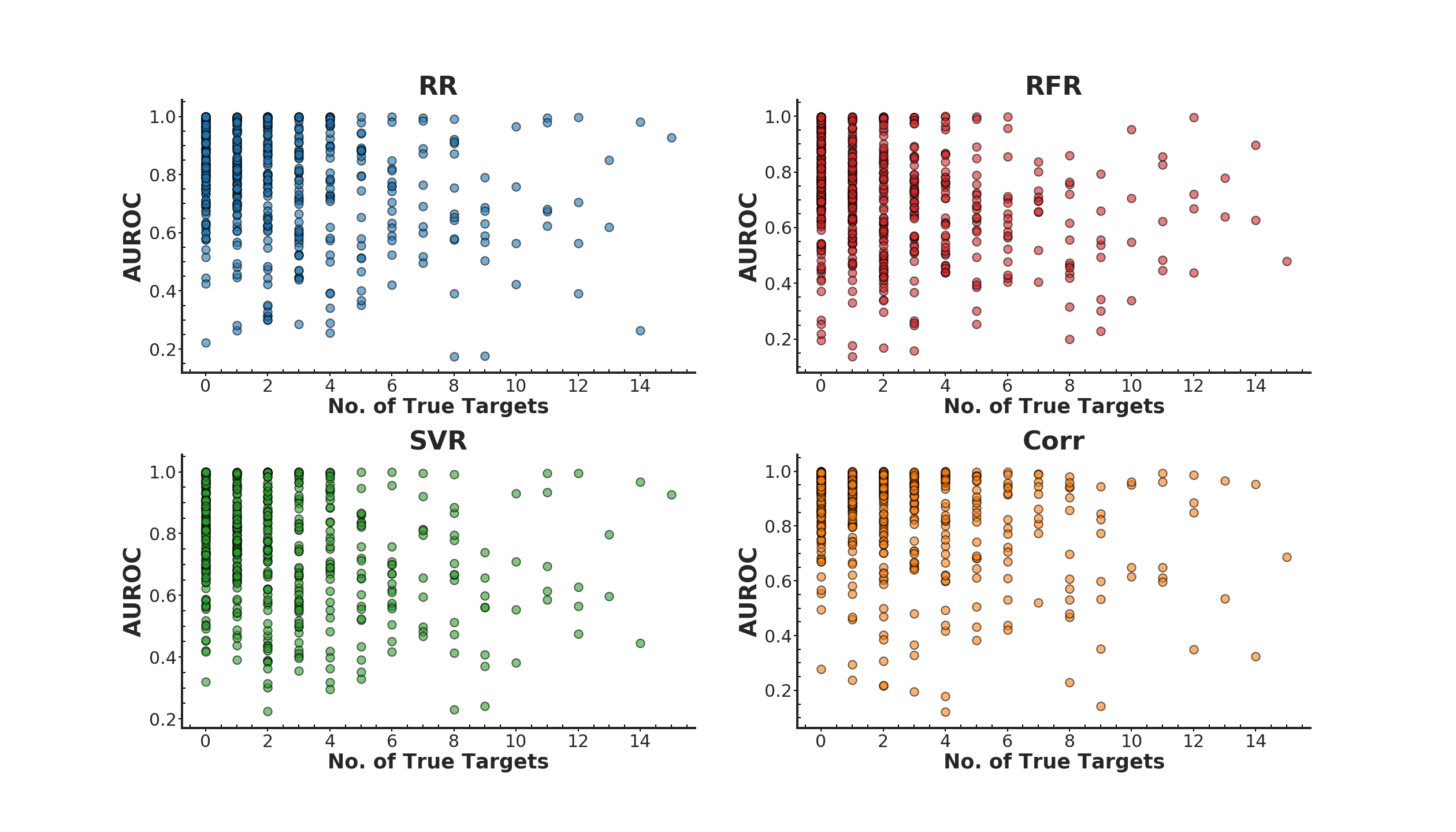}
  \end{minipage}
  \hfill
  \begin{center}
        \begin{minipage}[t]{.50\linewidth}
        \textbf{C}\\
      \includegraphics[width=\linewidth]{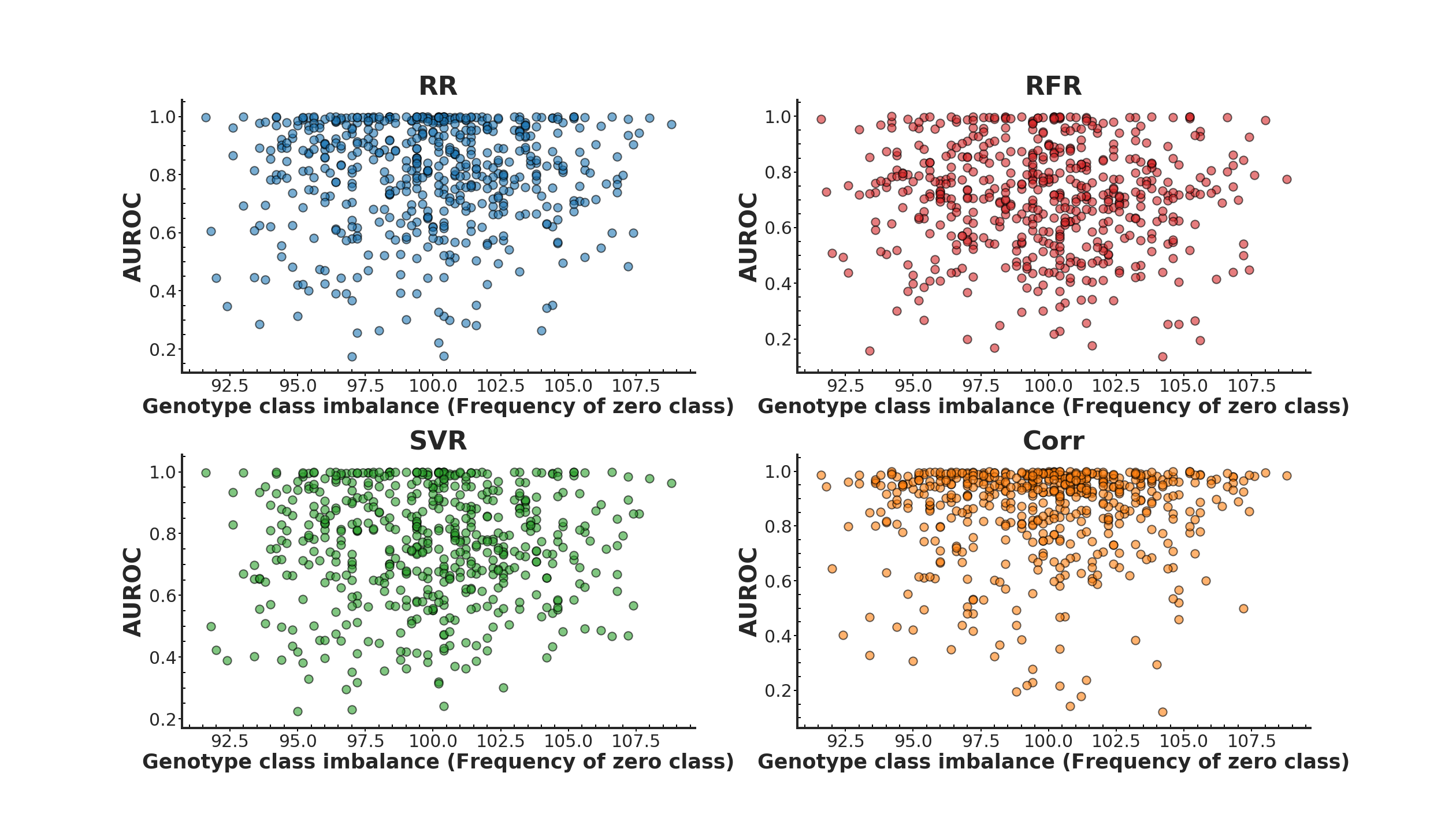}
  \end{minipage}
  \end{center}
  \caption{Scatter plots of trans-eQTL target prediction performance (AUROC) on DREAM5 simulated data against the number of selected model features (\textbf{A}), the number of true trans-eQTL targets in the ground-truth network (\textbf{B}), and the genotype class balance (frequency of the zero class) (\textbf{C}), for random forest regression (RFR), support vector regression (SVR), ridge regression (RR), and univariate correlation/naive Bayes (NB). The data shown are for \textbf{DREAM Network 2}.}
  \label{fig:eqtl-prediction-dream-control2}
\end{figure}

\begin{figure}[t!]
  \begin{minipage}[t]{.50\linewidth}
  \textbf{A}\\
      \includegraphics[width=\linewidth]{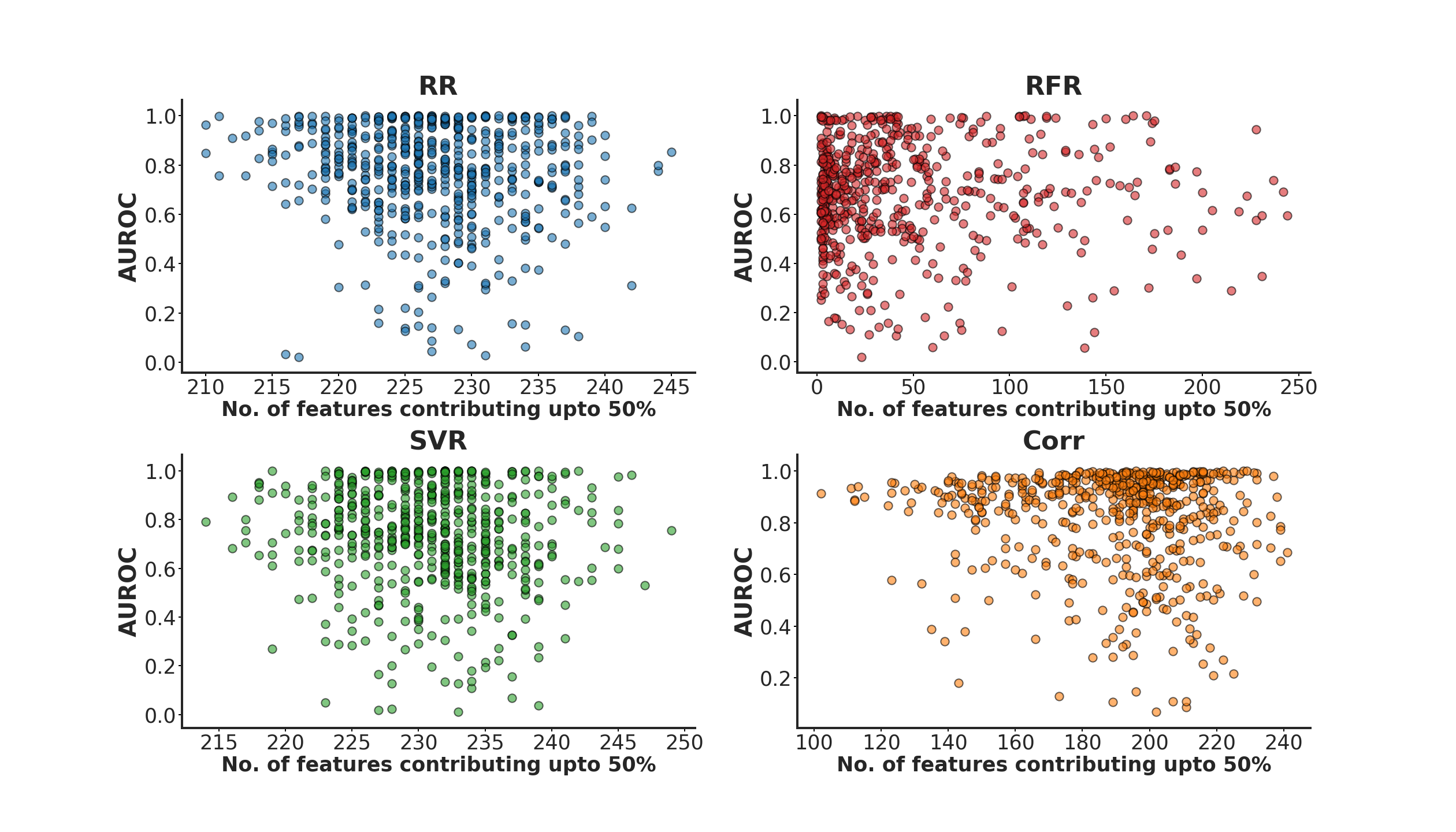}
  \end{minipage}
  \hfill
  \begin{minipage}[t]{.50\linewidth}
  \textbf{B}\\
      \includegraphics[width=\linewidth]{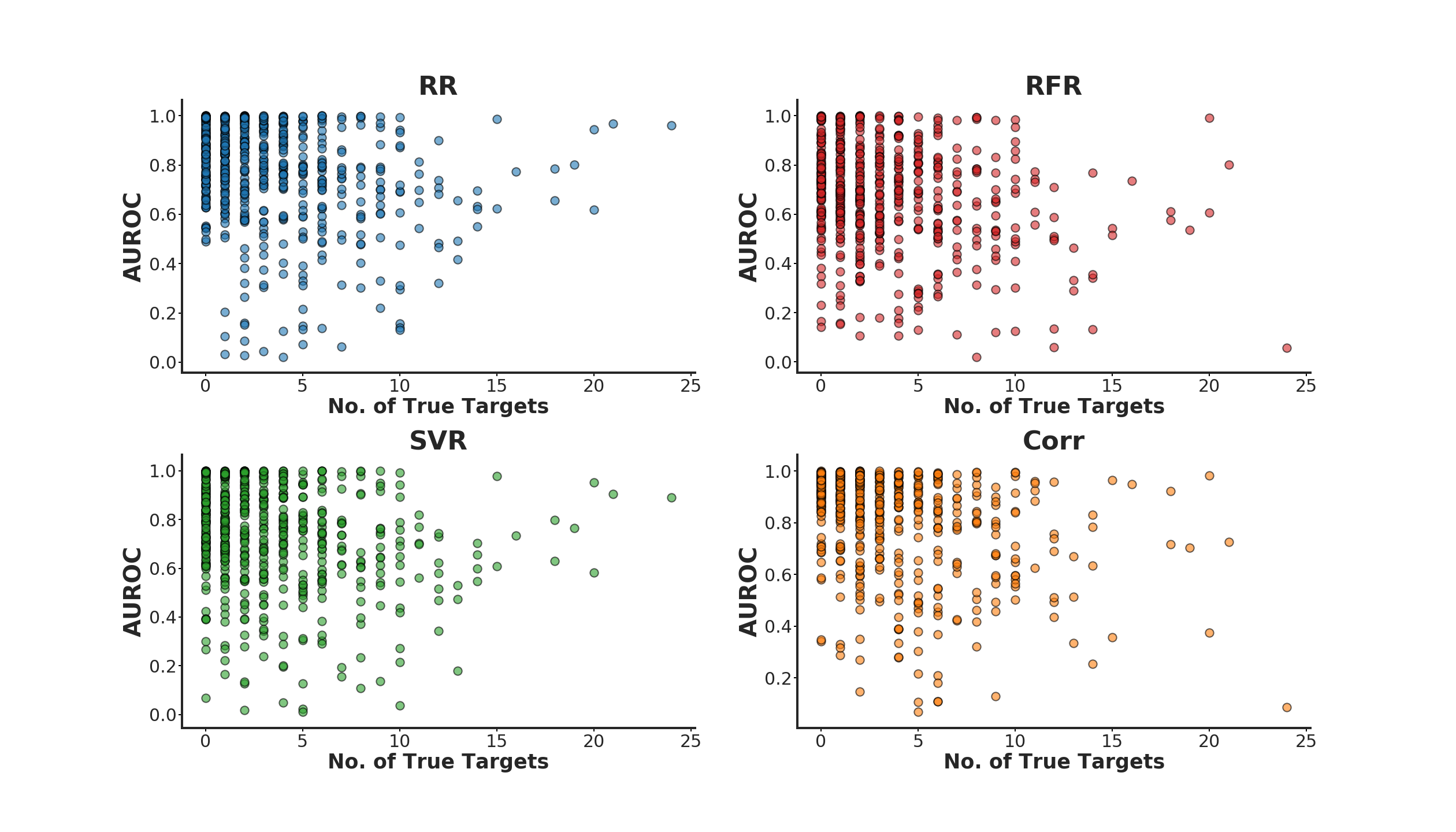}
  \end{minipage}
  \hfill
  \begin{center}
        \begin{minipage}[t]{.50\linewidth}
        \textbf{C}\\
      \includegraphics[width=\linewidth]{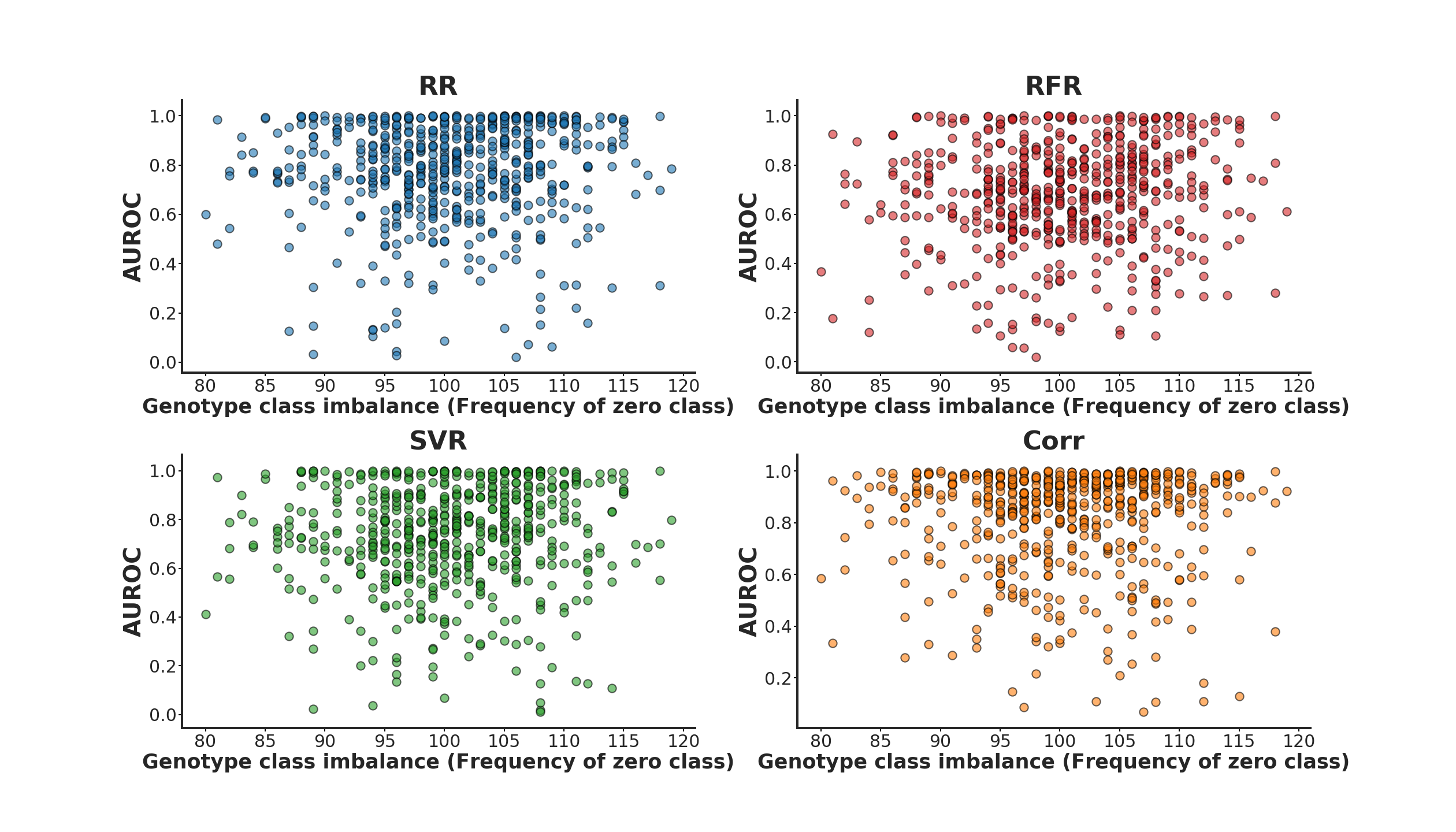}
  \end{minipage}
  \end{center}
  \caption{Scatter plots of trans-eQTL target prediction performance (AUROC) on DREAM5 simulated data against the number of selected model features (\textbf{A}), the number of true trans-eQTL targets in the ground-truth network (\textbf{B}), and the genotype class balance (frequency of the zero class) (\textbf{C}), for random forest regression (RFR), support vector regression (SVR), ridge regression (RR), and univariate correlation/naive Bayes (NB). The data shown are for \textbf{DREAM Network 3}.}
  \label{fig:eqtl-prediction-dream-control3}
\end{figure}

\begin{figure}[t!]
  \begin{minipage}[t]{.50\linewidth}
  \textbf{A}\\
      \includegraphics[width=\linewidth]{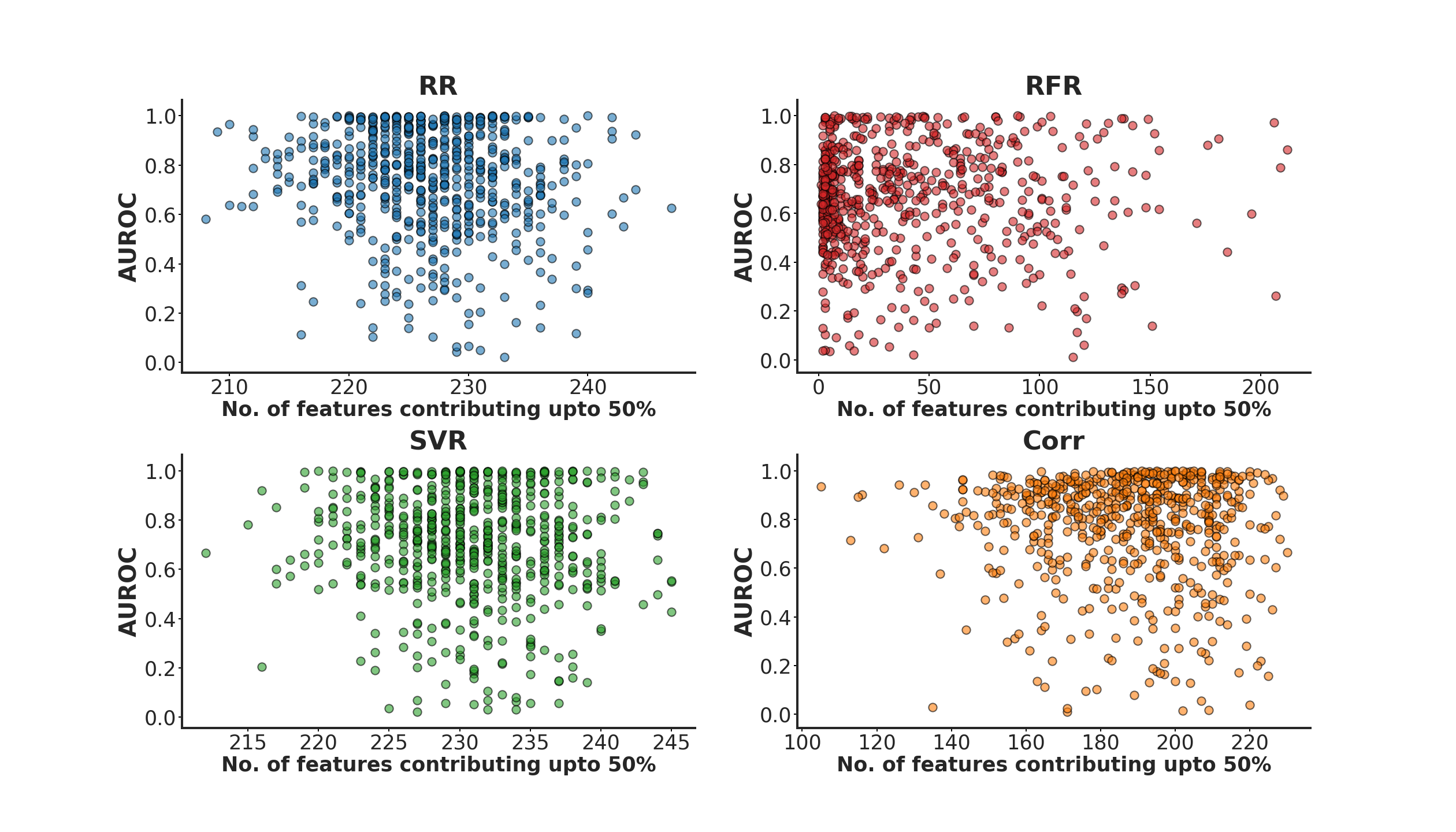}
  \end{minipage}
  \hfill
  \begin{minipage}[t]{.50\linewidth}
  \textbf{B}\\
      \includegraphics[width=\linewidth]{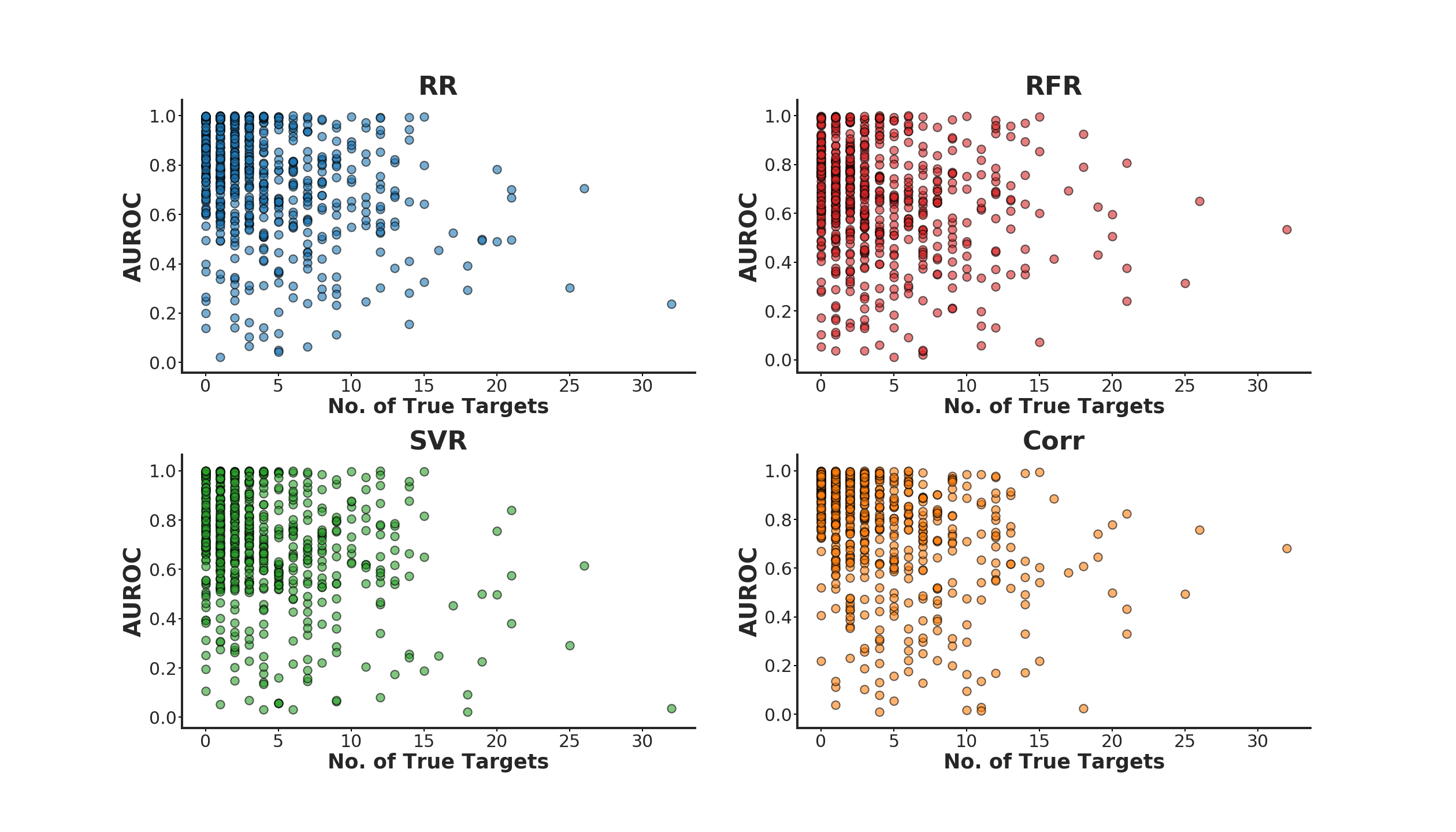}
  \end{minipage}
  \hfill
  \begin{center}
        \begin{minipage}[t]{.50\linewidth}
        \textbf{C}\\
      \includegraphics[width=\linewidth]{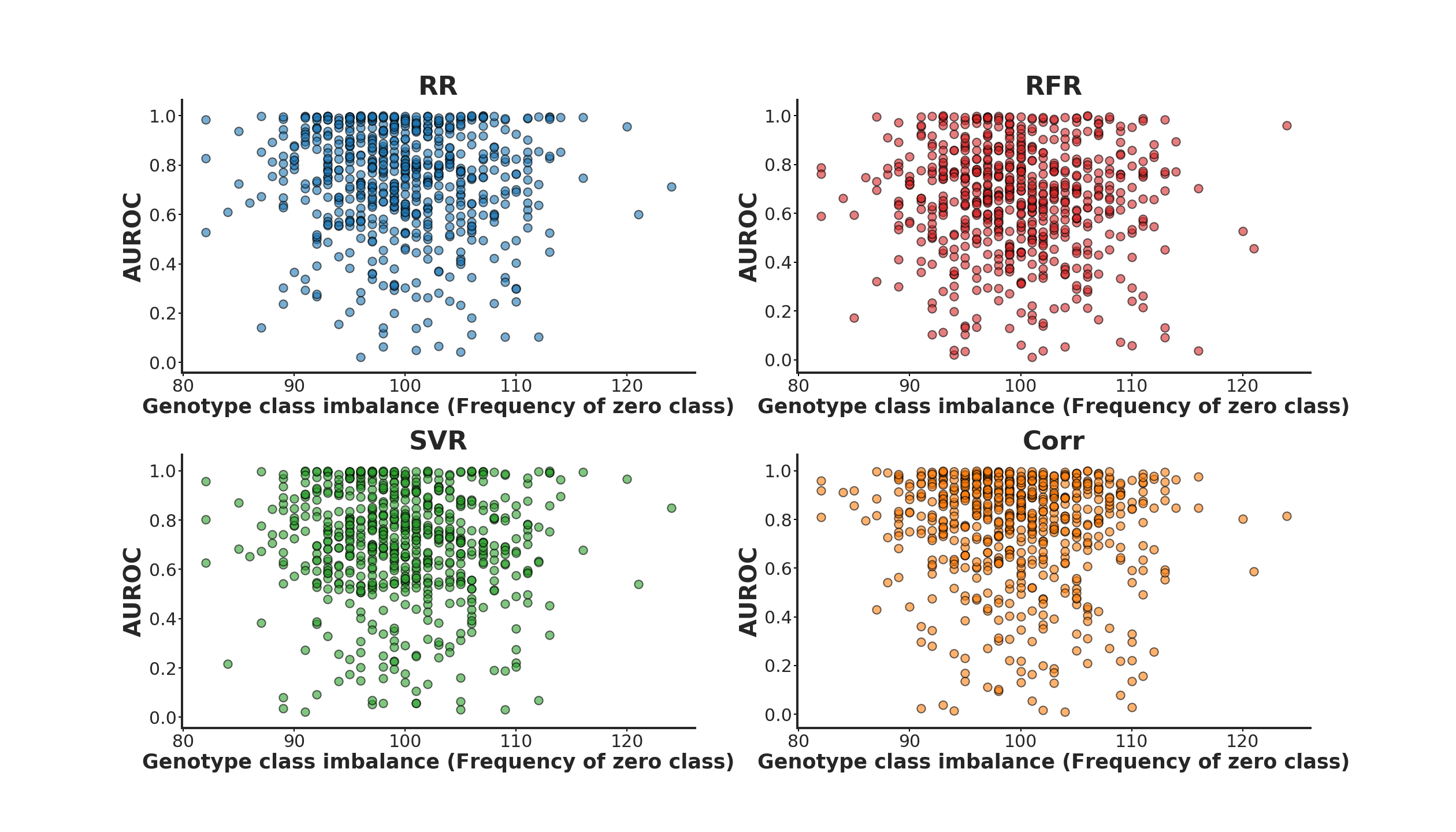}
  \end{minipage}
  \end{center}
  \caption{Scatter plots of trans-eQTL target prediction performance (AUROC) on DREAM5 simulated data against the number of selected model features (\textbf{A}), the number of true trans-eQTL targets in the ground-truth network (\textbf{B}), and the genotype class balance (frequency of the zero class) (\textbf{C}), for random forest regression (RFR), support vector regression (SVR), ridge regression (RR), and univariate correlation/naive Bayes (NB). The data shown are for \textbf{DREAM Network 4}.}
  \label{fig:eqtl-prediction-dream-control4}
\end{figure}

\begin{figure}[t!]
  \begin{minipage}[t]{.50\linewidth}
  \textbf{A}\\
      \includegraphics[width=\linewidth]{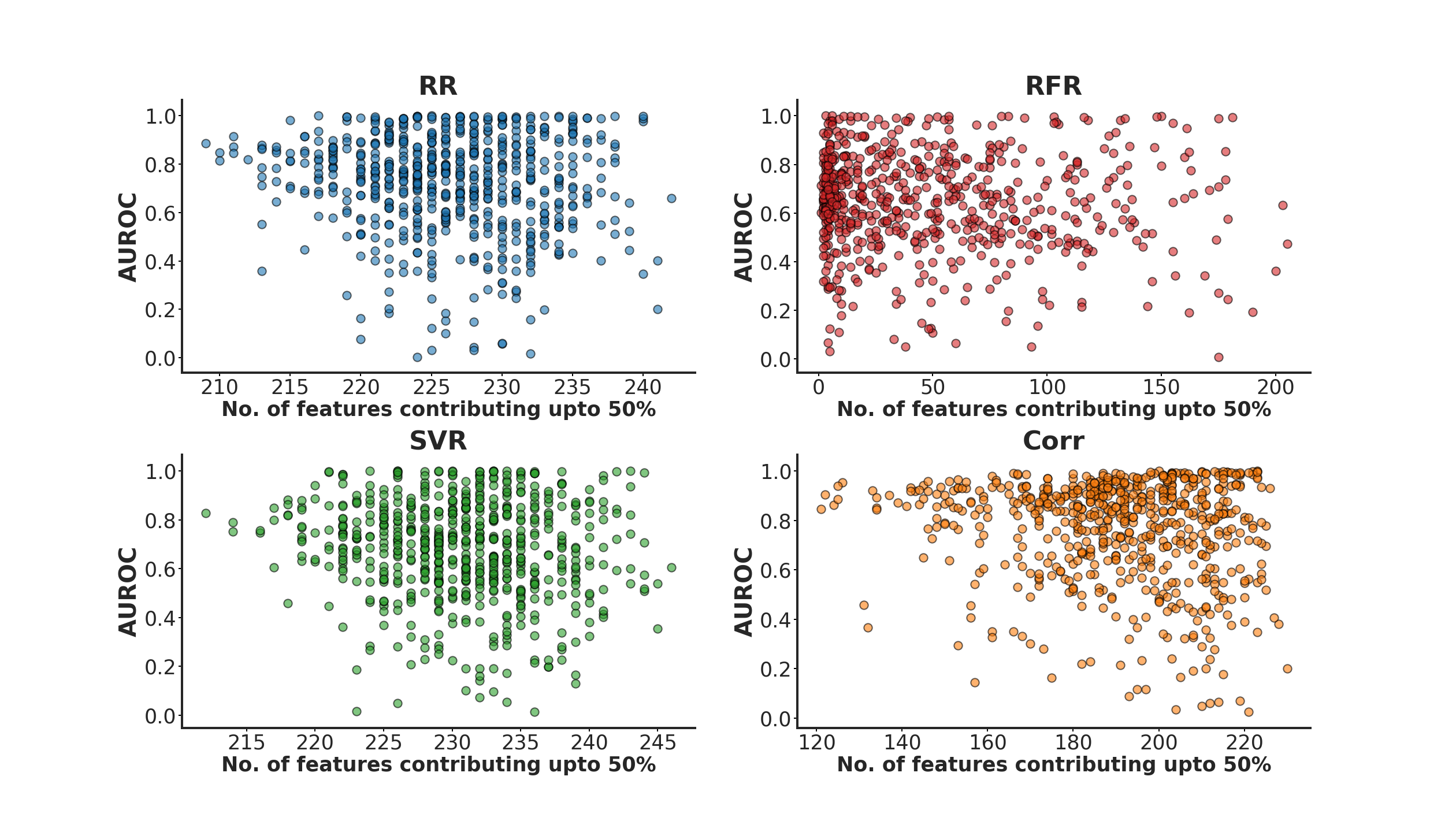}
  \end{minipage}
  \hfill
  \begin{minipage}[t]{.50\linewidth}
  \textbf{B}\\
      \includegraphics[width=\linewidth]{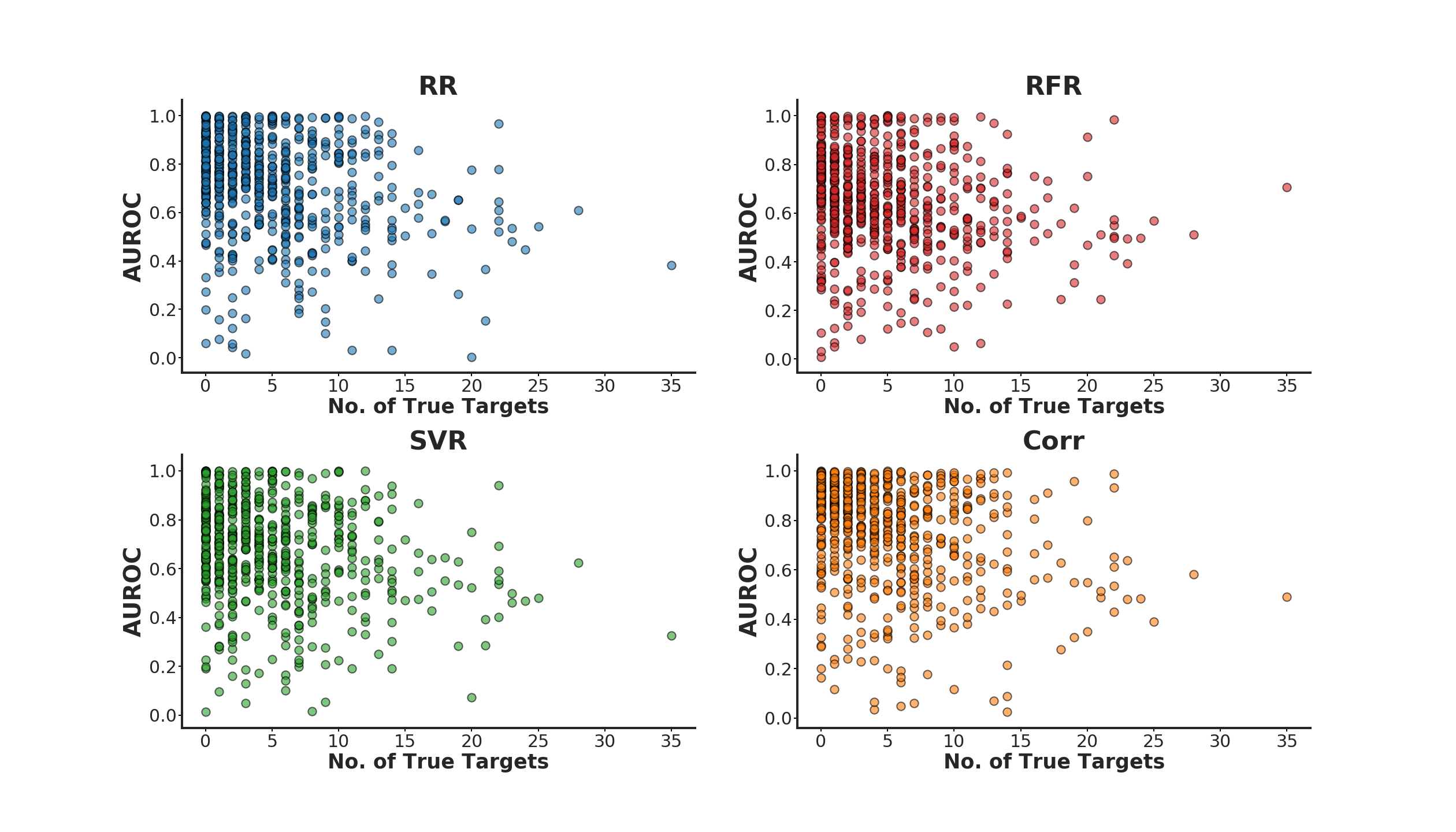}
  \end{minipage}
  \hfill
  \begin{center}
        \begin{minipage}[t]{.50\linewidth}
        \textbf{C}\\
      \includegraphics[width=\linewidth]{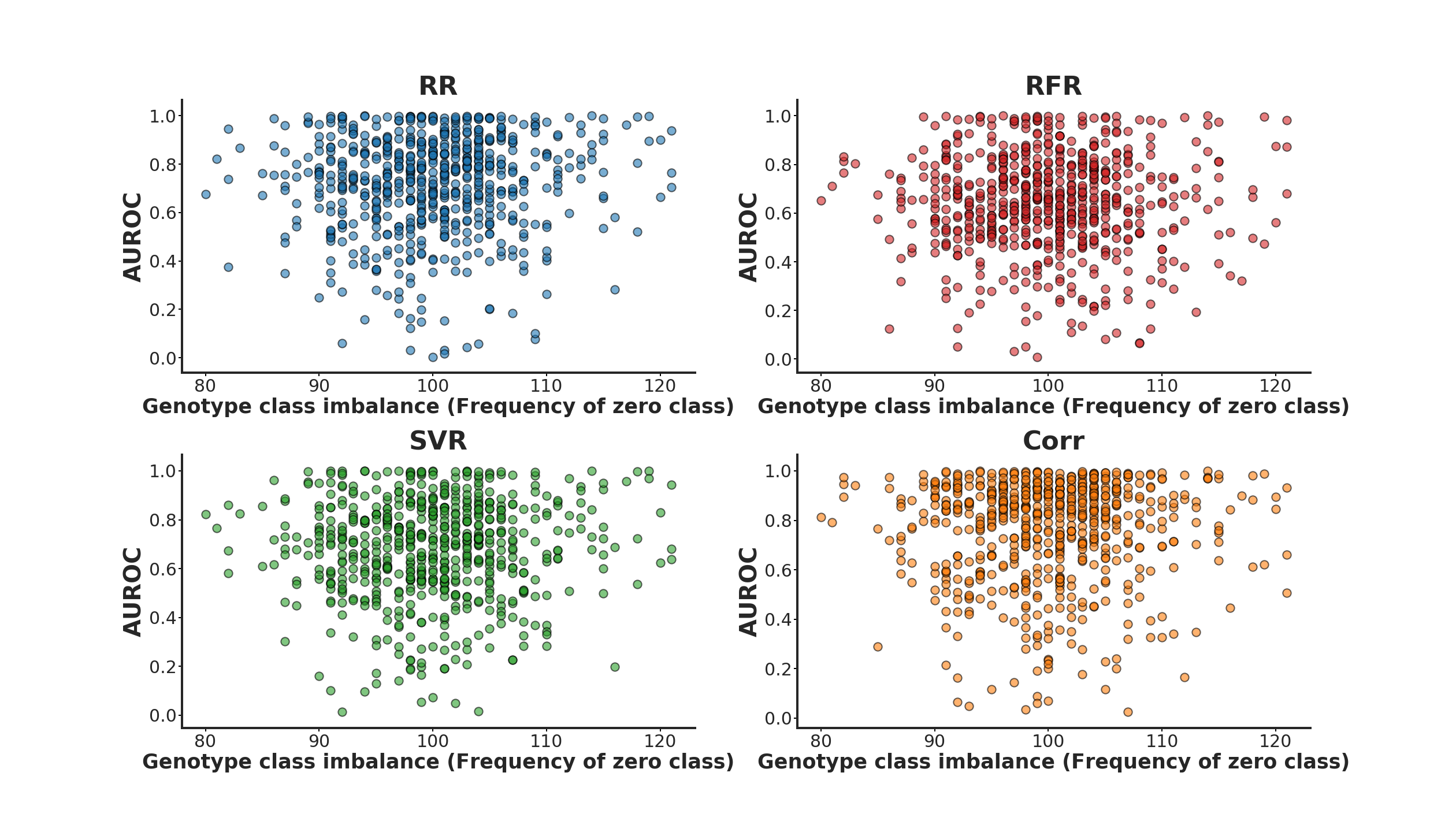}
  \end{minipage}
  \end{center}
  \caption{Scatter plots of trans-eQTL target prediction performance (AUROC) on DREAM5 simulated data against the number of selected model features (\textbf{A}), the number of true trans-eQTL targets in the ground-truth network (\textbf{B}), and the genotype class balance (frequency of the zero class) (\textbf{C}), for random forest regression (RFR), support vector regression (SVR), ridge regression (RR), and univariate correlation/naive Bayes (NB). The data shown are for \textbf{DREAM Network 5}.}
  \label{fig:eqtl-prediction-dream-control5}
\end{figure}

\begin{figure}[h!]
    \centering
    \includegraphics[width=1.00\linewidth]{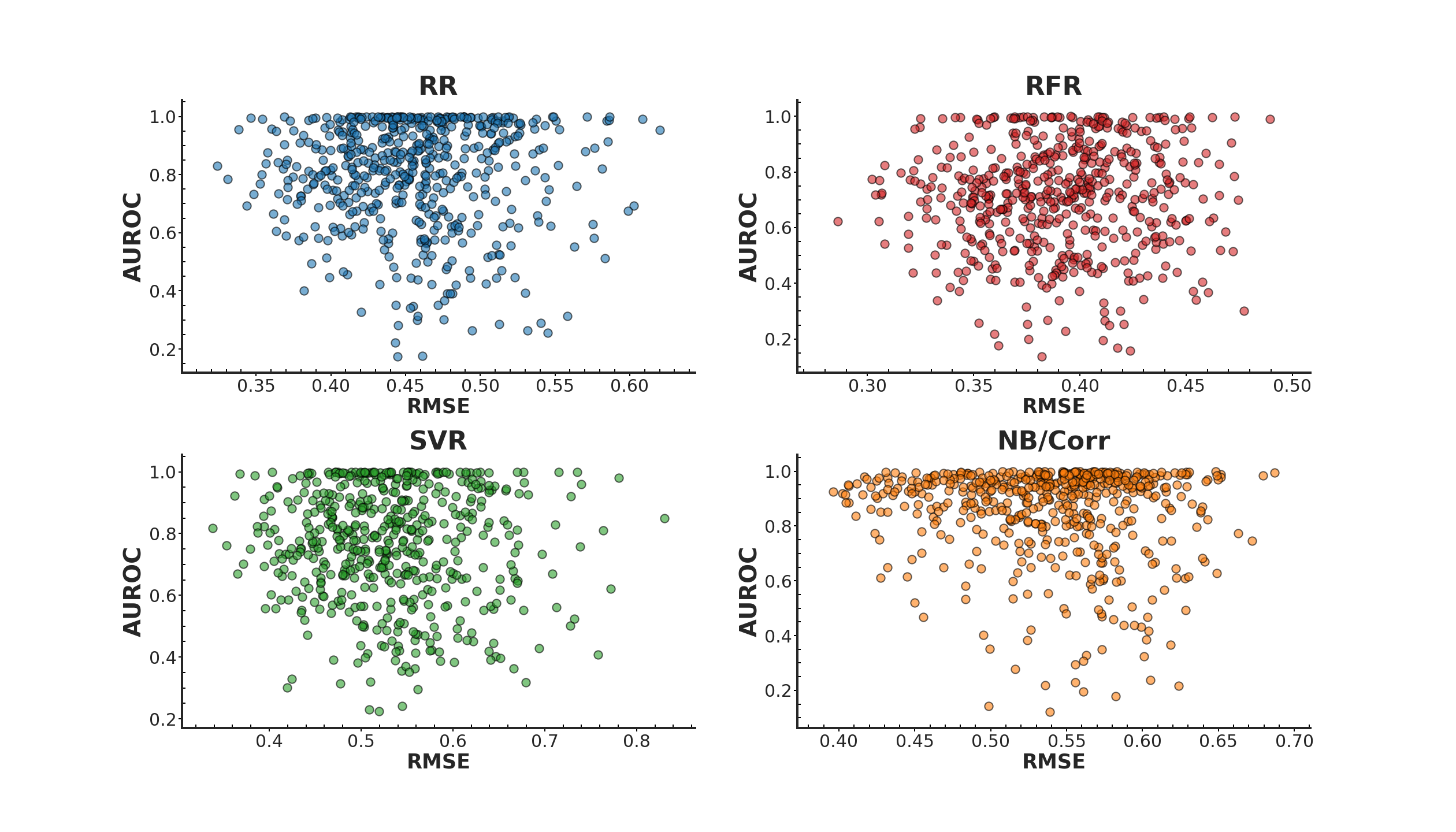}
    \caption{Scatter plots show trans-eQTL target prediction performance (AUROC) vs genotype prediction performance (RMSE) on DREAM5 simulated data for all genetic variants for random forest regression (RFR), support vector regression (SVR), ridge regression (RR), and univariate correlation/naive Bayes (NB/Corr). The data shown are for \textbf{DREAM Network 2}.}
  \label{fig:genotype-target-comparison-dream2}
\end{figure}

\begin{figure}[h!]
    \centering
    \includegraphics[width=1.00\linewidth]{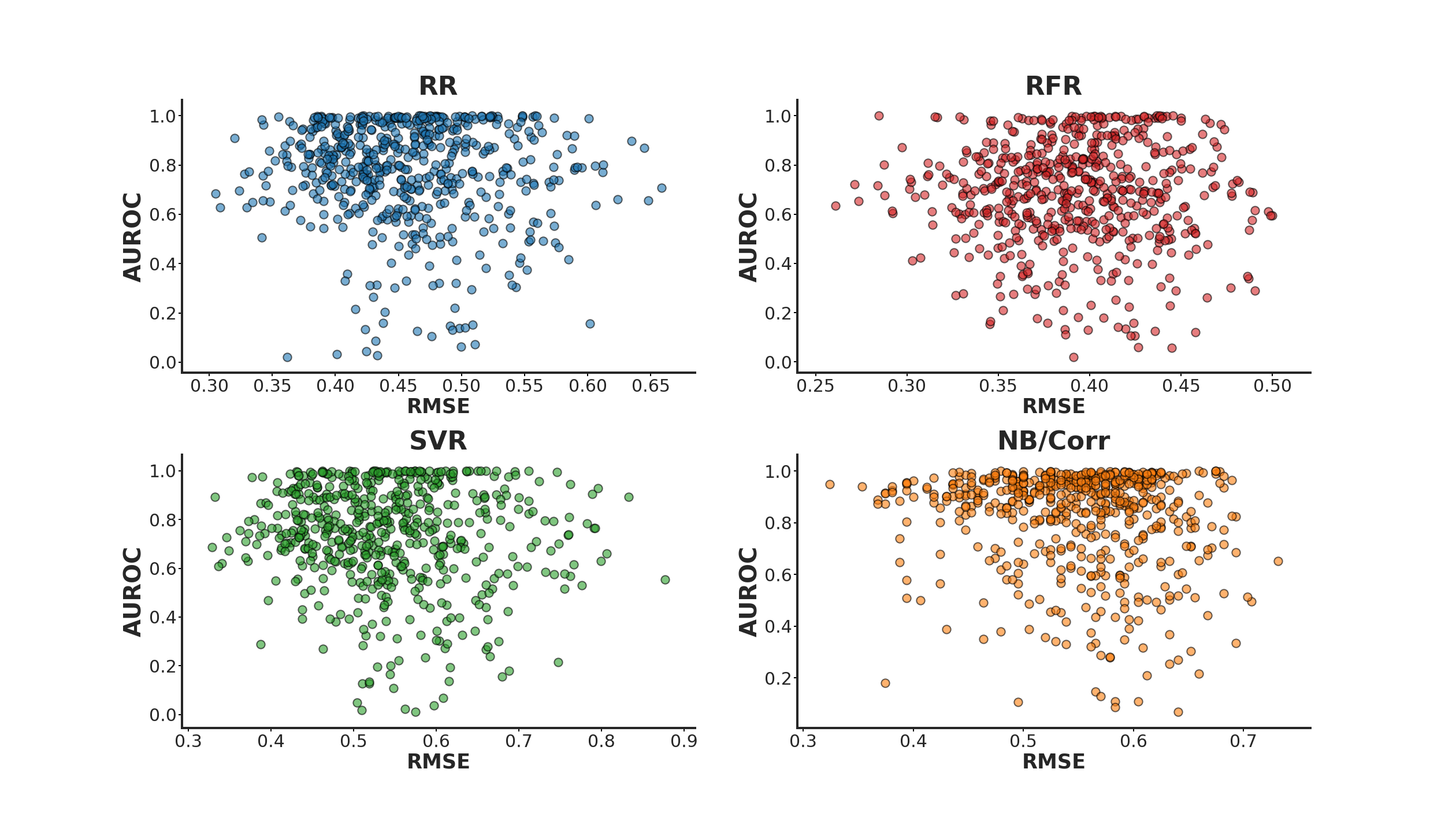}
    \caption{Scatter plots show trans-eQTL target prediction performance (AUROC) vs genotype prediction performance (RMSE) on DREAM5 simulated data for all genetic variants for random forest regression (RFR), support vector regression (SVR), ridge regression (RR), and univariate correlation/naive Bayes (NB/Corr). The data shown are for \textbf{DREAM Network 3}.}
  \label{fig:genotype-target-comparison-dream3}
\end{figure}

\begin{figure}[h!]
    \centering
    \includegraphics[width=1.00\linewidth]{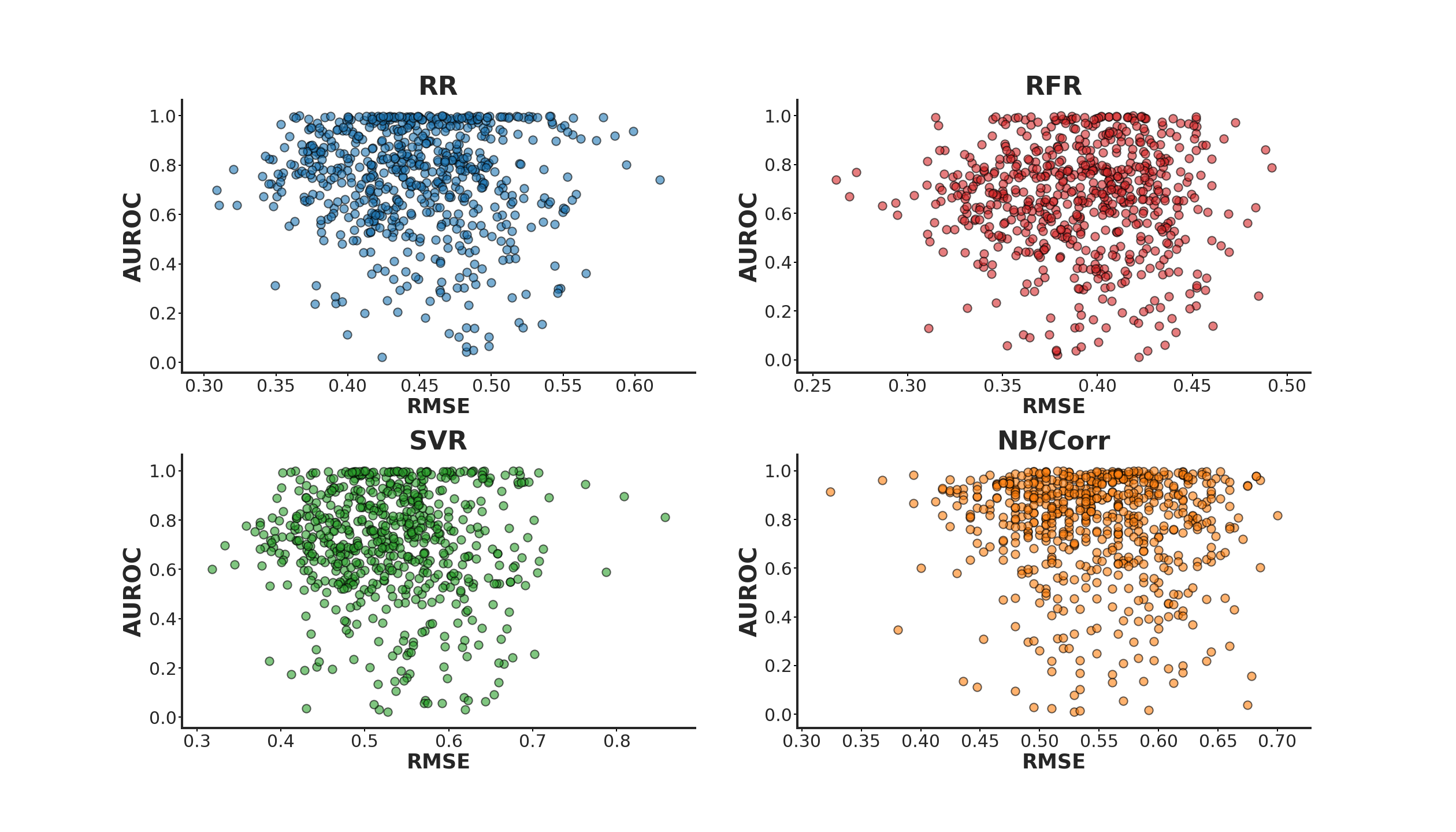}
    \caption{Scatter plots show trans-eQTL target prediction performance (AUROC) vs genotype prediction performance (RMSE) on DREAM5 simulated data for all genetic variants for random forest regression (RFR), support vector regression (SVR), ridge regression (RR), and univariate correlation/naive Bayes (NB/Corr). The data shown are for \textbf{DREAM Network 4}.}
  \label{fig:genotype-target-comparison-dream4}
\end{figure}

\begin{figure}[h!]
    \centering
    \includegraphics[width=1.00\linewidth]{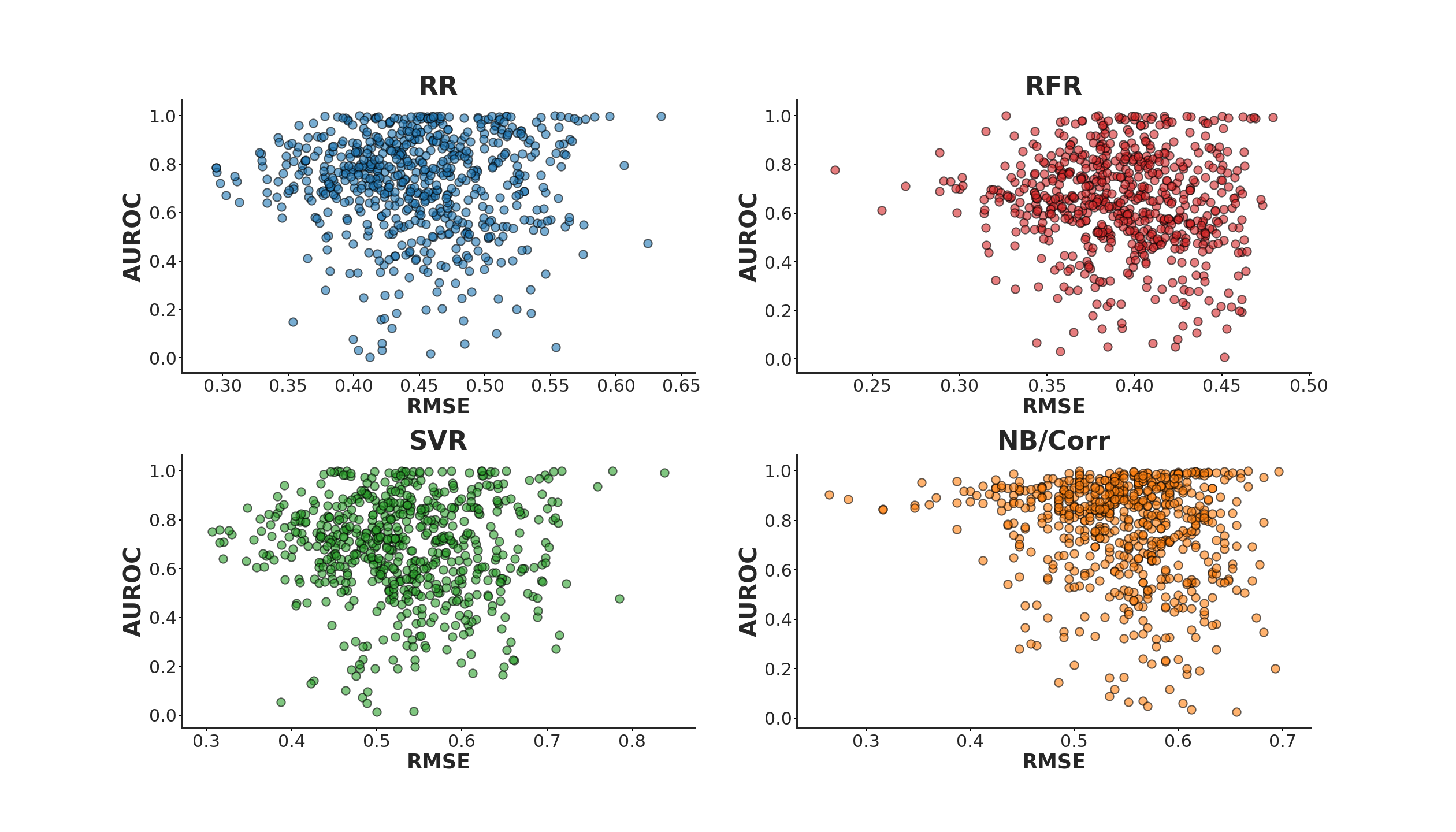}
    \caption{Scatter plots show trans-eQTL target prediction performance (AUROC) vs genotype prediction performance (RMSE) on DREAM5 simulated data for all genetic variants for random forest regression (RFR), support vector regression (SVR), ridge regression (RR), and univariate correlation/naive Bayes (NB/Corr). The data shown are for \textbf{DREAM Network 5}.}
  \label{fig:genotype-target-comparison-dream5}
\end{figure}

\newgeometry{}
\begin{figure}[h!]
  \centering
  \includegraphics[width=1.00\linewidth]{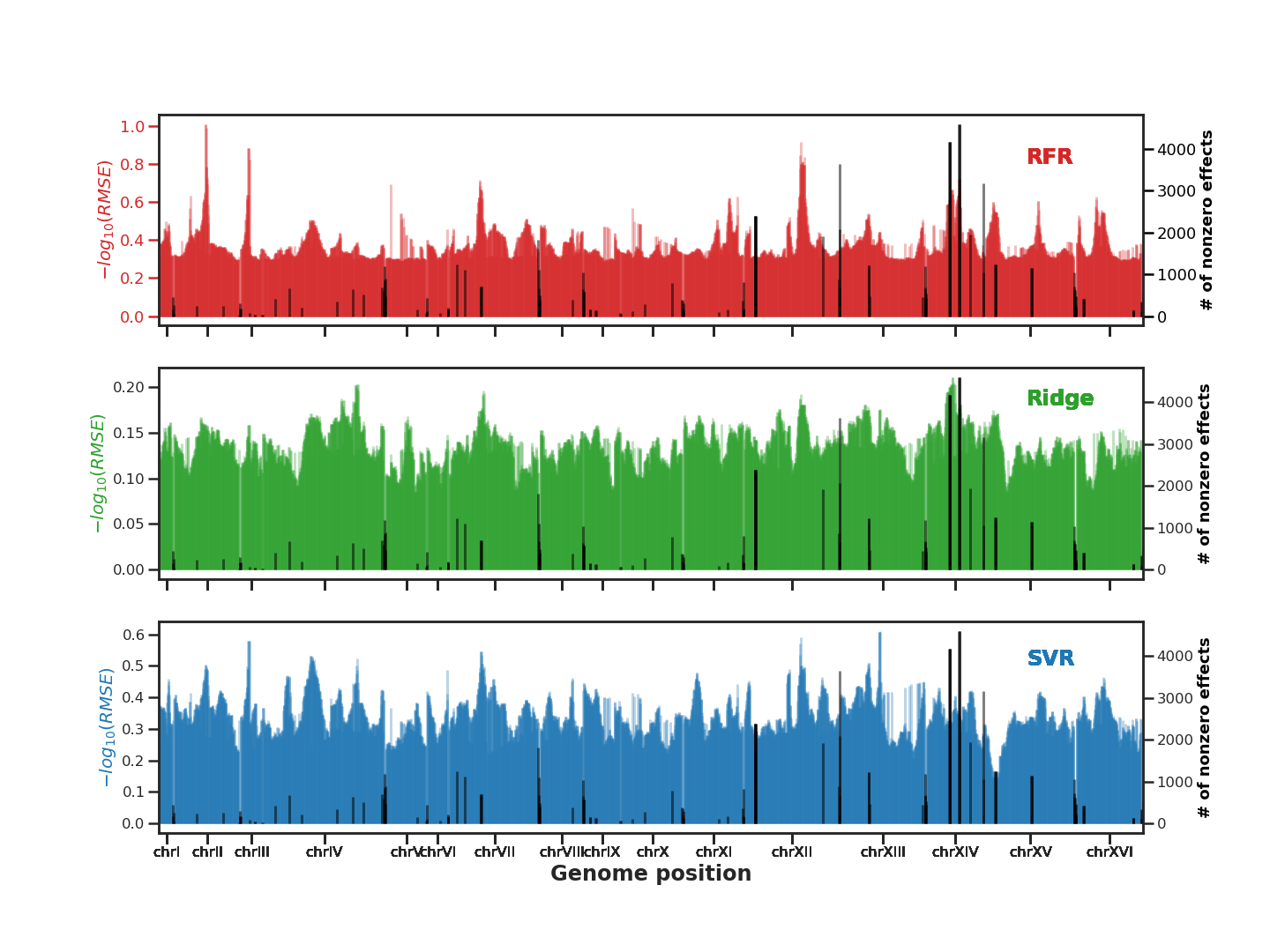}
  \caption[Expression hotspot maps SUP] {Expression hotspot maps showing the negative log transformed RMSE values vs genome position for 2884 SNPs in the yeast genome, for random forest (RF, top), ridge regression (Ridge, middle), and support vector regression (SVR, bottom). Genes on the same chromosome were excluded as predictors for each SNP.  Secondary axis on right shows number of non-zero effects of trans-regulatory hotspot variants from Albert et al. (2018)\footnotemark[2].}
  \label{fig:hotspots-rmse-yeastv2}
\end{figure}

\footnotetext[2]{Albert, F. W. et al. (2018). Genetics of trans-regulatory variation in gene expression. Elife, 7, e35471}

\restoregeometry

\end{document}